\documentclass[a4paper,reqno,oneside]{amsart}
\usepackage{amsmath,amsthm,amsfonts}
\usepackage{mathrsfs}
\usepackage[T1]{fontenc}
\usepackage[english]{babel}
\usepackage{amssymb,bbm,enumerate}
\usepackage[linktocpage=true,colorlinks=true, linkcolor=blue, citecolor=red, urlcolor=green]{hyperref}

\usepackage{MnSymbol} 

\usepackage[svgnames]{xcolor}

\usepackage{tikz}
\usepackage{tikz-cd}
\usepackage{pgfplots}
\usepackage{graphicx}
\usepackage{float}

\usetikzlibrary{decorations.markings}
\usetikzlibrary{hobby}

\usepackage{subcaption}

\usepackage{color}

\usepackage{soul}

\usepackage{float}
\restylefloat{table}

\newcommand{\R}{\mathbb{R}}
\newcommand{\C}{\mathbb{C}}
\newcommand{\Cb}{\overline{\mathbb{C}}}
\newcommand{\Cs}{\mathbb{C}^*}
\newcommand{\WCs}{\widetilde{\mathbb{C}^*}}
\newcommand{\N}{\mathbb{N}}
\newcommand{\Z}{\mathbb{Z}}
\newcommand{\Q}{\mathbb{Q}}
\newcommand{\la}{\lambda}

\newcommand{\mc}[1]{\mathcal{#1}}

\newcommand{\uu}{\underline{u}}

\newcommand{\uo}{\underline{0}}
\newcommand{\e}{\varepsilon}

\theoremstyle{plain}
\newtheorem{theorem}{Theorem}[section]
\newtheorem{lemma}[theorem]{Lemma}

\newtheorem{proposition}[theorem]{Proposition}
\newtheorem{corollary}[theorem]{Corollary}

\newtheorem{claim}{Claim}

\theoremstyle{definition}
\newtheorem{definition}[theorem]{Definition}

\theoremstyle{remark}
\newtheorem{remark}[theorem]{Remark}
\newtheorem{notation}{Notation}[section]

\title{A primer of the complex WKB method,  \\ With application to the ODE/IM correspondence}

\author{Gabriele Degano and Davide Masoero}
\address{Grupo de F\'isica Matem\'atica,
	Departamento de Matemática do Instituto Superior Técnico, Lisboa. \newline
	Departamento de Matemática da Faculdade de Ciências da Universidade de Lisboa}
\email{gabriele.degano.gd@gmail.com}

\address{Instituto Superior de Agronomia, Universidade de Lisboa \newline 
Grupo de F\'isica Matem\'atica,
Departamento de Matemática do Instituto Superior Técnico, Lisboa. 
}
\email{dmasoero@gmail.com}

\begin{document}

\pagestyle{plain}

\begin{abstract}
In these lectures, we provide an introduction to the complex WKB method, using as a guiding example a class of anharmonic oscillators that appears in the ODE/IM correspondence.
In the first three lectures, we introduce the main objects of the method, such as the WKB function, the integral equations of Volterra type, the quadratic differential and its horizontal/Stokes lines, the Stokes phenomenon, the notion of asymptotic values, the Fock-Goncharov coordinates and their WKB approximation. In the fourth and last lecture, we compute (and prove) the asymptotic behaviour of the spectrum of the anharmonic oscillators in two asymptotic regimes, when the momentum is fixed and the energy is large, and when the momentum (hence also the energy) is large.
\end{abstract}

\maketitle

\tableofcontents


\section*{Introduction}

The aim of these lectures is to introduce the reader to the complex WKB method, as a tool to deduce and prove asymptotic properties of solutions to linear differential equations in the complex plane.

In order to make our study concrete and innovative, we apply the complex WKB method to a specific family of differential equations, which are of independent interest in mathematical physics and of non-trivial nature (at least in the sense that the general solution cannot be obtained using known special functions), and for which the results we are going to obtain are not available in the mathematics literature.

The equations under study is 
\begin{equation}
	\label{eq:anharmonic}
	\frac{d^2\psi(x)}{dx^2}= U(x;E ,\ell) \psi, \quad U(x;E ,\ell)=
	x^{2 \alpha}+\frac{\ell(\ell+1)}{ x^2}-E, \quad x\in\WCs.
\end{equation}
The above equation is a Schr\"odinger equation in the complex plane with a potential $U$ depending on three parameters
$E,\ell,\alpha$. $E,\ell$ are complex parameters called the \textit{energy} and the \textit{angular momentum} \footnote{Since the parameter $\ell$ appears only in the expression $\ell(\ell+1)$, we restrict -- without loss in generality -- to the half-plane
$\Re \ell\geq - \frac12$.} and $\alpha>0$ is a positive real number, called the \textit{anharmonicity degree}, which we assume to have fixed. 
This family anharmonic oscillators appeared in the ODE/IM correspondence in relation with the ground state of the Quantum KdV model (see Appendix II for more details).

The independent variable $x$ takes value in
the universal cover of the punctured plane $\Cs$, which we denote by $\WCs$.
This is the natural domain of the coefficients of the equation because $U$ is multivalued on $\Cs$, whenever $2 \alpha$ is not a natural number.
Notice that $\WCs$ is also the natural domain for the solutions to the anharmonic oscillator \eqref{eq:anharmonic} even in the case $2 \alpha \in \N$, since the domain of the solutions to a linear ODE in the complex plane is the universal cover of the domain of the coefficients of the equation. 

In studying a linear ODE in the complex plane, the first question we must ask ourselves is about the nature and location of the singularities.
Assume for the moment that $2 \alpha$ is a natural number. In this case, $U$ is a meromorphic potential on the Riemann sphere, which we denote by $\Cb$, and it has two singular points $0$ and $\infty$.
\begin{itemize}
 \item[i)] The point $x=0$ is a regular singularity, and it has Frobenius indices $\ell+1$ and $-\ell$. This implies that there exists a basis
 of the space of solutions expressed in terms of Frobenius expansions:
$\chi_+=x^{\ell+1} \left(1+\sum_{n \geq 1} c_n x^n \right), x \to 0$ (subdominant Frobenius solution) and  $\chi_-= x^{-\ell} \left(1+\sum_{n \geq 1} d_n x^n \right)$ (dominant Frobenius solutions \footnote{The dominant solution must be possibly modified with a logarithmic term whenever the equation is resonant, namely $2 \ell+ 1 \in \N$}).
\item[ii)] On the contrary, $x=\infty$  is an irregular singularity. For any sector of the form $\Sigma_{k}=\{ |\arg x-\frac{ \pi \,k }{\alpha+1}| < \frac{\pi}{2\alpha+2}, \; k=0, \dots 2\alpha+1 \}$, called a Stokes sector, there exists a unique, up to a multiplicative constant, non-trivial solution $\Psi_k$ which converges to $0$ exponentially fast as $|x|\to \infty$ in any closed subsector of $\Sigma_k$. Moreover, any solution that is not a multiple of $\Psi_k$ diverges, exponentially fast, in any closed subsector of $\Sigma_k$.
In particular, the Sibuya solution $\Psi_0$ is (up to a multiplicative constant) the only solution vanishing as $x \to +\infty$.
\end{itemize}
The above picture can be extended to $\alpha>0$, as we will show in the Lecture 2, for any given $\alpha>0$: 
We can define the solution $\chi_+(x;E,\ell) \sim x^{\ell+1}$, subdominant at $0$, and
the Sibuya solution $\Psi_k(x;E,\ell)$ subdominant in the sector $\Sigma_k=\{ x \in \WCs |\arg x-\frac{ \pi \,k }{\alpha+1}| < \frac{\pi}{2\alpha+2} \}$, for any $k \in \Z$.

Our main interest is the \textbf{spectrum} of equation \eqref{eq:anharmonic}.
\begin{definition}
 The spectrum of the anharmonic oscillator is the subset of the space  of parameters $(E,\ell) \in   \C \times \{ \Re \ell\geq -\frac12 \}$ such that
 $\chi_+(x;E,\ell)$ and
$\Psi_0(x;E,\ell)$ are linearly dependent, namely if the solution subdominant as $x \to 0^+$ is also subdominant as $x \to +\infty$.
\end{definition}
Recall that the two solutions $f(x),g(x)$ to a linear second order ODE (namely a stationary Schr\"odinger equation in one dimension) are proportional if and only if their Wronskian $Wr_x[f(x),g(x)]$
vanishes, where 
\begin{align*}
& Wr_x[f(x),g(x)]=f(x)g'(x)-f'(x) g(x).
\end{align*}
It follows that the spectrum of equation \eqref{eq:anharmonic} is the zero locus of the Wronskian
of $\chi_+(x;E,\ell)$ and
$\Psi_0(x;E,\ell)$, which we denote by $Q_+(E,\ell)$, and it is called the \textbf{spectral determinant},
\begin{align} \nonumber
& Q_+(E,\ell)= Wr_x [\chi_+(x;E,\ell),\Psi_0(x;E,\ell)].
\end{align}

We are interested in studying the spectrum of equation \eqref{eq:anharmonic} in two asymptotic regimes
\begin{enumerate}
 \item The first asymptotic regime is defined by letting  $E \to \infty$ with $\ell$ fixed.
 \item The second asymptotic regime is defined by letting $E,\ell \to +\infty$  in such a way that
$E (\ell+\frac12)^{-\frac{2\alpha}{\alpha+1}} \to \nu \in (0,\infty)$.
\end{enumerate}
In order to make these two regimes \textit{visually intuitive}, we can introduce a small parameter,  which we denote by $\hbar$:
\begin{enumerate}
 \item In the first regime, letting
 \begin{equation}\label{eq:change1}
    x= \hbar^{-\frac{1}{\alpha+1}} \, y, \; E= \hbar^{-\frac{\alpha+1}{2\,\alpha}}
 \end{equation} 
 we obtain the equation
\begin{equation}
\label{eqn:intro_rescaled-fixed-l-infinite-e}
	\psi''(y)= \left( \hbar^{-2} \left( y^{2\alpha}-1 \right) + \frac{\ell(\ell+1)}{y^2} \right) \psi(y), \quad \hbar \to 0.
	\end{equation}
	\item In the second regime,  letting
    \begin{equation}\label{eq:change2}
    x= \hbar^{-\frac{1}{\alpha+1}} y , \; \hbar= \left(\ell+\frac12\right)^{-1}, \; \nu=E \left(\ell+\frac12\right)^{-\frac{2\alpha}{\alpha+1}},
 \end{equation} 
we obtain the equation
\begin{align}
\label{eqn:intro_rescaled-infinite-e-l}
 &	\psi''(y)  = \widetilde{U}(y;\nu,\hbar)  \psi(y), \quad \hbar \to 0 \\ \nonumber
 & \widetilde{U}(y;\nu,\hbar)=\left( \hbar^{-2} \left( y^{2\alpha}-\nu+ \frac{1}{y^2} \right) - \frac{1}{4\,y^2} \right) \psi(y).
\end{align}
\end{enumerate}
We could thus consider (\ref{eqn:intro_rescaled-fixed-l-infinite-e},\ref{eqn:intro_rescaled-infinite-e-l}) as the equations under study; however,
it is far more convenient to develop a general theory for \eqref{eq:anharmonic}, before specialising to an asymptotic regime.

\begin{remark}
When we fix the angular momentum $\ell$, the spectrum is the zero locus of a (non-constant) entire function of the energy, hence it is a discrete subset of the energy plane. In particular, in the case $\ell >-\frac12$, the spectrum coincides with the $L^2$ spectrum of a self-adjoint operator obtained from $\mathcal{L}=-\partial_x+x^{2\alpha}+ \frac{\ell(\ell+1)}{x^2}$, hence it is real; moreover, any point $E$ in the spectrum is strictly greater than $E_*=\alpha^{-\frac{\alpha}{1+\alpha}}(1+\alpha) \left(\ell+\frac12\right)^{\frac{2\alpha}{1+\alpha}}$ -- more details in Appendix II.
\end{remark}

\begin{remark}
When $\alpha=1$, we can compute the spectral determinant and the spectrum explicitly \cite[Chapter 10.15]{bateman2}:
One can choose a normalization of the subdominant solutions $\chi_+$ and $\Psi_0$ such that 
\begin{equation}
 Q_+(E,\ell)=\frac{1}{\Gamma\left(\frac{-E+2 \ell+ 3 }{4}\right)} 
\end{equation}
Since $1/\Gamma(-z)$ is an entire function whose zero locus coincides with  $\N=\{0,1,2, \dots\}$, then the spectrum is the disjoint union of the lines $E_n(\ell)= 4n + 3 + 2 \ell , n \in \N$. 
\end{remark}

\subsubsection*{Plan of the lectures}
The lectures are designed to be essentially self-contained, for a reader who has a good knowledge of complex analysis.

The main object of these lectures is the approximation of solutions to the anharmonic oscillators by the WKB function,
\begin{equation}\label{eq:psiwintro}
\Psi^W(x)=e^{\int_{x'}^x\left[ \sqrt{V(y)}-\frac{V'(y)}{4 V(y)}\right]dy},
\end{equation}
where $V$ is a function closely related to the potential $U$ but not necessarily coinciding with it. 
The logarithmic derivative of the function $\Psi^W$,  $\sqrt{V(x)}-\frac{V'(x)}{4 V(x)}$, is the truncation at the leading and first sub-leading term of the infinite, and notoriously divergent \footnote{In some specific cases the WKB series converges, see \cite{borrego24}. These are however exceptions.}, WKB series \cite{bender13}.

Despite its \textit{apparent} simplicity, the function $\Psi^W$ is a rich mathematical object, and it contains enough information to 
obtain a precise asymptotic description of the spectral determinant, in the regimes we are considering.
There is however quite a long road that one has to climb to start from \eqref{eq:psiwintro} and eventually get to the proof of the final theorems describing the asymptotics of the spectrum of the anharmonic oscillators, Theorem \ref{theorem:large-e-fixed-l} and Theorem \ref{theorem-large-e-l-bis}. We divided this road in four lectures, and we hope that the readers will enjoy the ride.

In Lecture I, we prove the fundamental Theorem of the WKB approximation on curves, Theorem \ref{thm:fundamental-theorem}. This theorem establishes sufficient conditions for a solution of a Schr\"odinger equation $\psi''(x)=U(x) \psi(x)$ to be approximated by the WKB function \eqref{eq:psiwintro}, along a fixed curve $\gamma$. To prove the theorem, we define the quantity $z(x)= \frac{\psi(x)}{\psi^W(x)}$, where $\psi$ is a solution of the Schr\"odinger equation, show that $z$ satisfies a Volterra integral equation, and provide a theoretical framework to address the well-posedness of the latter equation. We complete the first lecture by showing that, in the case of the anharmonic oscillator and assuming $\ell \neq 0$,
the function $V$ that appears in the WKB function \eqref{eq:psiwintro} is given by
\begin{equation}\label{eq:reducedintro}
  V(x)=U(x)+\frac1{4x^2}=  x^{2\alpha}-E+ \frac{(\ell+\frac12)^2}{x^2}.
\end{equation}
We call such a $V$ the \textbf{reduced potential} \footnote{Formula \eqref{eq:reducedintro} for the reduced potential lies beyonf our choice of the small parameter, $\hbar=\left(\ell+\frac12\right)^{-1}$, in the second asymptotic regime -- see  \eqref{eqn:intro_rescaled-infinite-e-l}; in fact, with such a definition of $\hbar$, the reduced potential scales exactly as $\hbar^{-2}$}.

In Lecture II, we address the asymptotic behaviour, at $0$ and $\infty$, of
solution to the anharmonic oscillator \eqref{eq:anharmonic}, by extending the Volterra integral equation to domains in the
complex plane. We will unveil that the asymptotic behaviour at $0$ is \textit{algebraic} and it does not depend on the way $0$ is approached; on the contrary the asymptotic behaviour at $\infty$ is \textit{exponential}, and it differs in different sectors about $\infty$, that are known as Stokes sectors. This is known as Stokes phenomenon. In the second lecture, 
we will also formulate the problem of the global asymptotic behaviour, namely how a solution or a basis of solutions with prescribed asymptotic behaviour in a region of the complex plane (either a neighbourhood of $0$ or a Stokes sector) behaves in a different region. We do so by introducing the concept of asymptotic values and Fock-Goncharov coordinates, which are an alternative to the notion of connection matrices and Stokes multipliers. We add to $\WCs$ the boundary points $0$ and $\infty_k,k \in \Z$, where $\infty_k$ `represents' a Stokes sector, in such a way that the ratio of any pair of solutions to the anharmonic oscillators extends continuously to the boundary. The value of the ratio of solutions at the boundary points are called asymptotic values, and the cross-ratio of any four asymptotic values is called a Fock-Goncharov coordinate.

In Lecture III, we develop a WKB theory of Fock-Goncharov coordinates. A Fock-Goncharov coordinate is by definition associated to a quadrilateral $\square$ whose vertices belong to the boundary of $\WCs$. In turn, via the WKB approximation, we associate to each quadrilateral a homotopy class $\gamma_{\square}$ of loops in $\WCs \setminus \{ V(x) =0 \}$ and we show that the corresponding Fock-Goncharov coordinate $R_{\square}$ is approximated by the expression
\begin{eqnarray}\label{eq:Rloopintro}
    R_{\square} \sim e^{\hbar^{-1}\oint_{\gamma_{\square}} \left[ \sqrt{V(x)} -\frac{V'(x)}{4V(x)}\right]dx} \left( 1+ O(\hbar) \right).
\end{eqnarray}
In particular, we study the above formula for the specific Fock-Goncharov coordinate that encodes the spectral problem $Q_+(E,\ell)=0$. As a result, we deduce the Bohr-Sommerfeld quantisation condition for the spectral problem, which, when $\ell$ is real, reads
\begin{equation}
\label{eqn:bohr-sommerfeldintro}
\frac{1}{\pi}\int_{x_-}^{x_+} \sqrt{E-x^{2 \alpha}-\frac{\ell+\frac{1}{2}}{x^2}}dx=n+\frac{1}{2},\quad n\in\N,
\end{equation}
where $x_-$ and $x_+$ are the two zeroes on the positive real axis of the reduced potential $V(x,E,\ell)$, for any $E$ above the theoretical minimum of the spectrum $E_*=\alpha^{-\frac{\alpha}{1+\alpha}}(1+\alpha) \left(\ell+\frac12\right)^{\frac{2\alpha}{1+\alpha}}$.

In Lecture IV, we prove that the spectrum of the anharmonic oscillator is well-approximated by the Bohr-Sommerfeld quantisation condition, in both asymptotic regimes we consider. This is the content of Theorem \ref{theorem:large-e-fixed-l} and Theorem \ref{theorem-large-e-l-bis} below. We notice that, even though some of the formulas that appear in the theorems are not novel, this is likely the first place where a full proof is given -- as we discuss in Remark \ref{rem:literaturethm} below.

Two appendices complement the lectures. In Appendix I, we collect a few results in analysis and complex analysis that are used in the lectures. In Appendix II,  we collect some general results on the anharmonic oscillator which are outside the scope of these lectures, including some notions on the ODE/IM correspondence.

\subsubsection*{Additional remarks and a brief note on the literature}

As we have stated above, these lectures are essentially self-contained, but not strictly self-contained. For lack of space, we do not cover the \textit{turning point analysis}, namely the analysis of the WKB approximation in a neighbourhood of a zero of the reduced potential. We make use of such an analysis at one point in the proof of theorems \ref{theorem:large-e-fixed-l} and \ref{theorem-large-e-l-bis}, and we refer the reader to a recent paper of the first author for details \cite{degano2024}.

These lectures are an introduction to the complex WKB method, but in the final lecture we specialise to the case of real $\ell$. In this case, the spectral problem is self-adjoint, then it reduces to the study of an ODE on the real line with real parameters. We could therefore dispense, at least in principle, of complex methods, and develop a purely real WKB method. However, this would cost us a lot of extra work, see e.g. \cite{maro18}, confirming the saying attributed to J. Hadamard  `the shortest path between two truths in the real domain passes through the complex domain'.

WKB approximation has a long and quite tortuous history. It keeps being reinvented, rediscovered, reanalysed. A thorough discussion of the literature is well-beyond the scope of this lecture \footnote{The literature on the WKB analysis of the quartic oscillator alone consists of hundreds of papers, since this is considered a toy model for the $\varphi^4$ quantum field theory \cite{bender69}.}. Here we limit ourselves to the following consideration. It seems to us that there are two fundamentally different approaches to the WKB analysis. One approach, that we do not follow here, is to represent the solution of a Schr\"odinger equation, as well as the spectral determinants, as the (Borel) resummation of (hopefully resummable) asymptotic series in the small parameter;  the research along this line has been quite strong and various techniques, some mathematically sound and some still heuristic, have been developed, which go under the name of resurgence, transasymptotic series, exact WKB approximation, etc..., see e.g. \cite{ecalle81,voros83,delabaere97,costin20,iwaki2014,sibuya75,kawai05,bertola24} and references therein. The 
second, possibly more classical, approach, that we follow here, is to truncate the asymptotic series at a finite order and then construct the actual solution of the Schr\"odinger equation by studying the integral equation satisfied by the ratio (or the difference) between the truncated series and the actual solution, see  e.g. \cite{erdelyi-book-wkb,olver97,fedoryuk93,eremenko18} and the authors' publications listed below. We call this approach the complex WKB method. Due to our quite extensive experience in the asymptotic analysis of anharmonic oscillators and Painlevé equations, see e.g. \cite{masoero2010poles,masoero10non,masoero10th,cinese,maro18,maro21,bridgeland-masoero-2023,masoero2024qfunctions,cotti23,degano2024}, we argue that -- in the analysis of anharmonic oscillators and similar equations -- this second approach has several advantages: it is simpler, since the first two terms of the series of WKB function are usually enough to obtain the desired asymptotic description of spectral quantities under study; it is mathematically safer, since it leads to reasonably simple and complete proofs; it is more flexible, since it allows studying the transition between different asymptotic regimes, in which the structure of the asymptotic series can happen to be radically different.

\subsubsection*{Acknowledgments}
The authors are members of the  COST Action CA21109 CaLISTA.
This research was supported by the FCT projects UIDB/00208/2020, DOI: \url{https://doi.org/10.54499/UIDB/00208/2020},  2021.00091.CEECIND, and 
 \\ 2022.03702.PTDC (GENIDE), DOI: \url{https://doi.org/10.54499/2022.03702.PTDC}. G. Degano is supported by the FCT Ph.D. scholarship UI/BD/152215/2021.  

These lectures were prepared for the Summer School `Contemporary Trends in Integrable Systems', Lisbon, 2024, promoted by the COST Action CaLISTA (CA21109) and the Group of Mathematical Physics of Lisbon University (GFM). Due to unforeseeable circumstances, D.M. was not able to participate in the school and deliver the lectures. He dedicates these lectures to the staff  of the neurology, infectious diseases, and intensive care units of the paediatric hospital Dona Estefânia.

\section{Lecture I. Fundamental Theorem of the WKB approximation}\label{chap:lecI}
We consider -- following closely \cite[Appendix]{bridgeland-masoero-2023} which in turn follows \cite{erdelyi-book-wkb} -- a scalar linear ODE of the form
\begin{equation}
\label{eq:schr}
\frac{d^2\psi}{dx^2}= U(x)\psi(x), \; x \in \C, 
\end{equation}
where the potential $U$ depends analytically on some unspecified complex parameters $\underline{u} \in \mathcal{U} \subset \C^M$.
We look for a putative approximate solution $\Psi(x)$, which we suppose to be of such a form that
\begin{equation}\label{eq:t}
 z(x)=\frac{\psi(x)}{\Psi(x)}
\end{equation}
is well-defined and approximately $1$ in a certain domain of $D$ to be later specified.

Defining the \textbf{forcing term}
\begin{equation}\label{eq:forcing}
 F(x)=U(x)-\frac{\Psi''(x)}{\Psi(x)}
\end{equation}
the equation \eqref{eq:schr} for $\psi(x)$, when rewritten in terms of the function $z(x)$ defined by \eqref{eq:t}, becomes
\begin{equation}\label{eq:usch}
 \frac{d}{dx}\big( \Psi^2(x) z'(x) \big)- \Psi^2(x) F(x) z(x)=0.
\end{equation}
We fix a point $x_0 \in \overline{D}$, $\overline{D}$ being the closure of $D$ in the Riemann sphere $\Cb$, the boundary conditions $z'(x_0)=0,z(x_0)=1$, and a smooth integration path
$\gamma$ connecting $x_0$ to another point $x \in \overline{D}$.
Integrating twice equation \eqref{eq:usch}, $z(x)$ is proven to solve the following integral equation
\begin{equation}\label{eq:integral}
 z(x)=1-\int_{\gamma,x_0}^x B(x,s) F(s) z(s) d s \,, \qquad
 B(x,s)=\int_{\gamma,s}^x \frac{\Psi^2(s)}{\Psi^2(r)} d r \; ,
\end{equation}
provided the above integral converges absolutely. The latter equation is a Volterra integral equation with
kernel $K(x,s)=B(x,s)F(s)$.

The previous computations prove the following
\begin{lemma}
\label{lemma:18-giugno-2024-1}
Let $D\subset \C^*$ be a domain, $\gamma\colon (0,1) \to D$ be an injective  smooth curve. Assume that the following two  conditions hold
\begin{align}\label{eq:boundedrho}
&    \int_0^1 |F(\gamma(t))| |\dot{\gamma}(t)| dt <+\infty\\
\label{eq:bounedB}
& \sup_{0<s\leq t<1}|B(\gamma(t),\gamma(s))|<+\infty,
\end{align}
 with $F(\cdot), B(\cdot,\cdot)$  as in (\ref{eq:forcing}, \ref{eq:integral}), and $\dot{\gamma}(t)=\frac{d}{dt}\gamma(t)$.

If the equation~\eqref{eq:schr} admits a solution $\psi(x;\underline{u})$ such that
\begin{equation}
\label{eqn:boundary-conditions}
\lim_{t\to 0^+} \frac{\psi(\gamma(t))}{\Psi(\gamma(t))}=1,\quad\lim_{t\to 0^+} \left.\frac{d}{dx}\frac{\psi(x)}{\Psi(x)}\right|_{x=\gamma(t)}=0
\end{equation} then the ratio $z(x):=\frac{\psi(x)}{\Psi(x)}$, restricted to $\gamma$, satisfies the Volterra integral equation~\eqref{eq:integral}.
\end{lemma}
In order to make use of the above result, we need three ingredients:
\begin{itemize}
\item[i)] Define/guess a meaningful approximating function;
\item[ii)] Find a curve such that the estimates (\ref{eq:boundedrho},\ref{eq:bounedB}) hold; namely, along $\gamma$
the forcing term $F$ is integrable and the kernel $B$ is bounded ;
\item[iii)] Eventually, invert the above lemma by showing that the
Volterra integral equation indeed admits a solution \eqref{eq:integral}.
\end{itemize}

In the complex WKB method the approximating function $\Psi$ is given by the formula
\begin{equation}
\label{wkb-approximant}
\Psi^W(x,x'; \uu):=\exp\left(\int_{x'}^x \left[\sqrt{V(y;\underline{u})}-\frac{1}{4}\frac{V'(y;\underline{u})}{V(y;\underline{u})}\right] dy\right),
\end{equation}
where $\sqrt{V}$ is the square-root of a never vanishing function $\sqrt{V}:D\to \C$, possibly depending analytically on the complex parameters $\underline{u}$, and not necessarily coinciding with the potential $U$\footnote{The reader who has already studied the WKB approximation may be puzzled by the fact that the function $V$ appearing in the WKB function $\Psi$ needs not
coincide with the potential $U$ of the Schr\"odinger equation. As we will show below, in the case of a regular singularity, $V$ must not coincide with $U$.}.

Given such an approximating function, a direct computation yields that
the forcing term $F$ and the kernel $B$, defined in (\ref{eq:forcing}, \ref{eq:integral}), have the following expression
\begin{align}
\label{eqn:forcing-term}
& F(x;\uu)dx=\frac{1}{\sqrt{V(x;\underline{u})}}\left[V(x;\uu)-U(x;\uu)+\frac{-4 V''(x;\uu) V(x)+5(V'(x;\uu))^2}{16(V(x;\uu))^2}\right]dx \\
\label{eq:kernelB}
& B(x,y;\uu)=\frac{1}{2}\left[\exp\left(- 2\int_{\gamma,y}^x\sqrt{V(s;\uu)}ds\right)-1\right],
\end{align}
where in the above expression $V',V''$ denotes the differentiation with respect to the complex variable $x$.

The conditions (\ref{eq:boundedrho},\ref{eq:bounedB}) of Lemma \ref{lemma:18-giugno-2024-1}, which state that along the smooth curve
$\gamma\colon (0,1) \to D$ the form $F dx$ is integrable  and the kernel $B$ is bounded, are expressed in terms of two related quantities
\begin{align}
& \rho_{\gamma}(t;\underline{u}):=  \int_0^t |F(\gamma(t);\underline{u})| |\dot{\gamma}(t)| dt ,
 \quad \rho_{\gamma}= \sup_{\uu \in \mathcal{U}} \rho_{\gamma}(1;\uu), \label{eq:rhogamma}\\ \label{eq:betagamma}
& \beta_{\gamma}=\inf_{\uu \in \mathcal{U}} \inf_{0<s\leq t <1} \Re \int_{s}^t \sqrt{V(\gamma(r))} \dot{\gamma}(t) dr.
\end{align}

We provide below sufficient conditions for the Volterra integral equation \eqref{eq:volterragamma} to be well-defined.
\begin{definition}\label{def:admissibletriple}
    We denote by $D$ a domain in $\Cs$, by $\overline{D} \subset \Cb$ its closure, by $\mathcal{U} \subset \C^M$ the product of $M$ domains $D_1 \times D_2 \times \dots D_M$ with $D_i \subset \C$, and by $\overline{U}$ its closure. Assume that
    $U,\sqrt{V}\colon D \times \mathcal{U} \to \C$ are analytic functions. Assume moreover that the analytic family of continuous curves $\gamma\colon [0,1] \times \mathcal{U} \to \Cb$  has fixed end-points (namely $\gamma(0;\uu)$ and $\gamma(1;\uu)$, which are $\uu$ independent), and  $\gamma(\cdot,\uu)\colon (0,1) \to D$  is smooth and injective, for every $\uu \in \mathcal{U}$.

The triple $(U,\sqrt{V},\gamma)$ is said to be admissible if
\begin{itemize}
\item[i)] $\sqrt{V(x;\uu)} \neq 0$, for all $(x,\uu) \in D \times \mathcal{U}$;
\item[ii)] $|F(\gamma(t);\underline{u})| |\dot{\gamma}(t;\uu)|$ is uniformly integrable in the following sense: There exists a positive $h\in L^1([0,1])$ such that $\sup_{\uu \in \mathcal{U}}|F(\gamma(t);\underline{u})| |\dot{\gamma}(t;\uu)| \leq h(t)$; consequently $\rho_{\gamma}<+\infty$, with $\rho_{\gamma}$ as per  \eqref{eq:rhogamma};
\item[iii)] The quantity $\beta_{\gamma}>-\infty$, where $\beta_{\gamma}$ is as per \eqref{eq:betagamma};
\item[iii,bis)] If $\lim_{t\to 1^-} \gamma(t;\uu) \in \overline{D}\setminus D$,
the following further requirement must be met \footnote{This is known as Levinson condition, \cite{cotti23}}: Fixed a $t_0 \in (0,1)$,
\begin{align}\label{eq:dichotomy}
\lim_{t\to 1^-}\inf_{u \in \mathcal{U}} \Re \int_{t_0}^t  \sqrt{V(\gamma(s;\uu)} \dot{\gamma}(s;\uu) ds  =\infty  .
\end{align}
\end{itemize}
\end{definition}
\begin{definition} $\mathcal{H}([0,1]\times \mathcal{U})$ denotes the linear space of continuous bounded functions $f$ on
$[0,1] \times \overline{\mathcal{U}}$ such that, for every $t \in[0,1]$, the restriction $f(t;\cdot)\colon \mathcal{U} \to \C$ is holomorphic. 
\end{definition}
The linear space $\mathcal{H}([0,1] \times\mathcal{U})$ is a Banach space  when equipped with the sup norm, $\|f \|_{\infty}:= \sup_{(t,\uu) \in [0,1] \times \mathcal{U}} \big| f(t;\uu)\big|$ -- see Proposition \ref{prop:closure} in Appendix I.
The definition of admissible triple is tailored to make the Volterra integral equation \eqref{eq:volterragamma} well-posed in this space.

\begin{proposition}\label{prop:Kgamma}
Let the triple $(U,\sqrt{V},\gamma)$ be admissible.

For any $f \in \mathcal{H}([0,1]\times \mathcal{U})$, define
\begin{equation}\label{eq:defKgamma}
 \mathcal{K}_{\gamma}[f](t;\underline{u})= \left. \begin{cases}
 \int_0^t B(\gamma(t),\gamma(s);\underline{u}) F(\gamma(s);\underline{u}) \dot{\gamma}(s) f(s;\underline{u}) d s, & 0 \leq t <1\\
\lim_{t \to 1^-} \int_0^t B(\gamma(t),\gamma(s);\underline{u}) F(\gamma(s);\underline{u}) \dot{\gamma}(s) f(s;\underline{u}) d s, & t=1
 \end{cases} \right.
\end{equation}
provided the integral converges and the limit exists. In the above formula $F(\cdot),B(\cdot,\cdot)$ are as in (\ref{eqn:forcing-term},\ref{eq:kernelB}). 

The operator $\mathcal{K}_{\gamma}$ is a continuous operator on $\mathcal{H}([0,1] \times \mathcal{U})$ and the following estimate holds:
\begin{align} \label{eq:Kkestimate}
& \left| \mathcal{K}_\gamma^k[f](t;\underline{u}) \right| \leq \frac{\left[\rho_{\gamma}(t;\underline{u})\right]^k}{k!} 
\left( \frac{1+e^{-\beta_{\gamma}}}{2} \right)^k\times \sup_{t \in [0,1]} |f(t;\uu) |,
\end{align}
where $\rho_{\gamma}(t;\uu)$ and $\beta_{\gamma}$ are defined in (\ref{eq:rhogamma},\ref{eq:betagamma}).
Consequently,
\begin{equation}\label{eq:Kknorm}
\|\mathcal{K}_{\gamma}^k \| \leq \frac{\rho_{\gamma}^k}{k!} \left( \frac{1+e^{-\beta_{\gamma}}}{2} \right)^k
\end{equation}
where $\| \cdot \|$ is the operator norm.
\begin{proof}
In this proof, in order to lighten the notation, we write $F_{\gamma}(t;\uu)=F(\gamma(t);\uu) \dot{\gamma}(t) $ and $B_{\gamma}(t,s;\uu)=B(\gamma(t),\gamma(s);\uu)$. We recall that, by hypothesis, $|B_{\gamma}(t,s;\uu) |\leq \frac{1+e^{-\beta_{\gamma}}}{2}$, and $|F_{\gamma}(t;\uu)|\leq h(t)$, with $h(t) \in L^1([0,1)]$
We divide the proof in two halves. In Part 1 we prove that $\mathcal{K}_{\gamma}$ is a well-defined continuous operator and estimates \eqref{eq:Kkestimate} when $k=1$. In Part 2,  we prove \eqref{eq:Kkestimate} when $k\geq 1$
\\

Part 1. \quad  When $k=1$, the estimate \eqref{eq:Kkestimate}, follows directly from the H\"older inequality:
If $f$ is an integrable function on $[0,1]$ and $g$ is a bounded function on $[0,1]$ then $f g$ is integrable
on $[0,1]$ and
\begin{align}\label{eq:holder}
\int_0^1 \left| f(t) g(t) \right| dt \leq \sup_{t \in [0,1]}| g(t)| \int_0^1 |f(t)|dt. 
\end{align}
Therefore, if $\mathcal{K}_{\gamma}$ is well-defined then \eqref{eq:Kknorm} holds, hence $\mathcal{K}_{\gamma}$ is a continuous operator.

Let us prove that $\mathcal{K}_{\gamma}$ is well-defined.
To this aim, we need to check that $\mathcal{K}_{\gamma}[f] $ is continuous and, fixed $t \in[0,1]$,  $\mathcal{K}_{\gamma}[f](t;\cdot) $ is analytic.

We first show that $\mathcal{K}_{\gamma}[f]$ is continuous at $(t,\uu)$, and analytic with respect to $\uu$, for any $t<1$.

We study 
$\mathcal{K}_{\gamma}[f](t+\e;\uu+\underline{\delta})-\mathcal{K}_{\gamma}[f](t;\uu+\underline{\delta})$ where, for simplicity of notation, we assume that $\e>0$. We have
\begin{align*}
& \left|\mathcal{K}_{\gamma}[f](t+\e;\uu+\underline{\delta})-\mathcal{K}_{\gamma}[f](t;\uu+\underline{\delta})\right|  =\int_{t}^{t+\e}
B_{\gamma}(t+\e,s;\uu+\underline{\delta})  F_{\gamma}(s;\uu+\underline{\delta})  f(s;\uu+\underline{\delta}) ds \\ +
& \int_0^{t} B_{\gamma}(t+\e,s;\uu+\underline{\delta})  F_{\gamma}(s;\uu+\underline{\delta})  f(s;\uu+\underline{\delta}) ds-
 \int_0^{t} B_{\gamma}(t,s;\uu)  F_{\gamma}(s;\uu)  f(s;\uu) ds.
\end{align*}
The first term on the right-hand side converges to zero, as 
$$
\lim_{\e \to 0} \left|\int_{t}^{t+\e}
B_{\gamma}(t+\e,s;\uu+\underline{\delta})  F_{\gamma}(s;\uu+\underline{\delta})  f(s;\uu+\underline{\delta}) ds\right| \leq \|f\|_{\infty} \frac{1+e^{-\beta_{\gamma}}}{2}  \int_{t}^{t+\e} h(s) ds \to 0.
$$
 The second term converges to zero since the function
$(\e,\underline{\delta}) \mapsto \int_0^{t} B_{\gamma}(t+\e,s;\uu+\underline{\delta})  F_{\gamma}(s;\uu+\underline{\delta})  f(s;\uu+\underline{\delta}) ds$ is continuous at $(0,\underline{0})$, see Proposition \ref{prop:dominatedintegral}. This proves continuity.
 Analyticity follows directly from Proposition \ref{prop:dominatedintegral}.

Let us now consider the limit $t \to 1^-$. In the case $\gamma(1;\uu) \in D$, the previous considerations can be extended to $[0,1]$ and the theorem is proven.
If on the contrary $\gamma(1;\uu) \notin D$, then by hypothesis (\ref{eq:dichotomy}) holds; we use the latter condition to prove that  
\begin{equation}\label{eq:limsupKft}
    \lim_{T \to 1^-} \sup_{t\geq T} \sup_{\uu \in \mathcal{U}}  |\mathcal{K}_{\gamma}[f](t)-\mathcal{K}_{\gamma}[f](T) | \to 0,
\end{equation}
for all $f \in H([0,1]\times \mathcal{U})$.
From the latter limit, it follows that the family of function  $\overline{f}_T(t;\uu)\colon [0,1] \times \mathcal{U} \to \C$, $T>1$,
\begin{equation*}
    \overline{f}_T(t;\uu)= 
    \begin{cases}  \mathcal{K}_{\gamma}[f](t;\underline{u}), &  0\leq t \leq T\\
\mathcal{K}_{\gamma}[f](T;\underline{u}), & T \leq t \leq 1
    \end{cases} 
\end{equation*}
converges in  $H([0,1] \times \mathcal{U})$. Therefore, we conclude that $K_{\gamma}[f] \in H([0,1] \times \mathcal{U})$.

Let us then prove \eqref{eq:limsupKft}. If $t\geq T$
\begin{align*}
\mathcal{K}_{\gamma}[f](t)-\mathcal{K}_{\gamma}[f](T) & =
\int_{T}^{t}  B_{\gamma}(t,s;\uu)  F_{\gamma}(s;\uu)  f(s;\uu) ds \\ 
+ & \int_0^{T} \left(B_{\gamma}(t,s;\uu) -B_{\gamma}(T,s;\uu) \right) 
  F_{\gamma}(s;\uu)  f(s;\uu) ds. 
\end{align*}
The first term on the right-hand side converges to $0$ since
$$\lim_{T \to 1^-} \sup_{t \geq T}  \sup_{\uu \in \mathcal{U}}\left|\int_{T}^{t}  B_{\gamma}(t,s;\uu)  F_{\gamma}(s;\uu)  f(s;\uu) ds \right| \leq \|f\|_{\infty} \frac{1+e^{-\beta_{\gamma}}}{2}  \int_{T}^{1} h(s) ds \to 0.$$
As for the second term on the right-hand side, we have
\begin{align*}
&\left| \int_0^{T} \left(B_{\gamma}(t,s;\uu) -B_{\gamma}(T,s;\uu) \right) 
  F_{\gamma}(s;\uu)  f(s;\uu) ds \right| \leq \\
& \frac12 \left(\sup_{t,\uu} |f|\right) \int_0^{T} \left| e^{\int_s^{T} \sqrt{V(\gamma(r);\uu)}\dot{\gamma}(r;\uu) dr } \left(1-e^{\int_{T}^{t}\sqrt{V(\gamma(r);\uu)}\dot{\gamma}(r;\uu)dr}\right) F(s) \dot{\gamma}(s)  \right| ds.
\end{align*}
Due to \eqref{eq:dichotomy}, we can define an increasing function $g(T)$, independent of $\uu$, 
such that $g(T) \to 1$ as $T \to 1$ and
$\lim_{T \to 1} \inf_{\uu} \Re \int_{g(T)}^{T} \sqrt{V(\gamma(r))}\dot{\gamma}(r) dr=\infty$.
Hence,
\begin{align*}
&  \int_0^{T} \left| e^{\int_s^{T} \sqrt{V(\gamma(r);\uu)}\dot{\gamma}(r;\uu) dr } \left(1-e^{\int_{T}^{t}\sqrt{V(\gamma(r);\uu)}\dot{\gamma}(r;\uu)dr}\right) F_{\gamma}(s)  \right| ds= \\
&  \int_0^{g(T)} \left| e^{\int_s^{T} \sqrt{V(\gamma(r);\uu)}\dot{\gamma}(r;\uu) dr } \left(1-e^{\int_{T}^{t}\sqrt{V(\gamma(r);\uu)}\dot{\gamma}(r;\uu)dr}\right)  F_{\gamma}(s)   \right| ds+ \\
&  \int_{g(T)}^{T} \left| e^{\int_s^{T} \sqrt{V(\gamma(r);\uu)}\dot{\gamma}(r;\uu) dr } \left(1-e^{\int_{T}^{t}\sqrt{V(\gamma(r);\uu)}\dot{\gamma}(r;\uu)dr}\right)  F_{\gamma}(s)   \right| ds.
\end{align*}
The first contribution converges to $0$ since $\lim_{T \to 1^-}\left|e^{\int_s^{T} \sqrt{V(\gamma(r))}\dot{\gamma}(r) dr }\right| = 0$ uniformly in $[0,g(T)]\times \mathcal{U}$,
by construction of the function $g(T)$. The second term converges to $0$ since $g(T) \to 1$.
\\

Part 2. \quad We now prove estimate \eqref{eq:Kkestimate} when $k \geq 2$.
Let $k\geq 2$ and $G(t,s)=B_{\gamma}(t,s;\uu) F_{\gamma}(s;\uu)$; we have
\[
\mathcal{K}_\gamma[f](z;\underline{u}) = \int_{0}^t\ldots \int_{0}^{s_3}\int_{0}^{s_2} f(s_1;\underline{u}) \prod_{j=1}^k G(s_{j+1},s_j;\underline{u}) ds_j
\]
where $s_{k+1}:= t$. It follows that
\begin{equation*}
\left|\mathcal{K}_\gamma[f](z;\underline{u})\right|\leq \left( \sup_{t \in [0,1] }\left| f(t;\uu)\right| \right) \int_{\Delta^k(t)} \left| G(t;s_1,\ldots,s_k;\underline{u})\right|ds_1\ldots ds_k
\end{equation*}
where
\[
\Delta^k(t):=\left\{(s_1,\ldots,s_k)\in[0,t]^{\times k}\,:\,0\leq s_1\leq \dots \leq s_k \leq t \right\}
\]
and
$$
G(t;s_1,\ldots,s_k;\underline{u})=\prod_{j=1}^k \left| G(s_{j+1},s_j;\underline{u})\right|, \; s_{k+1}=t.
$$

If $\mathcal{S}_k$ is the group of permutations on $k$ letters, then we can write
\[
[0,t]^{\times k}=\bigcup_{\sigma\in\mathcal{S}_k}\sigma(\Delta^k(t)),
\]
where 
for two distinct permutations $\sigma_1,\sigma_2\in\mathcal{S}_k$, the set $\sigma_1(\Delta^k(t))\cap\sigma_2(\Delta^k(t))$ has zero $k$-dimensional measure.
Using the definition of $B_{\gamma}$ and $F_{\gamma}$, we check that for any $\sigma\in\mathcal{S}_k$ we have 
\[
G(t,\sigma(s_1),\ldots,\sigma(s_k);\underline{u})=G(t,s_1,\ldots,s_k;\underline{u}).
\]
Therefore,
\begin{equation*}
\begin{aligned}
& \int_{\Delta^k(t)} \left|G(t,s_1,\ldots,s_k;\underline{u})\right|ds_1\ldots ds_k \\
& =\frac{1}{k!}\sum_{\sigma\in\mathcal{S}_k}\int_{\sigma\left(\Delta^k(t)\right)} \left|G(t,\sigma^{-1}(s_1),\ldots,\sigma^{-1}(s_k);\underline{u})\right| ds_1\ldots ds_k \\ 
& = \frac{1}{k!} \int_{[0,t]^k} \left| G(t,s_1,\ldots,s_k;\underline{u}) \right|ds_1\ldots ds_k \\\
& \leq
\frac{1}{k!} \left(\int_0^t \left|B_{\gamma}(t,s);\uu) F_{\gamma}(s;\uu) \right|ds \right)^k \leq \left(\frac{\rho_{\gamma}(t;\uu)}{k!}\right)^k \left( \frac{1+e^{-\beta_k}}{2}\right)^k.
\end{aligned}
\end{equation*}

\end{proof}
\end{proposition}

\begin{corollary}\label{cor:volterragamma}
Let the triple $(U,\sqrt{V},\gamma)$ be admissible.

For every $\underline{u} \in \mathcal{U}$, the Volterra integral equation
\begin{equation}\label{eq:volterragamma}
    z(t;\uu)=1-\int_0^t B(\gamma(t),\gamma(s);\uu)F(\gamma(s);\uu) \dot{\gamma}(s) z(s;\uu) ds
\end{equation}
where $F(\cdot;\cdot)$ and $B(\cdot,\cdot;\cdot)$ are as in (\ref{eqn:forcing-term},\ref{eq:kernelB}), admits a unique continuous solution $z(\cdot;\uu)\colon [0,1] \to \C$.

The solution $z$ satisfies the following estimates: for all $\uu \in \mathcal{U}$,
\begin{align} \label{eq:z0uu}
& z(0;\uu)=1 \\ \label{eq:ztuurho}
& \left|z(t;\uu)-1\right| \leq \exp\left( \rho_{\gamma}(t;\uu) \frac{1+e^{-\beta_{\gamma}}}{2}\right)-1.
\end{align}
In particular, if $\beta_{\gamma}=0$ then
\begin{align}
& \left|z(t;\uu)-1\right| \leq \exp\left( \rho_{\gamma}(t;\uu) \right)-1.
\end{align}
Moreover, the function $z\colon [0,1] \times \mathcal{U}\ni (t,\uu) \mapsto z(t;\uu)$ belongs to $\mathcal{H}([0,1] \times \mathcal{U})$ namely $z(t;\uu)$ is analytic with respect to the parameters $\uu \in \mathcal{U}$.
\end{corollary}
\begin{proof}
The integral equation \eqref{eq:volterragamma} reads
\begin{equation*}
z=1-\mathcal{K}_{\gamma}[z]
\end{equation*}
where $\mathcal{K}_{\gamma}$ is the operator defined in \eqref{eq:defKgamma}, and $1$ is the constant function with value one.

We can study the above equation with $\uu \in \mathcal{U}$ fixed, namely in the Banach space of bounded continuous functions on the unit interval $\mathcal{C}[0,1]$; or, we can let 
 $\uu$ vary and study the same integral equation on the Banach space $\mathcal{H}([0,1]\times \mathcal{U})$ introduced above.
In either case, due to Proposition \ref{prop:Kgamma} the solution exists and it is unique.
\\

Proof of Existence. \quad Due to (\ref{eq:Kknorm}), the series
$$ \hat{z}:=\sum_{k=0}^{\infty} (- \mathcal{K}_{\gamma})^k [1]$$ converges in  $C[0,1]$ and in $\mathcal{H}([0,1] \times \mathcal{U})$. Moreover, since
$$- \mathcal{K}_{\gamma}[\hat{z}]= \sum_{k=1}^{\infty} ( - \mathcal{K}_{\gamma})^k [1]=  \hat{z}-1,$$ then $\hat{z}$ satisfies the integral equation.
\\

Proof of Uniqueness. \quad Assume $w$ is a solution of the same equation. Then $\hat{z}-w= \mathcal{K}_{\gamma}(\hat{z}-w)$, hence $\hat{z}-w= \mathcal{K}^k_{\gamma}(\hat{z}-w)$ for any positive $k$. Due to \eqref{eq:Kknorm}, the norm of the operator $ \mathcal{K}^k_{\gamma}$ is strictly less than $1$ for $k$ large enough; therefore $\hat{z}-w$ is the zero function. 
\end{proof}

\begin{remark}
\label{remark:refined-estimate}
Sometimes it is convenient to consider the following upper bound
\begin{equation}
\label{eq:ztuurho-refined}
\left|z(t;\underline{u})-1\right|\le \exp\left(\tilde{\rho}_\gamma(t;\underline{u})\right)-1,
\end{equation}
where
\[
\tilde{\rho}_\gamma(t;\underline{u}):=\int_0^t \left|B(\gamma(t),\gamma(s);\uu)F(\gamma(s);\uu) \dot{\gamma}(s)\right| ds,
\]
which is slightly more refined than the one of inequality~\eqref{eq:ztuurho}. In Lecture IV, we will see a case in which we have to consider inequality~\eqref{eq:ztuurho-refined} to perform sufficiently refined estimates in WKB computations. 
\end{remark}

\begin{theorem}[Fundamental theorem of the WKB Approximation on curves]
\label{thm:fundamental-theorem}
Let the triple $(U,\sqrt{V},\gamma)$ be admissible, $\rho_{\gamma}(t,\uu)$ and $\beta_{\gamma}$ be as in (\ref{eq:rhogamma},\ref{eq:betagamma}), $x' \in D$ be an arbitrary but fixed point, and $\Psi^W(x,x';\uu)$ be the WKB function as per \eqref{wkb-approximant}.

For every $\uu \in\mathcal{U}$, 
there exists a unique solution $\psi(\cdot;\underline{u})\colon D \to \C$ to the differential equation \eqref{eq:schr} 
such that
\begin{equation}\label{eq:boundarytheorem}
    \lim_{t\to 0^+} \frac{\psi(\gamma(t);\underline{u})}{\Psi^W(\gamma(t),x';\uu)}=1 .
\end{equation}
Moreover, $\psi$ satisfies the following estimate
\begin{equation}
\label{eqn:fundamental-inequality}
\left|\frac{\psi(\gamma(t);\underline{u})}{\Psi^W(\gamma(t),x';\uu)}-1\right|\leq \exp\left( \rho_{\gamma}(t;\underline{u})
\left(\frac{1+e^{-\beta_{\gamma}}}{2} \right)\right)-1
\end{equation}
and $\psi$ depends analytically on the parameters $\uu$, namely $\psi\colon D \times \mathcal{U}\to\mathbb{C}$ is analytic.

Finally,
$\lim_{t \to 1^-}\frac{\psi(\gamma(t);\underline{u})}{\Psi^W(\gamma(t),x';\uu)}$ exists, it depends analytically on $\uu$,
and
\begin{equation}\label{eq:at1inrquality}
\lim_{t \to 1^-} \left|\frac{\psi(\gamma(t);\underline{u})}{\Psi^W(\gamma(t),x';\uu)} -1 \right| \leq
\exp\left( \rho_{\gamma}(1;\underline{u})
\left(\frac{1+e^{-\beta_{\gamma}}}{2} \right)\right)-1.
\end{equation}
In all inequalities above, whenever $\beta_{\gamma}=0$, the expression $\frac{1+e^{-\beta_{\gamma}}}{2}$ equals $1$.
\begin{proof}
By Lemma \ref{lemma:18-giugno-2024-1}, there is at most one solution that satisfies \eqref{eq:boundarytheorem}.

Let $z(t;\uu)$ be the unique solution to the Volterra equation \eqref{eq:volterragamma} whose existence was proved in Corollary \ref{cor:volterragamma}. By construction, the function $\psi(\gamma(t);\uu): = z(t;\uu) \Psi^W(\gamma(t),x';\uu)$ solves the differential equation \eqref{eq:schr} on $\gamma$ and it is analytic with respect to $\uu$.
Therefore, for every $\uu$, $\psi(\gamma(\cdot);\uu)$ extends to a global solution $\psi(\cdot;\uu)\colon D \to \C$ of \eqref{eq:schr}, and such an extension is analytic with respect to $\uu$ -- see Corollary \ref{cor:extension} in Appendix I.
Moreover, since $\lim_{t\to 0^+} z(t;\underline{u})=1$, it satisfies \eqref{eq:boundarytheorem}.

Equation \eqref{eq:boundarytheorem} follows directly from \eqref{eq:z0uu}, and equations (\ref{eqn:fundamental-inequality},\ref{eq:at1inrquality}) follow directly from \eqref{eq:ztuurho}.
\end{proof}
\end{theorem}

\subsection{WKB functions for the anharmonic oscillator}
Here we apply the theory developed in the previous section to the differential equation \eqref{eq:anharmonic}, which has potential $U=x^{2\alpha}+\frac{\ell(\ell+1)}{x^2}-E$, with $x \in \WCs$.
To do that,
we restrict $U$ to a domain $D \subset \WCs^*$ contained in a sector of finite amplitude, namely
$ D \subset \lbrace  x \in \WCs, a \leq \arg x \leq b \rbrace$ for some $a<b \in \R$, which can
 be naturally embedded in  $\Cb$ \footnote{The optimal amplitude of the sector and a better geometric setting will be discussed in the Lecture II.}. 

In such a geometric setting, the potential $U$ is regular on $D \subset \WCs \subset \C$, for every value of the parameters $E,\ell$. 
Recall that we are interested in the spectrum of \eqref{eq:anharmonic}, which is defined via solutions with a prescribed subdominant asymptotic behaviour as  $x \to 0$ and $x \to +\infty$. In order to study such solutions via the complex WKB analysis, we must find a function
$\sqrt{V}$ and a $(E,\ell)-$ family of curves $\gamma$ starting at $0$ or $+\infty$  such that the triples $(U, \sqrt{V},\gamma)$ are admissible, as per Definition \ref{def:admissibletriple}.
In particular, assuming that the curves $\gamma$ are tame enough, we must look for a $\sqrt{V}$ such that
the form $F(x) dx$, with $F(x)$ defined in \eqref{eqn:forcing-term}, is integrable at $0$ and $\infty$: there should exist an
$\e>0$ such that $|F(x)| \leq C |x^{-1+\e}|$ as $x \to 0$, and
    $|F(x)| \leq C x^{-1-\e}$ as $x \to \infty$.

As we show below, for $\ell \notin \{-\frac12,0 \}$, the convenient choice of the function $V$ is 
\begin{equation}\label{eq:reducedpotential}
    V(x;E,\ell) := U(x;E,\ell)+\frac{1}{4 x^2}= x^{2\alpha}-E+\frac{(\ell+\frac12)^2}{x^2},
\end{equation}
which we call the \textbf{reduced potential}.

\begin{lemma}\label{lem:Vlanger}
Let $\ell \notin \{-\frac12,0\}$.
\begin{itemize}
    \item[I)]
If $V(x)=U(x)+\frac{c}{x^2}$ for some $ c \in \C$, the one-form $F(x) dx$ is integrable at $0$ if and only if $c=\frac14$. In particular, the form $F \, dx$ is not integrable if
$V(x)=U(x)$.
\item[II)] Assume that $V(x)-U(x)=O(x^\beta),\,(V(x)-U(x))'(x)=O(x^{\beta-1}),\, (V(x)-U(x))''(x)=O(x^{\beta-2})$ as $x \to \infty$. If $\beta<\alpha-1$, the one-form $F(x) dx$ is integrable at $x=\infty$. 
\item[III)] In particular, if $V(x)=U(x)+\frac{1}{4x^2}$,
\begin{align}\label{eq:Fx0}
    &F(x)=O(|x|), \quad |x| \to 0 \\ \label{eq:Fxinfty}
    &F(x)= O(|x|^{-\alpha-2}), \quad  |x| \to \infty.
\end{align}
Estimate \eqref{eq:Fx0} holds uniformly on any compact of the space of parameters $\C \times \left\{ \ell \neq0, \ell \neq -\frac12 \right\}$, while estimate \eqref{eq:Fx0} holds uniformly on any compact of the space of parameters $\C \times \left\{\Re \ell \geq -\frac12 \right\}$.
\end{itemize}
\begin{proof}
I) \quad Expanding at $x=0$ we get
$$(-4  V''(x) V(x)+ 5 V'(x)^2)/(16 V^2(x))= -\frac{1}{4 x^2}- \frac{E}{\ell(\ell+1)+c} +o(1), x \to 0.$$
Hence $F(x)$ is integrable if and only if $c=\frac14$, in which case 
$F(x)= O(x)$.
\\

II) \quad Expanding at $\infty$, we obtain
$$
F(x)= O(x^{-\alpha-2})+ O(x^{\beta-\alpha}) .
$$
Hence the condition $\beta <\alpha-1$ is sufficient to imply that $f$ is integrable at $\infty$.
\\

III) \quad It follows from the proof of (I, II). Uniformity with respect to the parameters is left to the reader.
\end{proof}
\end{lemma}

\begin{remark}
In the WKB literature, the addition of the term $\frac{1}{4x^2}$ to the potential is called Langer modification and 
it is justified at times with shaky arguments. It appears naturally when one actually constructs a mathematical theory of the WKB approximation: it is the correction term that makes the Volterra operator well-defined in case of a regular singularity!
\end{remark}

Given a curve $\gamma:[0,1] \to \Cb$, we call
\begin{equation}\label{eq:sphericallength}
 \int_0^1 \frac{|\dot{\gamma}(t)|}{1+|\gamma(t)|^2} dt
 \end{equation}
  its spherical length.
  \begin{definition}
      We say that the family of curves $\gamma:[0,1] \times \mathcal{U} \to \Cb$ is unifomrly rectifiable (with respect to the spherical metric) if 
      the function $\frac{|\dot{\gamma}(t;\uu)|}{1+|\gamma(t;\uu)|^2}$ is uniformly integrable on $[0,1]$.
  \end{definition}
  We have the following corollary of  Lemma \ref{lem:Vlanger}.
\begin{corollary}\label{propo:Vlanger}
  Let $V(x)=U(x)+\frac{1}{4x^2}$ and 
$W$ be a compact subset of $ \C \times \{ \Re \ell >-\frac12, \ell \neq 0\}$. Given a $D \subset \C$ such that
$V (x) \neq 0$ for all $(x,E,\ell) \in D \times W$, fix a branch of $\sqrt{V}$.

If the family of curves $\gamma\colon [0,1] \times W \to \overline{D}$ is uniformly rectifiable with respect to the spherical metric, then $F(\gamma(t;\uu);\uu) \dot{\gamma}(t;\uu) $
    is uniformly integrable.
\begin{proof}
It follows from Lemma \ref{lem:Vlanger} that $| F(\gamma(t;\uu);\uu) | \leq \frac{C_W }{1+|\gamma(t;\uu)|^2}$
for some constant $C_W>0$ depending on $W$. The thesis follows.
\end{proof}
\end{corollary}

\begin{definition}\label{def:admissiblecurves}
Fix a compact $W$ in the space of parameters and a domain $D$.
The family of curves $\gamma\colon [0,1] \times W \to \overline{D}$ is called admissible if, given a branch of the square root of the reduced potential \eqref{eq:reducedpotential}, $(U, \sqrt{V},\gamma)$ is an admissible triple.
The family is said to be strictly admissible if moreover $\Re S$ is strictly monotone along $\gamma$, hence in particular  $\beta_{\gamma}=0$.
\end{definition}

We end the first lecture by noticing that 
we can write \eqref{eq:anharmonic} in the following form
\begin{equation}\label{eq:anharmoniclanger}
\psi''(x)= \left(V(x)-\frac{1}{4x^2} \right) \psi(x),\quad V(x)=x^{2\alpha}-E+\frac{(\ell+\frac12)^2}{x^2}.
\end{equation}
The above form of the equation justifies our choice of the small parameter $\hbar$ in the second asymptotic regime \eqref{eqn:intro_rescaled-infinite-e-l}, namely $\hbar= (\ell+\frac12)^{-1}$. In fact, with this choice, the main objects of the WKB approximation 
naturally scale with $\hbar$:
\begin{itemize}
    \item $V \to \hbar^{-2} V$;
    \item $F \to \hbar F$; 
    \item Consequently, if $\gamma$ is strictly admissible for a value of $\hbar>0$ then it is strictly admissible for all $\hbar >0$, and
    $\rho_{\gamma}= \hbar \rho_{\gamma}$.
\end{itemize}
Therefore, provided a strictly admissible family of curves $\gamma$ exists, by Theorem \ref{thm:fundamental-theorem} we conclude that there exists a $\hbar$ family of solutions, $\psi^{\hbar}(x)$ such that
\begin{equation}\label{eq:psihbar}
\left| \psi^{\hbar}(\gamma(t)) \big/ \exp\left( \int^x_{x'}\left[\hbar^{-1} \sqrt{V(y)}-\frac{V'(y)}{4 V(y)} \right]dy \right)-1\right| = 
O(\hbar) .
\end{equation}
This concludes Lecture I.
The problem of studying (strictly) admissible curves, as well as a proper definition of the solution \eqref{eq:psihbar}, will be addressed in the next Lecture.

\paragraph{\textbf{Exercises. I}}
1. Deduce equation \eqref{eq:integral} from equation \eqref{eq:usch}.

2. Deduce \eqref{eqn:forcing-term} and \eqref{eq:kernelB} from (\ref{eq:forcing},\ref{eq:integral},\ref{wkb-approximant}).

3. Assume that $\ell=0$ or $\ell=-\frac12$. Find $V(x)$ such that the form $F(x) dx$, given by expression \eqref{eqn:forcing-term}, is integrable both at $0$ and at $\infty$.

\newpage
\section{Lecture II. Asymptotic behaviour of solutions - Stokes Phenomenon}
In the first part of this lecture, we study the local (asymptotic) behaviour of solutions at $0$ and $\infty$ by extending the Volterra integral equation to domains in the complex plane foliated by strictly admissible curves. We will unveil that the asymptotic behaviour at $0$ is \textit{algebraic} and it does not depend on the way $0$ is approached; on the contrary, the asymptotic behaviour at $\infty$ is \textit{exponential}, and it differs in different sectors about $\infty$, that are known as Stokes sectors. This is known as Stokes Phenomenon.

In the second part of this lecture, we will formulate the problem of the global asymptotic behaviour, namely how a solution or a basis of solutions with prescribed asymptotic behaviour in a region of the complex plane (either a neighbourhood of $0$ or a Stokes sector) behaves in a different region. We formulate this problem in terms of Fock-Goncharov coordinates, and not in terms of connection matrices and Stokes multipliers, because Fock-Goncharov coordinates are more natural in the complex WKB method.

\subsection{Quadratic differential}

Recall the definition of a strictly admissible family of curves, Definition \ref{def:admissiblecurves}. This is essentially a family of curves with finite spherical length \eqref{eq:sphericallength} and such that 
$\Re \int^t \sqrt{V(\gamma(s))} \dot{\gamma}(s) ds$ is monotonically increasing, where $V$ is the reduced potential \eqref{eq:reducedpotential}. Along such curves, we can apply the fundamental theorem of the complex WKB method \ref{thm:fundamental-theorem} to find solutions approximated by the WKB function $e^{\int^x_{\gamma,x'} \left[\hbar^{-1} \sqrt{V(y)}-\frac{V'(y)}{4 V(y)} \right]dy}$. 

Since we are interested in the behaviour of solutions at $0$ and $\infty$, our next task is the study of strictly admissible curves at $0$ and $\infty$.

\begin{definition}\label{def:actionfunction} 
Let $D \subset \WCs$ be a domain such that $V(x) \neq 0, \forall x \in D$, and $x' \in D$ an arbitrary but fixed point. Given a choice of $\sqrt{V}$ on $D$,
we denote by $S$ the \textit{action function}
\begin{eqnarray}\label{eq:actionfunction}
    S(x;E,\ell)= \int^x_{x'} \sqrt{V(y;E,\ell)} \, dy
\end{eqnarray}
\end{definition}
We notice that $S: D \to \C$ is a conformal map since it is holomorphic and its derivative never vanishes. Therefore, the lines of steepest descent (or ascent, depending on the orientation) of $\Re S$ are the level curves of the
function $\Im S$. These are called horizontal trajectories of quadratic differential $V(x) dx^2$ and play an important role in complex analysis and mathematical physics. They also play a fundamental role in the complex WKB method and will be briefly studied here,
following the classical reference \cite{strebel}.
\begin{definition}
Let $-\frac{\pi}{2}<\theta \leq \frac\pi2$. A $\theta$-trajectory of the quadratic differential $V(x) dx^2$ is a curve along which $\Im \left( e^{-i\theta} S\right)$ is constant.

In particular, a $0$-trajectory, namely a trajectory along which $\Im  S$ is constant, is called a horizontal trajectory. A $\frac{\pi}{2}$-trajectory, namely a trajectory along which $\Re  S$ is constant, is called a vertical trajectory.
\end{definition} 
In order to study $\theta$-trajectories at $0$ and $\infty$, we need first to address the study of the local behaviour of the action function about these points. 
We have the following Lemma, which is left as an exercise to the reader
\begin{lemma}
The following asymptotic identities hold:
 \begin{align}  \label{eq:assrtV0}
  & \sqrt{V(x)}= \frac{\ell+\frac12}{x} + O(|x|), \quad x \to 0, \\ \label{eq:assrtVinf}
 & \sqrt{V(x)}= 
  x^{\alpha} \left(1+ \sum_{k=1}^{\left\lfloor \frac{1+\alpha}{2\alpha} \right\rfloor} c_k E^k x^{-2 k\,\alpha} \right) + 
  O(|x|^{-1-d_{\alpha}}) , \quad x \to \infty.
\end{align}
In the latter formula $c_k$ is the $k-th$ coefficient of the Taylor series of $(1-t)^{\frac12}$ at $t=0$,
and
\begin{equation}\label{eq:dalpha}
  d_{\alpha}=\alpha \left(1+ 2 \left\lfloor \frac{1+\alpha}{2\alpha} \right\rfloor \right)-1, \quad 0<d_{\alpha}\leq 2\alpha.  
\end{equation}
The asymptotic identity \eqref{eq:assrtV0} holds uniformly in any compact subset of the space of parameters $(E,\ell)\in \C \times \{\Re\ell\geq -\frac12, \ell \neq -\frac12\}$; the asymptotic identity \eqref{eq:assrtVinf} holds uniformly in any compact subset of the space of parameters $(E,\ell) \in \C \times \{\Re \ell \geq -\frac12 \}$. 
\begin{proof}
Straightforward computation.
\end{proof}
\end{lemma}
Due to the above expansions (\ref{eq:assrtV0},\ref{eq:assrtVinf}), the function $S$ has (up to a sign and an additive constant depending on the lower integration point $x'$) the following behaviour at $0$ and $\infty$,
\begin{align}
 \label{eq:expansionS0}
 & S(x)  = \big(\ell+\frac12\big) \log x + O(|x|), \quad x \to 0, \\[2ex]
\label{eq:expansionS}
 & S(x)  = R(x)+ O(|x|^{-d_{\alpha}}),  \quad x \to \infty,\\[2ex] \nonumber
  &  R(x)  = \begin{cases}
     \frac{x^{\alpha+1}}{\alpha+1} + \sum_{k=1}^{\left\lfloor \frac{1+\alpha}{2\alpha} \right\rfloor}  c_k E^k  \frac{ x^{\alpha(1-2k)+1}}{\alpha(1-2k)+1}, & \alpha \neq \frac{1}{2m-1},\, m \in \N^*, \\[2ex]
     \frac{x^{\alpha+1}}{\alpha+1} + \sum_{k=1}^{m-1}  c_k E^k \frac{ x^{\alpha(1-2k)+1}}{\alpha(1-2k)+1}+ c_m
     E^m \log x, & \alpha = \frac{1}{2m-1},\, m \in \N^*.
    \end{cases} \\ 
\end{align}
In the above equation $c_k$ and $d_{\alpha}>0$ are as in  \eqref{eq:assrtVinf}, and the terms $O(f(x))$ are uniform in compact subsets of the space of parameters.

Following Strebel, we can introduce local coordinates at $0$ and $\infty$ such that, in these coordinates, the action function coincides with the dominant term of the above expansions. 
\begin{proposition}\label{prop:localS}
\begin{itemize}
\item[I)] Assume $\alpha\notin \{ \frac{1}{2m-1}, m \in \N^* \}$.
Let $(E,\ell)$ belong to a compact subset $W$ of $ \C \times \{ \Re \ell\geq -\frac12\}$. There exists a holomorphic local change of coordinate
$x=\varphi(z)=z + O(|z|^{1-2\alpha})$ as $z \to \infty$ such
that
\begin{equation}
\widetilde{S}(z):= S(\varphi(z))= \frac{z^{\alpha+1}}{\alpha+1}.
\end{equation}
Moreover, $\varphi'(z )=1+O(|z|^{-2\alpha})$.

\item[II)] Assume $\alpha=\frac{1}{2m-1}$ with $m \in \N^*$. 
Let $(E,\ell)$ belong to a compact subset $W$ of $ \C \times \{ \Re \ell\geq -\frac12\}$.
There exists a holomorphic local change of coordinate
$x=\varphi(z)=z + O(|z|^{1-2\alpha})$ as $z \to \infty$ such
that
\begin{equation}
\widetilde{S}(z):= S(\varphi(z))= \frac{z^{\alpha+1}}{\alpha+1}+c_m E^m \log z,
\end{equation}
with $c_m$ the $m-th$ term in the Taylor expansion of $(1-t)^{\frac12}$. 
Moreover, $\varphi'(z )=1+O(|z|^{-2\alpha})$.

\item[III)] Let $(E,\ell)$ belong to a compact subset $W$ of $ \C \times \{ \Re \ell \geq -\frac12, \ell \neq -\frac12\}$. There exists a holomorphic local change of coordinate
$x=\varphi(z)= z + O(|z|^2)$ as $z \to 0$  such
that
\begin{equation}
\widetilde{S}(z):= S(\varphi(z))= \left(\ell+\frac12 \right) \log z.
\end{equation}
Moreover, $\varphi'(z)=O(|z|)$.
\end{itemize}
 
In all statements above, the terms $O(|f(z)|)$ are meant uniformly with respect to $W$.
\end{proposition}
\begin{proof}
This is proven in \cite[Theorem 7.2 and 7.4]{strebel}.

\end{proof}

Due to Proposition \ref{prop:localS}, we can compute the local behaviour of horizontal and vertical trajectories about $\infty, 0$. 
\\

\begin{figure}[H]
\includegraphics[width=8cm]{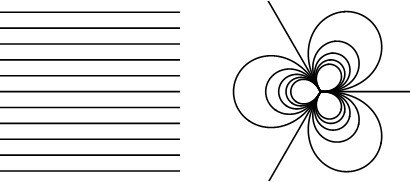}
\caption{Horizontal trajectories in a neighbourhood of a regular point and in a neighbourhood of $\infty$ when $\alpha=\frac12$.}
\label{fig:regularsimpleairy}
\end{figure}

\textbf{At $\infty$ when $\alpha \neq \frac{1}{2m-1}, m \in \N^*$.} In the coordinate $z$ introduced in point I) of Proposition \ref{prop:localS}, the trajectories defined by $\Im e^{-i \theta} S=c $ have polar
representation
\begin{align}\label{eq:standardhorinf}
\begin{cases}
\rho(\varphi)= \left(\frac{c}{\sin\big((\alpha+1)\varphi-\theta\big)}\right)^{\frac{1}{\alpha+1}},  &  \frac{\theta+2 k \pi}{\alpha+1}<\varphi < 
\frac{\theta+ (2k+1)\pi }{\alpha+1}
, \quad c>0 \\[2ex]
\varphi= \frac{\theta+ k \pi}{\alpha+1}, &  c= 0 \\[2ex]
\rho(\varphi)= \left(\frac{c}{\sin\big((\alpha+1)\varphi-\theta\big)}\right)^{\frac{1}{\alpha+1}}, & \frac{\theta+ (2k+1) \pi}{\alpha+1}<\varphi < 
\frac{\theta+ (2k+2) \pi }{\alpha+1}, \quad  c<0,
\end{cases} 
\end{align}
with $k \in \Z$ arbitrary. The horizontal trajectories at $\infty$ are represented in Figure \ref{fig:regularsimpleairy}.
\\

\begin{figure}[H]
\includegraphics[width=12cm]{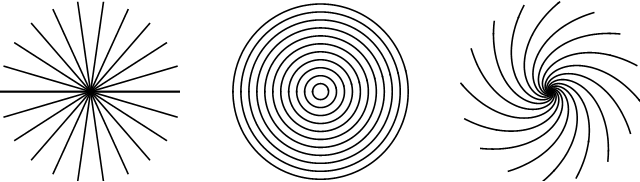}
\caption{$2\alpha \in \N$. Horizontal trajectories in a neighbourhood of $x=0$, when $\operatorname{Im}\left(\ell+\frac12\right)=0$, $\operatorname{Re}\left(\ell+\frac12\right)=0$, and when $\operatorname{Im}\left(\ell+\frac12\right),\operatorname{Re}\left(\ell+\frac12\right)\neq 0$.}
\label{fig:fuchsian}
\end{figure}

\textbf{At $0$.} In the coordinate $z$ introduced in point II) of Proposition \ref{prop:localS}, the trajectories defined by $\Im e^{-i \theta} S=c $ have the following polar representation. If $e^{-i\vartheta} (\ell+\frac12)= t e^{-i \beta}$ with $t>0$, then we have three different cases, depending on $\beta$:
\begin{align}\label{eq:standardhorzero}
\begin{cases}
\varphi= \frac{c}{t} 
, & \beta \in {0,\pi}, \\[2ex]
\log \rho= \frac{c}{t}, & \beta \in \{ \frac\pi2,-\frac\pi2 \}, \\[2ex]
\log \rho= -\big(\frac{c}{t} \cot \beta \big) \varphi, & \varphi \in \R, \beta \notin 
\{-\frac\pi2,0,+\frac\pi2,\pi\}.
\end{cases} 
\end{align}
In the first case, the trajectory is a straight line going through the origin, in the second case a circle centred at the origin, and in the third case an infinite spiral entering in the origin. The horizontal trajectories at $0$ are represented in Figure \ref{fig:fuchsian}.

\begin{definition}\label{def:standardtrajectories}
We name the $\theta$-trajectories of the quadratic differential $x^{2\alpha} dx^2$, whose explicit form is given in \eqref{eq:standardhorinf}, the standard $\theta$-trajectories at $\infty$.
We name the $\theta$-trajectories of the quadratic differential $\frac{\left(\ell+\frac12\right)^2}{x^2} dx^2$, whose explicit form is given in \eqref{eq:standardhorzero}, the standard $\theta$ trajectories at $0$.
\end{definition}

From the above computations, we immediately deduce the following result, which mimics \cite[Theorem 7.2, Theorem 7.4]{strebel}, and describes the qualitative behaviour of $\theta$-trajectories locally at $0$ and $\infty$. 
\begin{proposition}\label{prop:localhor}
\begin{itemize}
\item[I)] For every $(E,\ell)$, the $\theta$-trajectories of the quadratic differential $V(x;E,\ell) dx^2$ enters into $\infty$ in the distinguished directions $\arg x= \frac{\theta+ k\,\pi}{\alpha+1} $, with $k \in \Z$.
There exists $M>0$ such that any trajectory that enters the domain $D_M^{\infty}=\{ x \in \WCs, |x| >M \}$  tends to $\infty$. Moreover, any trajectory that lies in $D_M$ tends to $\infty$ into two consecutive distinguished directions.

\item[II)] For every $(E,\ell),\, \ell \neq -\frac12$, in a domain of the form $D_M^0= \{ x \in \WCs, |x|< M \}$ with $M$ sufficiently small, the $\theta$-trajectories of the quadratic differential $V(x;E,\ell) dx^2$, are either `circles' (namely, lines whose polar representation is $\rho(\varphi)$ with $\rho$ a quasi-periodic function), or half-lines entering into the origin, or logarithmic spirals. The three cases depends on whether $e^{-i\vartheta} \left(\ell+\frac12\right)$ is real, or imaginary, or complex.
\end{itemize}
\end{proposition}

In order to apply the Fundamental Theorem of the WKB Approximation on curves \ref{thm:fundamental-theorem}, we substantiate the above qualitative discussion, with quantitative estimates. 

\begin{lemma}\label{lem:horatinftyzero}
\begin{itemize}
\item[I)] Fix a compact subset $W$ of the space of parameters $\C \times \{\Re \ell \geq -\frac12 \}$, a number $\theta$, $|\theta|<\frac\pi2$, and  $k \in \Z$. Let $D^{\infty}_M= \{ x \in \WCs,\, |x| >M \}$. For all $M$ sufficiently large so that $V(x,E,\ell)\neq 0,\,  \forall (x,E,\ell) \in D_M \times W$,
choose  $\sqrt{V}$ in such a way that $\lim_{t \to +\infty }\Re S(e^{i \frac{ k \pi}{\alpha+1}})=-\infty$.
Let finally $\gamma_{\theta}:(0,1) \to D^{\infty}_M$ be a parameterised  branch of a standard $\theta$-trajectory at $\infty$ 
lying in $D^{\infty}_M$ and such that $\lim_{t \to 0^+} \arg \gamma_{\theta}=\frac{\theta+ k \pi}{\alpha+1}$.

If $M$ is sufficiently large, $\gamma_{\theta}$ is strictly admissible, and 
\begin{equation}\label{eq:rhogammatheta}
    \rho_{\gamma_{\theta}}(t)= O \left(\big|\gamma_{\theta}\big|^{-\alpha-1}\right)
\end{equation}
uniformly in $W$.

\item[II)] Fix a compact subset $W$ of the space of parameters $\C \times \{ \Re \ell > -\frac{1}{2}, \ell \neq 0\}$.
Let $D^0_M= \{ x \in \WCs,\, |x| >M \}$. For all $M$ sufficiently small so that $V(x,E,\ell)\neq 0,\, \forall (x,E,\ell) \in D^0_M \times W$, choose  $\sqrt{V}$ in such a way that $\lim_{x \to 0 }\Re S(x)=-\infty$.
Finally,  let $\gamma_{\varphi}:(0,1) \to D^0_M$, $\varphi \in \R$, be a parametrisation of the segment, $\{ x \in \WCs,\, \arg x= \varphi,\, 0<|x| < M\}$, so that $\lim_{t \to 0^+} \gamma_{\varphi}(t)=0$.

If $M$ is sufficiently small, then $\gamma_{\varphi}$ is strictly admissible, and 
\begin{equation}
\rho_{\gamma_{\varphi}}(t)= O\left(\big|\gamma_{\varphi}(t)\big|\right) ,
\end{equation}
uniformly in $W$.
\end{itemize}
\end{lemma}
\begin{proof}
We prove I) and leave II) as an exercise to the reader.
\\

I).\quad By Definition \ref{def:admissiblecurves}, a curve $\gamma$ is strictly admissible if $(U,\sqrt{V},\gamma)$ is an admissible triple, and $\Re S$ is strictly monotone along $\gamma$. We therefore need to prove the following: if $M$ is sufficiently large then
\begin{itemize}
    \item[i)] $\Re S$ is strictly monotone increasing along $\gamma_{\theta}$ for every $(E,\ell) \in W$.
    \item[ii)] $|F(\gamma(t);E,\ell)||\dot{\gamma}_{\theta}|$ is uniformly integrable on $W$.
\end{itemize}
Property i) follows from \eqref{eq:expansionS}, and property ii) follows from corollary \ref{propo:Vlanger}. By Lemma \ref{lem:Vlanger}, $F=O(x^{-\alpha-2})$ as $|x| \to \infty$ - from this \eqref{eq:rhogammatheta} follows.
\end{proof}
Since we have found strictly admissible curves, starting at $0$ and $\infty$, from the Fundamental Theorem of the WKB Approximation on curves, Theorem \ref{thm:fundamental-theorem}, we immediately deduce the following
\begin{theorem}
\begin{itemize}
  \item[I)]  For every $(k,\theta,c) \in \Z \times (-\frac\pi2,\frac{\pi}{2}) \times \C^*$, there exists a unique subdominant solution $\Psi_k(x;E,\ell)$  to \eqref{eq:anharmonic} on the ray $\left\{\arg(x)=\frac{\theta+k\pi}{\alpha+1}\right\}$ such that
    \begin{align}
     &  \lim_{t \to + \infty}  \left| \Psi_k\left(t e^{i \frac{\theta+ k \pi}{\alpha+1}}; E ,\ell\right)  e^{(-1)^{k+1} R\left(t \, e^{i \frac{\theta+ k \pi}{\alpha+1}}\right)}-c \right| = O(t^{-1-e_\alpha}), \\
     & \; e_{\alpha}= \min\{d_{\alpha},\alpha+1\},
    \end{align}
    where $d_{\alpha}$ is as in \eqref{eq:dalpha}. In the above estimate, $ O(t^{-1-\alpha}) $ is uniform with respect to any compact subset of the space of parameters $\C \times \{\Re \ell \geq -\frac12 \}$. Moreover, the function $\Psi_k(x;E,\ell)$ is entire with respect to $(E,\ell)$.

    \item[II)] Let $\ell \in \{ \Re \ell > -\frac12, \Re \ell \neq 0 \}$. For every $\theta \in  \R$, there exists a unique subdominant solution $\chi_+(x;E,\ell)$ to \eqref{eq:anharmonic}  on the ray $\left\{\arg(x)= \theta\right\} $ such that
    \begin{eqnarray}
      \lim_{t \to 0^+}  \left| \chi_+ (t e^{i \theta}; E ,\ell)  t^{-(\ell+1)} e^{-i \theta (\ell+1)}-c\right| = O(t).
    \end{eqnarray}
    In the above estimate $ O(t^{-1}) $ is uniform with respect to any compact subset of the space of parameters $\C \times \{ \Re \ell > -\frac{1}{2}, \ell \neq 0\}$. Moreover, the function $\chi_+(x;E,\ell)$ is holomorphic with respect to $(E,\ell)$.

\end{itemize}

\end{theorem}
Using the local behaviour at $0$ and $\infty$ of $\theta-$trajectories and the Fundamental Theorem of the WKB Approximation on curves, we have deduced the existence of:
\begin{itemize}
    \item A unique subdominant solution on any ray entering $0$;
    \item A unique subdominant solution on any ray entering $\infty$, but on the forbidden rays (sometime called \textit{Stokes lines})  $ \left\{\arg x= \frac{\frac{\pi}{2}+ k \pi}{\alpha+1}, k \in \Z \right\}$.
\end{itemize}
We should now ask ourselves: Are these solutions the restriction of the same global solution?
The situation is rather different in the two cases, at $x=0$ and at $x=\infty$. In the case at $x=0$, there exists a unique global solution that vanishes along any ray approaching $0$.
In the case $x=\infty$, we divide $\WCs$ in an infinite number of sectors of amplitude $\frac{\pi}{\alpha+1}$, called Stokes sectors, whose boundaries are the forbidden rays described above. For every such a sector, there exists a unique global subdominant solution.
\begin{definition}
\label{stokes-sectors}
For any $k \in \Z$, the $k$-th Stokes sector $\Sigma_k$ is defined as,
\begin{equation}
   \Sigma_k= \left\lbrace x \in \WCs, |\arg x- \frac{k\,\pi}{\alpha+1}| < \frac{\pi}{2\alpha+2} \right\rbrace \subset \WCs.  
\end{equation}   
\end{definition}
If we restrict to  $\{ |x|>M \}$ where $M$ is large enough so that $V$ is never-vanishing, the WKB function is of the form 
$$
\Psi^{W}(x)= C\, V(x)^{-\frac14}e^{\pm S(x)},  \quad S(x)= \int^x_{x'} \sqrt{V(y)} dy, \quad C \neq 0,
$$
where $S(x)$ is the integral of $\sqrt{V}$ with asymptotics $S(x)= \frac{x^{\alpha+1}}{\alpha+1}+o(x^{\alpha+1})$ -- see equation \eqref{eq:expansionS} -- and the sign $\pm $
is fixed by requiring that $\Re S(\gamma(t)) \to -\infty$ as $t \to 0$.

We see that the eventual sign of $\Re S(x)$, for $|x|$ large, is alternating: it is $+1$ in the even Stokes sectors, it is $-1$ in the odd Stokes sectors, and indeterminate on the boundary between sectors. 
$\Re S$ changes sign when crossing the rays $\arg x= \frac{\frac{\pi}{2}+ k \pi}{\alpha+1},\, k \in \Z$. This corresponds to two important facts, which we will prove below and constitute the essence of the Stokes phenomenon:
\begin{itemize}
    \item[i)] For every Stokes sector $\Sigma_k$, there exists a unique (up to a scalar multiple) solution $\Psi_k$ -- called Sibuya solution -- vanishing exponentially fast along any ray contained in the sector. Such a solution diverges exponentially fast along any ray contained in the neighbouring sectors.
    \item[ii)] Along the rays $\arg x=\frac{(2k +1)\pi }{2\alpha+2}$, $k \in \Z$, the function $x^{\frac\alpha2} \psi(x)$, with $\psi$ an arbitrary non-zero solution, is oscillatory.
\end{itemize}
We provide a proof of the above claims in Theorem \ref{thm:sibuya} and Theorem \ref{thm:chip} below, starting from the analysis at $x=\infty$, which is more interesting.

\subsubsection{Behaviour of solutions at $\infty$: Stokes phenomenon}
In order to study the WKB asymptotic in a domain in the complex plane, we extend the theory that we have developed for the case of curves.
Given the approximate solution
\begin{equation*}
\Psi^W(x, x'):=\exp\left(\int_{x'}^x \left[\sqrt{V(y)}-\frac{1}{4}\frac{V'(y)}{V(y)}\right] dy\right),
\end{equation*}
we fix a point $x_0 \in \overline{D}$, and look for a solution $\psi: D \to \C$ of equation \eqref{eq:anharmonic} with the boundary conditions 
$$
\lim_{x \to x_0 } \frac{\psi(x)}{\Psi^W(x,x')}=1,\quad \lim_{x \to x_0 } \frac{d}{dx}\frac{\psi(x)}{\Psi^W(x,x')}=0, \quad x_0 \in \overline{D}.
$$
Reasoning as in Lecture I, we deduce that  $z(x)= \frac{\psi(x)}{\Psi^W(x,x')}$ must solve the integral equation 
\begin{align*}\nonumber
&  z=1-\mathcal{K}[z],\\
&  \mathcal{K}[f](x)=\int_{\gamma,x_0}^x \left(\frac{e^{-2 \int^x_{\gamma,y}\sqrt{V(w)} dw}-1}{2} \right) F(y) f(y) dy ,
\end{align*}
where $\gamma$ is a curve connecting $x_0$ with $x$, and the forcing term $F$ is as per \eqref{eqn:forcing-term}. We aim to solve the above integral equation on the space on the Banach space $\mathcal{H}(D)$, of bounded continuous functions on $\overline{D}$, which are analytic on $D$. To this aim, we need to define a domain $D$ and a set of admissible curves, starting at $x_0$, such that
\begin{itemize}
\item Any point in $\overline{D}$ is connected to $x_0$ by an admissible curve;
\item The integral $\mathcal{K}[f](x)$ does not depend on the admissible curve;
\item The integral $\mathcal{K}[f](x)$ is analytic and bounded.
\end{itemize}

Let us now fix $x_0 =\infty$. A convenient domain is chosen as follows.
\begin{definition}\label{def:volterradomains}
For any $M>0$, we let $$Q_M = \{ z \in \C, \Re z >M \} \bigcup \{z \in \C, \Im z < - M \} \bigcup  \{z \in \C, \Im z > M \}.$$
Moreover, given $(k,\theta) \in \Z  \times (-\frac{\pi}{2},\frac\pi2)$ we let
$$ D_{M,k,\theta} = \left\{ x \in \WCs, e^{-i\theta}x^{\alpha+1} \in Q_M,  | \arg{x} -\frac{k\pi +\theta}{\alpha+1}|< \frac{\pi}{\alpha+1} \right\}.$$
For every $\e \in \left(0,\frac{\pi}{2}\right)$, we let $$D^{\e}_{M,k} = \bigcup_{|\theta|\leq \frac{\pi}{2}-\e} D_{M,k,\theta} \subset \WCs .$$
Finally, we denote by  $\overline{D}^{\e}_{M,k}$  the closure of $D^{\e}_{M,k}$ in $\widetilde{\mathbb{C}^*}$.
\end{definition}
Before proceeding further, we explain the choice of the domain $\overline{D}^{\e}_{M,k}$. The domain $D_{M,k,\theta}$ with is foliated by standard $\theta$-trajectories which starts at $\infty$ inside the sector $\Sigma_k$. In fact, $D_{M,k,\theta}$ are the connected (and simply connected) components of the counterimage, under the map $e^{-i\theta}x^{\alpha+1}$, of the domain $Q_M$.  
The domain $Q_M$ is foliated by the lines $\Im z=c$ with $|c|>M$, and by the half-line $\Im z=c$, $\Re z >M$ with $|c| \leq M$, whose counterimages are standard $\theta$-trajectories. Therefore,
\begin{itemize}
\item $\overline{D}^{\e}_{M,k}$  \textit{coincides asymptotically} with the closed sector 
 $ \left|\arg x - \frac{k\,\pi}{\alpha+1} \right| \leq \frac{3 \pi-\e}{2\alpha+2}$. This is the union of the closure of the Stokes sector $\Sigma_k$ with the neighbouring sectors $\Sigma_{k\pm1}$,
 but for two $\e-$small subsectors, localised at the boundary between $\Sigma_{k\pm1}$ and $\Sigma_{k\pm2}$.
 \item Through every point in $\overline{D}^{\e}_{M,k}$ passes at least a standard $\theta$-trajectory (with $|\theta|\leq\frac\pi2-\e$) which starts at $\infty$ inside the Stokes sector $\Sigma_k$. These, as we know from Lemma \ref{lem:horatinftyzero}, are strictly admissible curves if $M$ is large enough. 
\end{itemize}
For these two properties, $\overline{D}^{\e}_{M,k}$ is a good candidate for a domain on which we can define the Volterra integral equation. Given such a domain, and fixed a compact $W$ in the space of parameters, the natural linear space on which the Volterra operator acts is $\mathcal{H}(D^{\e}_{M,k} \times \operatorname{int} W)$: the space of  bounded continuous functions on $\overline{D}^{\e}_{M,k} \times W$, which restricted to $D^{\e}_{M,k} \times \operatorname{int} W $ are holomorphic; equipped with the sup norm $\| \cdot\|_{\infty}$, this is a Banach space, see Proposition \ref{prop:closure}, in Appendix 1. We have collected all ingredients to define the Volterra operator.
\begin{definition}\label{def:volterraondomain}
Fix a compact $W$ in the space of parameters $\C \times \{\Re \ell \geq -\frac12 \}$, an integer $k \in \Z$ and an $0 < \e < \frac\pi2$. Given $M$ big enough so that $V(x,E,\ell) \neq 0 $ for all $(x,E,\ell) \in D^{\e}_{M,k} \times W$, choose the branch of $\sqrt{V}$ in such a way that $\lim \Re S = -\infty$ as $x \to \infty$ along the ray $\arg x= \frac{k  \pi}{\alpha+1}$.
For every $f \in \mathcal{H}(D^{\e}_{M,k}\times \operatorname{int} W)$, we define
\begin{equation}\label{eq:Kdomaindef}
 \mathcal{K}[f](x;E,\ell)=\int_{\gamma,\infty}^x \left(\frac{e^{-2 \int^x_{\gamma,y}\sqrt{V(w;E,\ell)} dw}-1}{2} \right) F(y;E,\ell) f(y;E,\ell) dy .
\end{equation}
In the above formula, $\gamma$ is a standard $\theta$-trajectory, with $|\theta|\leq \frac{\pi}{2}-\e$ connecting $x$ with $\infty$.
\end{definition}
The Volterra operator is well-defined.
\begin{lemma}\label{lem:KDinfty}
Fix a compact $W$ in the space of parameters $\C \times\left\{\Re \ell \geq -\frac12 \right\}$, an integer $k \in \Z$ and an $ 0 < \e < \frac\pi2$. If $M>0$ is large enough, the expression \eqref{eq:Kdomaindef} defines a bounded operator $\mathcal{K}\colon \mathcal{H}({D}^{\e}_{M,k} \times \operatorname{int} W) \to \mathcal{H}({D}^{\e}_{M,k} \times \operatorname{int} W)$. Moreover,
\begin{equation}\label{eq:Kfxf}
|\mathcal{K}[f](x)| \leq C_{W,\e,k} |x|^{-1-\alpha} \|f\|_{\infty}.
\end{equation}
\begin{proof}
We provide a sketch the proof, since a full proof has recently appeared in \cite[Lemma 2.8]{cotti23} in a much more general setting. 
The few steps, that are needed to reduce the general case studied in \cite[Lemma 2.8]{cotti23}  to the particular case studied, are an exercise to the reader.

We have to prove the following: If $M$ is large enough, then
\begin{itemize}
    \item[i)] The value of $\mathcal{K}[f](x;E,\ell)$ does not depend on the integration curve.
    \item[ii)] $\mathcal{K}[f] \in \mathcal{H}({D}^{\e}_{M,k} \times \operatorname{int} W)$, namely 
    \begin{itemize}
        \item $\mathcal{K}[f] $ is bounded and continuos on $\overline{D}^{\e}_{M,k}\times W$;
        \item $\mathcal{K}[f] $ restricted to $D^{\e}_{M,k} \times \operatorname{int} W $is holomorphic;
    \end{itemize}
    \item[iii)] Estimate \eqref{eq:Kfxf} holds.
\end{itemize}
Proof of i). \quad Let $\gamma_{\theta_1}$ and $\gamma_{\theta_2}$ with  $-\frac{\pi}{2}+\e\leq \theta_1<\theta_2\leq\frac{\pi}{2}-\e$ be two standard $\theta$-trajectories passing through $x$. Fix a number $\theta_3$ with $\theta_2<\theta_3 <\frac\pi2$ and choose
a sequence of points $x_n$ on $\gamma_{\theta_1}$ converging to $\infty$. For $n$ large enough, through every $x_n$ passes a $\theta_3$-trajectory; we let $l_n$ denote the segment of such a trajectory connecting the point $x_n$ with the $\theta_2$-trajectory. Denoting by $\tilde{x}_n$ the intersection of $l_n$ with $\gamma_{\theta_2}$, we have that
\begin{align*}
& \int_{\gamma_{\theta_1},\infty}^x \left(\frac{e^{-2 \int^x_{\gamma,y}\sqrt{V(w;E,\ell)} dw}-1}{2} \right) F(y;E,\ell) f(y;E,\ell) dy \\
- & \int_{\gamma_{\theta_2},\infty}^x \left(\frac{e^{-2 \int^x_{\gamma,y}\sqrt{V(w;E,\ell)} dw}-1}{2} \right) F(y;E,\ell) f(y;E,\ell) dy \\
= &\int_{\gamma_{\theta_1},\infty}^{x_n} \left(\frac{e^{-2 \int^x_{\gamma,y}\sqrt{V(w;E,\ell)} dw}-1}{2} \right) F(y;E,\ell) f(y;E,\ell) dy \\
- &\int_{\gamma_{\theta_2},\infty}^{\tilde{x}_n} \left(\frac{e^{-2 \int^x_{\gamma,y}\sqrt{V(w;E,\ell)} dw}-1}{2} \right) F(y;E,\ell) f(y;E,\ell) dy \\
+ & \int_{l_n,x_n}^{\tilde{x}_n} \left(\frac{e^{-2 \int^x_{\gamma,y}\sqrt{V(w;E,\ell)} dw}-1}{2} \right) F(y;E,\ell) f(y;E,\ell) dy .
\end{align*}
Since, for $M$ large enough, the function $ \Re \int^x_{\gamma_{\theta},y}\sqrt{V(w;E,\ell)} dw$ is monotonically increasing along a $\theta$-trajectory, then we can estimate from above the absolute value of the right-hand side by
\begin{align*}
 &  \|f\|_{\infty}\left( \int_{\gamma_{\theta_1},\infty}^{x_n} |F(y;E,\ell)||dy|+\int_{\gamma_{\theta_2},\infty}^{\tilde{x}_n} |F(y;E,\ell)||dy| + \int_{l_n,x_n}^{\tilde{x_n}} |F(y;E,\ell)||dy| \right).
\end{align*}
By Lemma \ref{lem:Vlanger}, $|F(x)| \leq C_W \frac{1}{1+ |x|^2}$, therefore
the latter expression is estimated from above by 
\begin{align*}
 & C_W  \|f\|_{\infty}\left( \int_{\gamma_{\theta_1},\infty}^{x_n} \frac{|dy|}{1+ |y|^2}+\int_{\gamma_{\theta_2},\infty}^{\tilde{x}_n}\frac{|dy|}{1+ |y|^2} + \int_{l_n,x_n}^{\tilde{x}_n}\frac{|dy|}{1+ |y|^2} \right).
\end{align*}
Since the three $\theta$-trajectories have finite spherical length, then the three terms converge to $0$ as $n \to \infty$.
\\

Proof of ii). \quad The first part is essentially identical to the proof of Proposition \ref{prop:Kgamma}. The second part is straightforward.
\\

Proof of iii). \quad To prove estimate \eqref{eq:Kkestimate}, we make the change of variable $w=x^{-\alpha-1}$. In this coordinate, the standard $\theta$-trajectories \eqref{eq:standardhorinf} become simple curves passing through centred at $w=0$ with maximal distance from $w=0$ equals to  $\sup_{t}|\gamma_{\theta}(t)|^{-\alpha-1}$; moreover, since $|F(x)|=O(|x^{-\alpha-2}|)$, as proven in Lemma \ref{lem:Vlanger}, $|F(w)||dx| \leq C |dw|$ for some positive $C$.
\end{proof}
 \end{lemma}

Having extended the Volterra operator to sectorial neighbourhoods of $\infty$, we can prove the existence of Sibuya solutions. 

\begin{theorem}\label{thm:sibuya}[\cite[Proposition 4.5]{masoero2024qfunctions}]
Let $d_{\alpha}>0$ and $R(x)$ as in \eqref{eq:dalpha}. For every $k \in \Z$, there exists a unique solution $\Psi_k(x;E,\ell)$, called k-th Sibuya solution, such that, for every $\e>0$,
\begin{eqnarray}\label{eq:psiktheorem}
    \left|\Psi_k(x;E,\ell) x^{\frac{\alpha}{2}}e^{-(-1)^k R(x)}-1 \right| = O (|x^{-e_{\alpha}}|), \; e_{\alpha}= \min\{d_{\alpha},\alpha+1\},
\end{eqnarray}
as $x \to \infty$ in $ \left|\arg x - \frac{k\,\pi}{\alpha+1} \right| \leq \frac{ \pi}{2\alpha+2}-\e$. The term $O (|x^{-e_{\alpha}}|)$ is uniform in any compact subset of the space of parameters. Moreover, the solution $\Psi_k(x;E,\ell)$ is entire with respect to $E,\ell$.
\begin{proof}
Proof of Existence. \quad We fix a compact $W$ in the space of parameters and an $\e'<\e$, so that $ \left|\arg x - \frac{k\,\pi}{\alpha+1} \right| \leq \frac{ \pi}{2\alpha+2}-\e$ is eventually contained in $\overline{D}^{\e'}_{M,k}$. 
We consider the Volterra integral equation
\begin{equation*}
    z=1-\mathcal{K}[z],\quad z \in \mathcal{H}(D^{\e}_{M,k} \times \operatorname{int} W).
\end{equation*}
Due to \eqref{eq:Kfxf}, if $M$ is sufficiently large then $\|\mathcal{K}\| <1$, hence the integral equation admits a unique solution $\tilde{z}$, which can be written as $\sum_{n=0}^{\infty} K^n[1] $ with $\mathcal{K}^0[1]$ the constant function $1$. Hence,
\begin{equation}\label{eq:z1oxa}
    \tilde{z}(x;E,\ell)=1+O(|x|^{-1-\alpha})
\end{equation}

Fixed $x' \in   D^{\e'}_{M,k}$, by construction the function 
\[
\begin{aligned}
&\psi_k\colon D^{\e'}_{M,k} \times \operatorname{int} W \to \mathbb{C}, \\
&\psi_k(x;E,\ell) = \tilde{z}(x;E,\ell) e^{\int_{x'}^x\left[ \sqrt{V(y;E,\ell)}-\frac{V'(y;E,\ell)}{V(y;E,\ell)}\right]dy}
\end{aligned}
\]
satisfies the anharmonic oscillator equation \eqref{eq:anharmonic}, and it is analytic. Due to \eqref{eq:expansionS} and \eqref{eq:z1oxa}, $\psi_k(x;E,\ell)$ satisfies the estimate  
\begin{eqnarray*}
    \left|\psi_k(x;E,\ell) x^{\frac{\alpha}{2}}e^{-(-1)^k R(x)}-c(x',E,\ell) \right| = O (|x^{-e_{\alpha}}|), \; e_{\alpha}= \min\{d_{\alpha},\alpha+1\},
\end{eqnarray*}
where $c(x',E,\ell)$ is a never vanishing analytic function. Therefore, the analytic extension to $\WCs \times W$ of the function $\Psi_k(x;E,\ell):=\frac{\psi_k(x;E,\ell)}{c(x',E,\ell)}$ is a solution of \eqref{eq:anharmonic} satisfying the estimate \eqref{eq:psiktheorem}, and it is analytic.
\\

Proof of Uniqueness. \quad Uniqueness follows from existence. Since $\Psi_{k+1}$ diverges exponentially in $\Sigma_k$, any solution not-proportional to  $\Psi_k$ diverges in $\Sigma_k$. Therefore, $\Psi_k$ is uniquely determined by \eqref{eq:psiktheorem}.

\end{proof}
\end{theorem}

\begin{corollary}\label{cor:stokesmult}
For every $k$, the solutions $\{\Psi_k,\Psi_{k+1} \}$ forms a basis of the space of solutions.

In particular,
\begin{equation}\label{eq:stokesmult}
    \Psi_{k+1}(x;E,\ell)= \Psi_{k-1}(x;E,\ell) +\sigma_{k}(E,\ell) \Psi_k 
\end{equation}
where $\sigma_k(E,\ell)$ is an entire function of $E,\ell$.
\end{corollary}

\begin{proof}
Since $\{\Psi_{k-1},\Psi_{k} \}$ is a basis, then  $\Psi_{k+1}(x;E,\ell)= A(E,\ell) \Psi_{k-1}(x;E,\ell) +\sigma_{k}(E,\ell) \Psi_k$
for some functions $A$ e $\sigma$. After \eqref{eq:psiktheorem}, 
$$ \lim_{x \to +\infty, x \in \Sigma_k } \frac{\Psi_{k-1}(x)}{\Psi_k(x)} =1 , \quad \lim_{x \to +\infty, x \in \Sigma_k } \frac{\Psi_{k}(x)}{\Psi_k(x)} =0 $$
whenever the limit is taken along a ray inside $\Sigma_k$. Hence, $A=1$.
\end{proof}

\begin{definition}\label{eq:stokesmultiplier}
The quantity $\sigma_k$ defined in equation \eqref{eq:stokesmult} is called the $k$-th Stokes multiplier of the anharmonic oscillator.
\end{definition}

\subsubsection{Behaviour of solutions at $0$}
We can repeat the same construction to obtain a subdominant solution at $0$, that we call $\chi_+$. Notice that another more algebraic construction of $\chi_+$ is given in the Appendix 2.

We start by defining the domain of $\WCs$ and the functional space where we want to solve the Volterra integral equation.
 \begin{definition}\label{def:D0Mtheta}
Fix a compact $W$ in the space of parameters $ \C \times \{ \Re \ell>-\frac12,\, \ell \neq 0 \} $ and a $\theta>0$. For every  $M>0$,
let $D^{0}_{M,\theta}=\{ x \in \WCs,\, |\arg x|< \theta,\, |x|<M\}$, and $\overline{D}^{0}_{M,\theta}=\{ x \in \WCs,\, |\arg x|\leq \theta,\, |x|\leq M \} \cup \{0\} $. Let also $ \mathcal{H}(D^{0}_{M,\theta}\times \operatorname{int} W)$ denote the Banach space -- with respect to the sup norm -- of bounded continuous functions on $\overline{D}^{0}_{M,\theta}\times W$, which restricted to $D^{0}_{M,\theta} \times \operatorname{int} W$  are analytic. Fixed  $M$ so small that $V(x;E,\ell) \neq 0$ for all $(x,E,\ell) \in \overline{D}^{0}_{M,\theta} \times W $, choose
the branch of $\sqrt{V}$ in such a way that $\lim \Re S = -\infty$ as $x \to 0$.  
For every  $f \in \mathcal{H}(D^{0}_{M,\theta}\times \operatorname{int} W)$, define
\begin{equation}\label{eq:KD0Mtheta}
 \mathcal{K}[f](x;E,\ell)=\int_{0}^x 
 \left(\frac{e^{-2 \int^x_{\gamma,y}\sqrt{V(w;E,\ell)} dw}-1}{2} \right) F(y;E,\ell) f(y;E,\ell) dy .
\end{equation}
\end{definition}
Reasoning as in Lemma \ref{lem:KDinfty}, we deduce that \eqref{eq:KD0Mtheta} defines a bounded operator on $\mathcal{H}({D}^{0}_{M,\theta}\times\operatorname{int} W)$.
\begin{lemma}\label{lem:KD0}
 Fix $W$ and $\theta$ as in the definition \ref{def:D0Mtheta}. If $M$ is small enough $\mathcal{K}: \mathcal{H}(D^{0}_{M,\theta}\times \operatorname{int} W) \to  \mathcal{H}(D^{0}_{M,\theta}\times \operatorname{int} W)$ is a well-defined bounded operator, and
\begin{equation}
|\mathcal{K}[f](z;E,\ell)| \leq C_{W,\theta} |x| \|f\|_{\infty},
\end{equation}
for some constant $C_{W,\theta}>0$, depending on $W,\theta$ only.
\begin{proof}
 The proof is left to the reader.
\end{proof}
 \end{lemma}
As a corollary of the above lemma, we construct the subdominant solution $\chi_+$ with asymptotic behaviour $x^{\ell+1}\left( 1+ O (x) \right)$ as $x\to0$. 
\begin{theorem}\label{thm:chip}
There exists a unique solution $\chi_+$  such that or any $\e >0$
\begin{eqnarray}\label{eq:chipbehaviour}
    |\chi_+(x;E,\ell) x^{-\ell-1}-1| \leq C_{W,\e} |x|,\quad x \to 0 \mbox{ and } |\arg x|\leq \frac{1}{\e},
\end{eqnarray}
for any compact $W$ subset of the space of parameters $\C\times \left\{\Re \ell >-\frac12, \ell \neq 0 \right\}$. 
Such a solution is analytic with respect to the parameters $(E, \ell) \in \C\times \left\{\Re \ell >-\frac12, \ell \neq 0 \right\}$.
\begin{proof}
The proof of I. is a corollary of Lemma \ref{lem:KD0}, following the very same steps as in the proof of Theorem \ref{thm:sibuya}

\end{proof}
\end{theorem}
In the theorem above, we have constructed a function with subdominant asymptotic behaviour at $0$. What about solutions with
the dominant behaviour? When $2\alpha \in \N^*$, the potential is meromorphic, and one can construct a Frobenius solution with dominant behaviour $\chi_-(x)=x^{-\ell}\left( 1+ O(x) \right)$ by standard methods. In fact, as proven in \cite{masoero2024qfunctions}, a generalised Frobenius solution exists for every $\alpha$; see \eqref{eq:chi-ser} in Appendix 2. From that result, we deduce that for every $(E, \ell) \in \C\times \left\{\Re \ell >-\frac12, \ell \neq 0 \right\}$,
the dominant asymptotic behaviour
    \begin{eqnarray}\label{eq:chimbehaviour}
    \lim_{x \to 0,\,  |\arg x|\leq \frac{1}{\e}}  \chi(x;E,\ell) \; x^{\ell}=1, \mbox{ for every } \e>0,
 \end{eqnarray}
determines a one dimensional affine space, whose associated vector space in the linear span of the dominant solution $\chi_-$.

\begin{remark}
Notice that it is possible to construct the solution $\chi_-$ with a dominant behaviour at $0$ using the complex WKB method, see \cite{degano2024}. In fact, if for one sign of $\sqrt{V}$ the WKB function behaves as $x^{\ell+1}$ as $x \to 0$, with the opposite sign, the WKB function behaves as $x^{-\ell}$. However, since we do not need to work with these solutions in the sequel, we omit such a construction here.
\end{remark}

Having constructed the Sibuya solution $\Psi_0$, subdominant at $+\infty$, and the solution $\chi_+$, subdominant at $0^+$, we can define the spectral determinant.
\begin{definition}
The spectral determinant is the following functions of the parameters $E,\ell$,
\begin{equation}\label{eq:spectraldet}
Q_+: \C \times \left\{\Re \ell >-\frac12,\, \ell \neq 0\right\} \to \C, \quad Q_+(E,\ell):=Wr_x [\chi_+(x;E,\ell),\Psi_0(x;E,\ell)] 
\end{equation}
\end{definition}

\begin{theorem}
   The spectral determinant $Q_+$ is holomorphic in $\C\times \left\{\Re \ell >-\frac12 \right\}$.
    \begin{proof}
    From Theorems \ref{thm:sibuya} and \ref{thm:chip}, $Q_+$ holomorphic on $\C\times \left\{\Re \ell >-\frac12, \ell \neq 0 \right\}$. From the Frobenius series of $\chi_+$, equation \eqref{eq:chi+ser}, we deduce that, for fixed $E$, $\ell=0$ is a removable singularity of $Q_+$. Therefore, $Q_+$ extends holomorphically to $\C\times \left\{\Re \ell >-\frac12 \right\}$.
    \end{proof}
    \end{theorem}

 \subsection{Fock-Goncharov Coordinates}
 We introduce here an important concept in the theory of ODEs in the complex plane, and more in general of the theory of meromorphic functions, which is the notion of asymptotic values - see \cite{bergweiler95}. This allows us to substitute the spectral determinant 
$Q_+$ and the Stokes multipliers $\sigma_k$ with another set of spectral functions which are best suited to WKB analysis (and have a natural geometric interpretation) called Fock-Goncharov coordinates.

The idea behind the notion of asymptotic values is that we can enlarge the space $\WCs$ by adding boundary points which represent the asymptotic regimes/directions that emerged from the local study of solutions of the anharmonic oscillator, at $0$ and $\infty$.

\begin{definition}
We extend $\WCs$ by adding the boundary points $0$ and $\infty_k, k \in \Z$:
\begin{itemize}
    \item A sequence  $\{ x_m \}_{m \in \N}$ is said to converge to $\infty_k$ if $|x_m| \to +\infty$ and $\exists \e>0$ such that $|\arg x_m -\frac{k \pi}{\alpha+1}| \leq \frac{\pi-\e}{2\alpha+2}$ for $m$ large enough; 
    \item A sequence $\{x_m\}_{m\in\mathbb{N}}$ is said to converge to $0$ 
if $|x_m| \to 0$ and $\exists \e >0$ such that $ |\arg x_m | \leq \e$ for $m$ large enough.
\end{itemize}
\end{definition}
The great advantage of this new topological space is that  the ratio of any two solutions of the anharmonic oscillator, which is a holomorphic function $f:\WCs \to \Cb $, extends continuously to the boundary. 
\begin{definition}[Asymptotic values]
Let $\{ \psi, \tilde{\psi} \}$ be a basis of solutions of \eqref{eq:anharmonic}.
We let
\begin{align}\label{def:asymptotic-value-zero}
 &   W_{\underline{0}}\left(\psi, \tilde{\psi}\right)=\lim_{x\to 0} \frac{\psi(x)}{\widetilde{\psi}(x)} \; \in \Cb, \\ 
\label{def:asymptotic-values-infinity}
& W_{k}\left(\psi, \tilde{\psi}\right)= \lim_{x\to\infty_k} \frac{\psi(x)}{\widetilde{\psi}(x)} \; \in \Cb,  \; k \in \Z.
\end{align}  
$W_{\underline{0}}$ and $W_k$ are called the asymptotic values of the anharmonic oscillator.
\end{definition}
In the following lemma, we collect the main properties of the asymptotic values.
\begin{lemma}\label{lem:Ws}
Assume that $\Re \ell >-\frac12$.
Let $a,b$ belong to the set of symbols $ \left\{\underline{0}\right\} \cup \Z$, 
\begin{itemize}
\item[I)] The asymptotic value $W_a\left(\psi, \tilde{\psi}\right)$ is well-defined for all $a$.

\item[II)] If
$\varphi=e_1 \psi+ e_2 \tilde{\psi}$, $\tilde{\varphi}=e_3 \psi+ e_4 \tilde{\psi}$ with $e_1 e_4- e_2 e_3 \neq 0$, then
\begin{eqnarray}\label{eq:asymptoticchange}
    W_a\left(\phi, \tilde{\phi}\right)= \frac{e_1 W_a\left(\psi, \tilde{\psi}\right) +e_2}{e_3W_a\left(\psi, \tilde{\psi}\right)+e_4}.
\end{eqnarray}

\item[III)] Let $a'\in \left\{\underline{0}\right\} \cup \Z$, and let $\psi_{a'}=\chi_+$, the subdominant solution at zero, for  $a'=\uo$, and $\psi_a'=\Psi_k$, the k-th Sibuya solution, for $a'=k \in \Z$. The following holds
\begin{eqnarray}
 W_a\left(\psi, \tilde{\psi}\right)=W_{b}\left(\psi, \tilde{\psi}\right)\Longleftrightarrow \psi_a \propto \psi_b,
\end{eqnarray}
where, in the above equation, $\psi_a \propto \psi_b$ means that $\psi_a$ and $\psi_b$ are proportional.
In particular, $W_{k} \neq W_{k+1}$ for all $k \in \Z$.
\end{itemize}
\end{lemma}

\begin{proof}
I). \quad By theorem \ref{thm:sibuya}, the pair of consecutive Sibuya solutions  $\{\Psi_k,\Psi_{k+1}\}$ form a basis.
Hence $\psi= e_1 \Psi_{k}+ e_2 \Psi_{k+1}$ and $\tilde{\psi}= e_3 \Psi_{k}+ e_4 \Psi_{k+1}$. Therefore, 
since $\lim_{x \to \infty_k} \frac{\Psi_k(x)}{\Psi_{k+1}(x)}=0$, it follows that 
$W_k(\psi, \tilde{\psi})$ exists, and
$$
W_k\left(\psi, \tilde{\psi}\right)= \begin{cases}
 \frac{e_2}{e_4}, & e_4 \neq 0,  \\
\infty, & e_4 = 0. 
\end{cases}
$$
The same consideration hold for the limit at $0$. The subdominant solution $\chi_+(x)= x^{\ell+1}(1 + O(x) )$ and any another solution $\phi(x)$ with the asymptotics $\phi(x)= x^{-\ell} (1 + O(x))$ form a basis, and its ratio $\frac{\chi_+(x)}{\varphi(x)}= x^{2\ell+1} (1 + O(x)) \to 0 $ since $\Re \ell > -\frac12$.
\\

II). \quad It is a direct consequence of I.
\\

III). \quad Due to II), it is sufficient to prove the equivalence for a particular basis of solutions. We choose the basis $\{\psi_a,\tilde{\psi}\}$, with $\tilde{\psi}$ not proportional to $\psi_a$. In this case, we have that $W_a=0$, and that $W_b=0$ if and only if $\psi_a \propto \psi_b$.
\end{proof}

By the above Lemma, the action of the group $GL(2,\C)$ on the basis of solutions of the anharmonic oscillators descends to a $PGL(2,\C)$ action on asymptotic values.
Therefore, the cross-ratio of four asymptotic values is invariant under change of basis. These invariant quantities were introduced in \cite{dtba} in the context of anharmonic oscillators, but they are often called Fock-Goncharov coordinates, see \cite{bridgeland-masoero-2023}, since they had already appeared in a much more general context in \cite{fock2006moduli}.
\begin{definition}
Let $E,\ell$ be fixed.
Let $a,b,c,d$ be distinct indices in $ \left\{\underline{0}\right\} \cup \Z$ such that $\# \{ W_a,W_b,W_c,W_d\}\geq 3$.
The cross-ratio
\begin{equation}\label{eq:fockgoncharov}
R_{(a,b,c,d)}(E;\ell)= \frac{(W_a-W_b)(W_c-W_d)}{(W_a-W_d)(W_b-W_c)}
\end{equation}
is called a Fock-Goncharov co-ordinate.
\end{definition}

\begin{lemma}\label{lem:fockgoncharov}
Let $a,b,c,d$ be distinct indices in $\left\{ \underline{0}\right\} \cup \Z$. For every $a' \in \{a,b,c,d\}$,  let $\psi_{a'}=\chi_+$, the subdominant solution at zero, if $a'=\uo$, while let $\psi_{a'}=\Psi_k$, the k-th Sibuya solution, if $a'=k \in \Z$.

On the region $$D= \left\{ (E,\ell) \in \C \times \left\{ \Re \ell >-\frac12\right\}, \# \{ W_a,W_b,W_c,W_d\}\geq 3 \right\},$$ the Fock-Goncharov coordinate $R_{(a,b,c,d)}$ is a meromorphic function, and
the following identity holds
\begin{equation}\label{eq:RasW}
R_{(a,b,c,d)}=- \frac{W_a(b,d)}{W_c(b,d)},
\end{equation}
with
\begin{equation*}
W_{a}(b,c) := \begin{cases}
& W_a(\psi_b,\psi_c), \; \mbox{ if } Wr_x[\psi_b,\psi_c] \neq 0 \\
& \frac{\psi_a}{\psi_b}, \quad \mbox{if }\psi_b \propto \psi_c = 0 .
\end{cases}
\end{equation*}
Moreover,
\begin{equation} \label{eq:valuecross}
     R_{(a,b,c,d)} = \begin{cases}   0 &  \Longleftrightarrow \psi_a \propto \psi_b \mbox{ or } \psi_c \propto \psi_d  \\
      - 1  & \Longleftrightarrow \psi_a \propto \psi_c \mbox{ or } \psi_b \propto \psi_d \\
      \; \infty &  \Longleftrightarrow \psi_a \propto \psi_d \mbox{ or } \psi_b \propto \psi_c
    \end{cases} 
\end{equation}
In the formulas above, $\psi_a \propto \psi_b$ means that $\psi_a$ and $\psi_b$ are proportional.
\end{lemma}
\begin{proof}
Identities \eqref{eq:RasW} and \eqref{eq:valuecross} are straightforward to check.

By the action of the group $S_4$ on the indices $\left\{a,b,c,d\right\}$ we can always reduce to the case $a,c \in \Z$. The Sibuya solutions $\psi_{a},\psi_{a+1}$ form a basis of solutions which are entire in the parameters $E,\ell$. Therefore, we can write $\psi_b= C_{b,a}(E,\ell) \psi_a+ \tilde{C}_{b,a}(E,\ell) \psi_{a+1} $ and $\psi_d=C_{d,a}(E,\ell) \psi_a+ \tilde{C}_{d,a}(E,\ell) \psi_{a+1} $ with the coefficients $C_{b,a},\tilde{C}_{b,a},C_{d,a},\tilde{C}_{d,a}$ being holomorphic on $\C \times \left\{\Re \ell >\frac12\right\}$, which is the domain of analyticity of $\chi_+$ with respect to the parameters. Therefore, $W_a(b,d)$ is a meromorphic function on $\C \times \left\{\Re \ell >\frac12\right\}$. Reasoning as above, we deduce that also $W_c(b,d)$ is a meromorphic function on $\C \times \left\{\Re \ell >\frac12\right\}$. Therefore, $R_{(a,b,c,d)}$ is meromorphic. 
\end{proof}

\begin{corollary}
\label{Rzero-properties}
\begin{itemize}
\item[I)] On the locus $\{ (E,\ell) \mbox{ s.t. }\Psi_{1} \mbox{ is not proportional to } \Psi_{-1} \}$ -- with $\Psi_{\pm1}$ being the $\pm1$ Sibuya solution -- the cross-ratio 
\begin{equation}
\label{eqn:rzero-psi-pm1-indep}
R_{\uo}:=R_{(\uo,-1,0,1)}=-\frac{W_{\uo}(1,-1)}{W_0(1,-1)}
\end{equation}
is a meromorphic function. Furthermore, $Q_+(E,\ell)=0$ if and only if $R_{\underline{0}}(E,\ell)+1=0$.

\item[II)] The function $R_{(k,k+2,k+1,k-1)}$ is an entire function of $E,\ell$, and the following identity holds:
\begin{equation}\label{eq:Rksigma}
R_{(k,k+2,k+1,k-1)}= \sigma_k \sigma_{k+1},
\end{equation}
where $\sigma_k$ is the $k-th$ Stokes multiplier.
\end{itemize}
\end{corollary}
\begin{proof}
I). \quad The thesis is a special case of Lemma \ref{lem:fockgoncharov}.
\\

II). \quad By definition of Stokes multipliers \eqref{eq:stokesmult}, $\Psi_k=\Psi_{k+2}+\sigma_{k+1}\Psi_{k+1}$ and $\Psi_{k-1}=\psi_{k+1}+\sigma_k\Psi_k$. Inserting this identities in \eqref{eq:RasW}, we obtain \eqref{eq:Rksigma}. Since, by Corollary \ref{cor:stokesmult}, the Stokes multipliers $\sigma_k,\sigma_{k+1}$ are entire functions, then $R_{(k,k+2,k+1,k-1)}$ is an entire function too.
\end{proof}

The above corollary marks the end of the second lecture. The importance of the Fock-Goncharov coordinates is two-fold:
\begin{itemize}
\item The Fock-Goncharov coordinates encode the global asymptotic behaviour of solutions to the anharmonic oscillator, in a basis independent way. In particular, the Fock-Goncharov coordinate $R_{\uo}=R_{(\uo,-1,0,1)}$ encodes the spectral problem $Q_+(E,\ell)=0$, as we have shown above.
\item The Fock-Goncharov coordinates can be computed via the complex WKB approximation, as an explicit leading term (expressed in terms of a period of the differential $\sqrt{V(x)}dx$) times a correcting factor which converges to $1$ in the small $\hbar$ limit. 
\end{itemize}  
In the next lecture we will address the latter statement, namely we will illustrate the WKB theory of the Fock-Goncharov coordinates.

\begin{remark}
Identity \eqref{eq:Rksigma} was observed in \cite{dtba} in the case of the quantum cubic oscillator, namely the Schr\"odinger equation with potential $U(x)=x^3+a x +b$. 
\end{remark}

\paragraph{\textbf{Exercises. II}}

1. Complete the proof of Lemma \ref{lem:KDinfty}.

2. Construct, via the complex WKB method, a dominant solution at $0$ for all values of parameters $(E,\ell) \in \C \times \left\{\Re \ell >-\frac12,\, \Re \ell \neq0 \right\}$; see \cite{degano2024}.

3. Define a suitable topology on the space $\WCs \cup \{0\} \cup \{\infty_k, k \in \Z \}$.

4. Compute the action of the symmetric group $S_4$ on $R_{(a,b,c,d)}$. Have you seen this action before?

\newpage

\section{Lecture III. WKB approximation of Fock-Goncharov co-ordinates}
In this lecture, we develop a WKB theory of Fock-Goncharov coordinates.
In the first part of this lecture, we show that $R_{(a,b,c,d)} \sim e^{\hbar^{-1}\oint_{\gamma} \left[ \sqrt{V(x)} -\frac{V'(x)}{4V(x)}\right] dx}$ where $\gamma$ is a closed loop in $\WCs \setminus \{V(x)=0 \}$ associated to the coordinate.
In the second part of the lecture, we study the above formula for the particular case of the coordinate $R_{\uo}=R_{(\uo,-1,0,1)}$ when $E>0,\ell>-\frac12$ -- since  $R_{\uo}$ encodes the spectral problem that we are interested in. 
As a result, we deduce the Bohr-Sommerfeld quantisation conditions for the spectral problem, equations \eqref{eqn:bohr-sommerfeld} below.

Recall that a Fock-Goncharov coordinate is defined by choosing four distinct ordered points on the boundary of $\WCs$, namely the vertices of an oriented quadrilateral in this space.
\begin{definition}[Admissible Quadrilateral]\label{def:admissiblequadrilateral}
For any quadruple of distinct indices $a,b,c,d$ in $ \underline{0} \cup \Z$, we consider the oriented quadrilateral $x_a\,x_b\,x_c\,x_d$, with vertices $x_a,x_b,x_x,x_d$, where
\begin{equation}
    x_a= \begin{cases}
     0, & \mbox{ if } a= \underline{0} \\
    \infty_a, & \mbox{ if }a \in \Z
    \end{cases}
\end{equation}
Let $W$ be a compact set in $\C \times \left\{ \Re \ell >-\frac12,\, \ell \neq 0  \right\}$. 
We say that the quadrilateral is WKB-admissible if there exist
four strictly admissible curves, $\gamma_{a,b}, \gamma_{b,c},\gamma_{c,d},\gamma_{d,a}$, connecting $x_a$ to $x_b$, $x_b$ to $x_c$, $x_c$ to $x_d$, and $x_d$ to $x_a$. If these exist, they are said to provide a WKB-realization of the sides of the quadrilateral.
\end{definition}
Let $x_a, x_b$ be vertices of an admissible quadrilateral. By definition, there exists a strictly admissible curve $\gamma_{a,b}$ connecting $x_a$ to $x_b$ (the same curve with the opposite orientation connects $x_b$ to $x_a$). Therefore, by the Fundamental Theorem of the WKB Approximation on curves, Theorem \ref{thm:fundamental-theorem}, the Volterra integral equation \eqref{eq:volterragamma} admits a unique solution $z_{\gamma_{a,b}}\colon [0,1] \to \C$; moreover, denoting by $\psi_a$ the normalised solution subdominant at $x_a$ (if $a=k \in \Z$, $\psi_a$ is the k-th Sibuya solution $\Psi_k$, while if $a=\uo$,  $\psi_a$ the solution subdominant at $0$, $\chi_+$), we have that 
 $$\psi_a(\gamma_{a,b}(t)) = C e^{\int_{t_0}^t\left[ \sqrt{V(\gamma(s)}-\frac{V'(\gamma(s)}{4 V(\gamma(s))} \dot{\gamma}(s) \right]ds} z_{\gamma_{a,b}}(t) .$$
In the formula above, $C \neq 0$ is a normalization constant that depends on $t_0$, and $\sqrt{V}$ is chosen in such a way that $\Re \int_{t_0}^t \sqrt{V(\gamma(s)}\dot{\gamma}(s) ds$ is increasing.
Since we are interested in the asymptotic behaviour of $\psi_a$ at $x_b$, we define
    \begin{equation}\label{eq:zab}
        z_{a,b}:=z_{\gamma_{a,b}}(1) \in \C.
    \end{equation}

We also have the following result, which is quite useful to simplify the computations.
\begin{lemma}\label{lem:zkk1}
\begin{itemize}
\item[I)] Let $z_{a,b} \in \C$ denote, as per \eqref{eq:zab}, the end-point value of the solution of the Volterra integral equation along a curve $\gamma_{a,b}$ connecting the boundary point $x_a$ to the boundary point $x_b$, with $a,b \in \{\uo\} \cup \Z$. The quantity $z_{a,b}$ depends only on the
the homotopy class in $\WCs \setminus \{ V(x)=0 \}$ of the admissible path along which we compute it. 

\item[II)] For all $k \in \Z$, there exists a homotopy class of strictly admissible curves connecting $\infty_k$  with $\infty_{k+1}$ such that $z_{k+1,k}=z_{k,k+1}=1$. This is the class of standard horizontal-trajectories at $\infty$, which do not encircle any zero of $V$.
\end{itemize}
\end{lemma}
\begin{proof}
I) \quad The thesis follows from the fact that $\psi_a$ is single valued in $\WCs \setminus \{ V(x)=0 \}$.
\\

II) \quad Let $E,\ell$ be fixed.
Recall the definition of standard horizontal trajectory at $\infty$, see equation \eqref{eq:standardhorinf}. These are parameterised by the real numbers $c$, $\gamma_c$. If $c$ is positive and sufficiently large, the standard horizontal trajectory $\gamma_c$ is strictly admissible, namely $\Re S$ is monotone increasing along it, and connects $\infty_k$ with $\infty_{k+1}$.  From Theorem \ref{thm:fundamental-theorem}, we have that 
\begin{equation*}
 |z_{k,k+1} -1| \leq  e^{\rho_{\gamma_c}}-1,
\end{equation*}
where, as per \eqref{eq:rhogamma}, $\rho_{\gamma_c}=\int_0^1| F(\gamma_c(t))|\dot{\gamma}_c(t)|dt$.
Since by Lemma \ref{lem:Vlanger}, $F(x)=O(x^{-2-\alpha})$ then  $\lim_{c \to +\infty} \rho_{\gamma_c} \to 0$. Therefore, $z_{k,k+1}=1$. By choosing the same path with the reversed orientation, we deduce that $z_{k+1,k}=1$.
\end{proof}

\paragraph{\textbf{From a quadrilateral to a cycle through surgery}} 
At a given vertex of an admissible quadrilateral, for example $x_a$, there is an outcoming strictly admissible curve $\gamma_{a,b}$ and an outcoming strictly admissible curve $\gamma_{d,a}$. Due to the local behaviour of horizontal trajectories at the singular points, see Lemma \ref{lem:horatinftyzero} \footnote{A detailed proof for the case of the cubic potential can be found in \cite[Lemma A.8]{bridgeland-masoero-2023}}, for every $\e>0$, we can deform $\gamma_{a,b}$ and $\gamma_{d,a}$  to obtain two new strictly admissible curves 
$\tilde{\gamma}_{d,a},\tilde{\gamma}_{a,b}$ such that
\begin{itemize}
\item The deformation is local: $\gamma_{a,b}-\tilde{\gamma}_{a,b}$ and $\gamma_{d,a}-\tilde{\gamma}_{d,a}$ are loops with base point $x_a$, not encircling any zero of $V(x)$, such that $\rho_{\gamma_{a,b}-\tilde{\gamma}_{a,b}}< \e$ and $\rho_{\gamma_{d,a}-\tilde{\gamma}_{d,a}}< \e$, where $\rho$ is as defined in \eqref{eq:rhogamma}.
\item $\tilde{\gamma}_{d,a},\tilde{\gamma}_{a,b}$ overlap in a neighbourhood of $x_a$ and there coincide with a standard horizontal trajectory.
\end{itemize}
Performing the surgery described above at all four vertices, as a result, we obtain a closed curve $\gamma_{a,b,c,d}$ in $\WCs \setminus \{V=0\}$; see Figure \ref{figure:surgery} for a pictorial representation. Notice that even though the 
curve depends on the deformations, its homotopy class in $\WCs \setminus \{V=0\}$ is, by construction, independent on the deformation.
\begin{definition}[Loop Associated to an Admissible Quadrilateral]\label{def:surgery}
The loop associated to a WKB realization of an admissible quadrilateral is the homotopy class in $\WCs \setminus \{V=0\}$ of the closed curve obtained by the surgery operation described above. 
\end{definition}
Given a WKB realization,  we define the WKB or semiclassical Fock-Goncharov coordinate as
\begin{align} \nonumber
 R^{W}_{(a,b,c,d)}(E,\ell)  & 
 := -e^{-\oint_{\gamma_{a,b,c,d}} \left[\sqrt{V(x;E,\ell)}-\frac{V'(x;E,\ell)}{4V(x;E,\ell)}\right]dx} \\
 & =-  e^{\frac{i\pi}{2} \mbox{ind}_{\gamma_{a,b,c,d}}(V) }e^{-\oint_{\gamma_{a,b,c,d}} \sqrt{V(x; E,\ell)}dx}.   \label{eq:RW}
\end{align}
In the above equation,  $\mbox{ind}_{\gamma_{a,b,c,d}}(V)= \sigma \times (Z - P)$, where $\sigma$ is the orientation of $\gamma_{a,b,c,d}$, $Z$ the number of zeroes (counting multiplicities) of $V$ encircled by $\gamma_{a,b,c,d}$, and $P$ the number of poles  (counting multiplicities) of $V$ encircled by $\gamma_{a,b,c,d}$.

We can now prove what constitutes the essential result in the complex WKB method: if a quadrilateral $x_ax_bx_cx_d$ is admissible, then the corresponding Fock-Goncharov coordinate is well approximated by
formula \eqref{eq:RW}, the exponential of a period of the differential $\sqrt{V}dx$, 

\begin{lemma}\label{lem:essential}
Let the quadrilateral $x_ax_bx_cx_d$ be admissible, and $R^W_{(a,b,c,d)}$ be the corresponding  WKB Fock-Goncharov coordinate  \eqref{eq:RW}.

The following identity holds
\begin{equation}\label{eq:RRWzzzz}
    \frac{W_a(b,d)}{W_c(b,d)}=-R^W_{(a,b,c,d)} \frac{z_{b,a} z_{d,c}}{z_{b,c} z_{d,a}}.
\end{equation}
In the equation above, the quantities $z_{a',b'}$, with $a',b' \in\{a,b,c,d \}$, are as in \eqref{eq:zab}.
\begin{proof}
We assume that the surgery was made so that the strictly admissible curves overlap in a neighbourhood of the boundary points. We then choose points $\tilde{x}_a,\tilde{x}_b,\tilde{x}_c,\tilde{x}_d$ in the intersections of the overlapping curves and call $\gamma_{a,b,c,d}$ the loop obtained by making the surgeries at these four points; see Figure \ref{figure:surgery}.

We consider the solution $\psi_b$, subdominant at $x_{b}$, and we approximate it with the WKB function $e^{\int_{\tilde{x}_c}^x \left[\sqrt{{V}(y)}-\frac{V'(y)}{4 V(y)} \right] dy} $. By the Fundamental Theorem of WKB Approximation, Theorem \ref{thm:fundamental-theorem}, we can choose a normalisation of $\psi_b$ such that
\[
\begin{aligned}
\psi_{b}\left(\gamma_{b,a}(t)\right) = & \exp\left\{\int_{\gamma_{c,b},\tilde{x}_c}^{\tilde{x}_b} \left[ \sqrt{V(y)}-\frac{V'(y)}{4 V(y)} \right] dy\right\} \times \\
& \times \exp\left\{\int_{\gamma_{b,a},\tilde{x}_b}^{\gamma_{b,a}(t)} \left[ \sqrt{V(y)}-\frac{V'(y)}{4 V(y)} \right] d y\right\} \;z_{\gamma_{b,a}}(t) 
\end{aligned}
\]
and
\[
\psi_{b}\left(\gamma_{b,c}(t)\right) =   \exp\left\{\int_{\tilde{x}_c,\gamma_{b,c}}^{\gamma_{b,c}(t)} \sqrt{V(y)}-\frac{V'(y)}{4 V(y)}d y\right\} \; z_{\gamma_{b,c}}(t).
\]
In the above equation, $\gamma_{b,a}$ (resp. $\gamma_{c,b}$) denotes the curve $\gamma_{a,b}$ (resp. $\gamma_{b,c}$) with the opposite orientation: $\gamma_{b,a}(t)=\gamma_{a,b}(1-t)$.
We choose the same normalisation for the solution $\psi_d$ subdominant at $x_d$, namely, we approximate it with $e^{\int_{\tilde{x}_c}^x \left[ \sqrt{{V}(y)}-\frac{V'(y)}{4 V(y)} \right] dy}  $.

By doing so, we obtain the following representations
\begin{align*}
& W_a(b,d) = \lim_{x \to x_a} \frac{\psi_a(x)}{\psi_b(x)} = 
e^{-\oint_{\gamma_{a,b,c,d}} \left[ \sqrt{V(y)}-\frac{V'(y)}{ 4 V(y)} \right] dy } \frac{z_b(a)}{z_d(a)} \\
& W_c(b,d) = \lim_{x \to x_c} \frac{\psi_a(x)}{\psi_b(x)} = \frac{z_b(c)}{z_d(c)}.
\end{align*}
By definition of $R^W$, equation (\ref{eq:RW}), the thesis follows.

\end{proof}

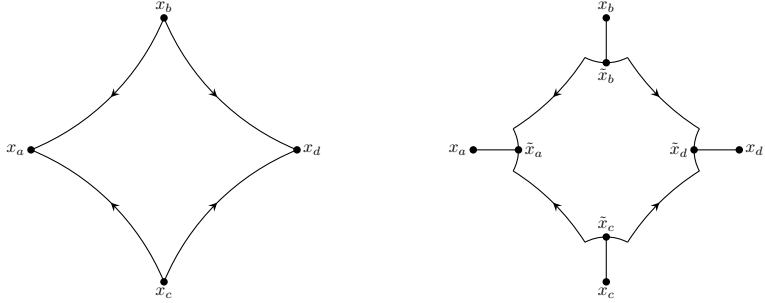
\begin{figure}[H]
\centering
\begin{subfigure}[t]{0.45\textwidth}
\begin{tikzpicture}[decoration={markings, mark= at position 0.5 with {\arrow{stealth}}},scale=0.35, every node/.style={scale=0.65}]

\draw[postaction=decorate] (5,5)  to [curve through={(3,2)}] (0,0);
\draw[postaction=decorate] (5,5) to [curve through={(7,2)}] (10,0);
\draw[postaction=decorate] (5,-5)  to [curve through={(3,-2)}] (0,0);
\draw[postaction=decorate] (5,-5) to [curve through={(7,-2)}] (10,0);

\fill (0,0) circle[radius=4pt];
\fill (5,5) circle[radius=4pt];
\fill (10,0) circle[radius=4pt];
\fill (5,-5) circle[radius=4pt];

\draw (0,0) node[left] {$x_a$};
\draw (5,5) node[above] {$x_b$};
\draw (5,-5) node[below] {$x_c$};
\draw (10,0) node[right] {$x_d$};

\end{tikzpicture}
\end{subfigure}
\begin{subfigure}[t]{0.45\textwidth}
\begin{tikzpicture}[decoration={markings, mark= at position 0.5 with {\arrow{stealth}}},scale=0.35, every node/.style={scale=0.65}]

\draw[postaction=decorate] (4.2,3.5)  to [curve through={(3,2)}] (1.5,0.8);
\draw[postaction=decorate] (5.8,3.5) to [curve through={(7,2)}] (8.5,0.8);
\draw[postaction=decorate] (4.2,-3.5)  to [curve through={(3,-2)}] (1.5,-0.8);
\draw[postaction=decorate] (5.8,-3.5) to [curve through={(7,-2)}] (8.5,-0.8);

\fill (0,0) circle[radius=4pt];
\fill (5,5) circle[radius=4pt];
\fill (10,0) circle[radius=4pt];
\fill (5,-5) circle[radius=4pt];

\fill (1.7,0) circle[radius=4pt];
\fill (8.3,0) circle[radius=4pt];
\fill (5,3.3) circle[radius=4pt];
\fill (5,-3.3) circle[radius=4pt];

\draw (0,0) node[left] {$x_a$};
\draw (5,5) node[above] {$x_b$};
\draw (5,-5) node[below] {$x_c$};
\draw (10,0) node[right] {$x_d$};

\draw (1.7,0) node[right] {$\tilde{x}_a$};
\draw (8.3,0) node[left] {$\tilde{x}_d$};
\draw (5,3.3) node[below] {$\tilde{x}_b$};
\draw (5,-3.3) node[above] {$\tilde{x}_c$};

\draw (1.5,0.8) arc[start angle=28.07, end angle=0, radius=1.7];

\draw (1.5,-0.8) arc[start angle=-28.07, end angle=0, radius=1.7];

\draw (1.7,0) -- (0,0);

\begin{scope}[shift={(10,0)}]
\draw (-1.5,0.8) arc[start angle=151.93, end angle=180, radius=1.7];

\draw (-1.5,-0.8) arc[start angle=208.07, end angle=180, radius=1.7];

\draw (-1.7,0) -- (0,0);
\end{scope}

\begin{scope}[shift={(5,5)}]
\draw (0.8,-1.5) arc[start angle=-61.93, end angle=-90, radius=1.7];

\draw (-0.8,-1.5) arc[start angle=-118.07, end angle=-90, radius=1.7];

\draw (0,-1.7) -- (0,0);
\end{scope}

\begin{scope}[shift={(5,-5)}]
\draw (0.8,1.5) arc[start angle=61.93, end angle=90, radius=1.7];

\draw (-0.8,1.5) arc[start angle=118.07, end angle=90, radius=1.7];

\draw (0,1.7) -- (0,0);
\end{scope}

\end{tikzpicture}
\end{subfigure}

\caption{\small Topological representation of a quadruple of admissible lines joining the vertices of an admissible quadrilateral $x_ax_bx_cx_d$ before and after the surgery. }
\label{figure:surgery}
\end{figure}

\end{lemma}

Here we show how this result can be used in practice. Let us consider the anharmonic oscillator in the second asymptotic regime, defined by the change of variable $y =  \hbar^{\frac{1}{\alpha+1}} x$, and the change of parameters $\nu=E \left(\ell+\frac12\right)^{-\frac{2\alpha}{\alpha+1}}, \hbar=\left(\ell+\frac12\right)^{-1}$ \eqref{eqn:intro_rescaled-infinite-e-l}. As it was already noticed in the end of the first lecture, in this regime the reduced potential scales as $\hbar^{-2}$ and the forcing term \eqref{eqn:forcing-term} (hence the error $\rho_{\gamma}$) scales as $\hbar$. To be more precise, defining $\widetilde{V}(y,\nu,\hbar)=\widetilde{U}(y;\nu,\hbar)+\frac{1}{4y^2} $, where $\widetilde{U}$ is the potential of \eqref{eqn:intro_rescaled-infinite-e-l}, and by $\widetilde{F}$ the corresponding forcing term, we have
\begin{align}\label{eq:VhV}
& \widetilde{V}(y;\nu,\hbar)=\hbar^{-2} \left( x^{2\alpha}-\nu + \frac1{x^2}\right)= \hbar^{-2}\widetilde{V}(x;\nu,1), \\
\label{eq:FhF}
& \tilde{F}(y;\nu,\hbar)= \hbar  \tilde{F}(y;\nu,1).
\end{align}
Assume that, as $\nu$ varies in some set $N$ while $\hbar$ is equal to $1$, a given quadrilateral $x_ax_bx_cx_d$ is admissible.
Then the quadrilateral is admissible for all $(\nu,\hbar) \in N \times (0,\infty)$: in fact, since
the points $0$ and $\infty_k, k \in \Z$ are invariant under the change of variable $y =  \hbar^{\frac{1}{\alpha+1}} x, \hbar >0$, the end-points of the curves $\gamma_{a,b},\gamma_{b,c},\gamma_{c,d},\gamma_{d,a}$ do not change with $\hbar>0$; by \eqref{eq:VhV}, they are strictly admissible for any $\hbar>0$. 
Furthermore, from \eqref{eq:FhF} we immediately deduce that if $\gamma$ is any of these four curves, the quantity $\rho_{\gamma}$ which controls the WKB approximation vanishes as $O(\hbar)$. Consequently,
\begin{equation}\label{eq:zabh}
\left| z_{a',b'} -1\right| =O(\hbar) \mbox{  as  } \hbar \to 0, \; \forall a',b' \in \{a,b,c,d \}
\end{equation}
Therefore, after Lemma \ref{lem:essential}, 
   $ \frac{W_a(b,d)}{W_c(b,d)}=-R^W_{(a,b,c,d)} \left(1 + O (\hbar) \right)$. Recalling that, by Lemma \ref{lem:fockgoncharov}, $R_{a,b,c,d}=- \frac{W_a(b,d)}{W_c(b,d)}$, we have arrived to the following theorem, which constitutes the essence of the complex WKB method.
\begin{theorem}
\label{theorem:wkb-cross-ratio}
Assume that, as $\nu$ varies in a given set $N$ and $\hbar$ is equal to $1$, a given quadrilateral $x_ax_bx_cx_d$ is admissible, and moreover the following restriction holds: $W_{a} \neq W_c$ or $W_b \neq W_d$ for $\hbar$ small enough.
Let $\gamma$ be the loop associated to the admissible quadrilateral.

Then if $\hbar>0$ is small enough
\begin{equation}\label{eq:RRw}
\left|  R_{(a,b,c,d)}(\nu,\hbar) e^{\hbar^{-1}\oint_{\gamma} \sqrt{V(x;\nu,1)}dx} +e^{-\frac{i\pi}{2}\mbox{ind}_{\gamma}(V) } \right| \leq C_N  \hbar  ,
\end{equation}
where $C_N>0$ is a constant depending on the set $N$. 

Assume moreover that there exists a $\theta_0$, with $\theta_0 <\frac\pi2 $, such that, for all $\theta$ with $ |\theta| \leq \theta_0$, $\Re e^{-i \theta} \int^x \sqrt{V(w)} dw $ is strictly monotone increasing along the four curves which realize the sides of the admissible quadrilateral.  Then if $|\hbar|$ is small enough, on the sector $|\arg \hbar| \leq \theta_0$, the following estimate holds
\begin{equation}\label{eq:RRwcomplex}
\left|  R_{(a,b,c,d)}(\nu,\hbar) e^{\hbar^{-1}\oint_{\gamma} \sqrt{V(x;\nu,1)}dx} +e^{-\frac{i\pi}{2}\mbox{ind}_{\gamma}(V) } \right| \leq C_{N,\theta_0}  |\hbar| .
\end{equation}
In particular, whenever all sides are realized by horizontal trajectory, the above estimate holds for any $\theta_0$ with $0\leq \theta_0<\frac{\pi}{2}$.
\end{theorem}
\begin{proof}
Equation \eqref{eq:RRw} was proven above.

To prove \eqref{eq:RRwcomplex} we reason as follows.
After the surgery, any admissible curve in a neighbourhood of either end-points coincide with a horizontal trajectory. In particular, if the end point of a curve is the point $\infty_k$, then the argument of the admissible curve tends to $\frac{k \pi}{\alpha+1}$, see Proposition \ref{prop:localhor}. Hence, if we apply to a curve a rotation of $\theta$ radians with $|\theta|<\frac{\pi}{2\alpha+2}$, then the curve still ends at $\infty_k$. The point $0$ is on the contrary invariant under any rotation.
 We conclude that all four end-points $x_a,x_b,x_c,x_d$ are invariant under the transformation $y=  \hbar^{\frac{1}{\alpha+1}} x$ if and only if $|\arg \hbar | < \frac\pi2$.
 Therefore, if there exists a $\theta_0, \theta_0 <\frac\pi2 $ such that, for all $\theta$ with $ |\theta| \leq \theta_0$, $\Re e^{-i \theta} \int^x \sqrt{V(w)} dw $ is monotone increasing, then the four curves realize the quadrilateral $x_ax_bx_cx_d$ for all $\hbar$, with $|\arg \hbar| \leq \theta_0$. Therefore $\rho_{\gamma}=O(\hbar)$, hence \eqref{eq:zabh} follows.
\end{proof}

\begin{remark}
It follows from  \eqref{eq:RRw} that if $\gamma$ and $\gamma'$ are the homotopy classes of loops associated with two possibly distinct WKB realizations of the same quadrilateral, then $\oint_{\gamma} \sqrt{V(x)}dx=\oint_{\gamma'} \sqrt{V(x)}dx$.
\end{remark}

\subsection{Stokes Complex}
\label{subsec:stokes-complex}
As we have proved above, we can compute the small $\hbar$ limit of Fock-Goncharov coordinates via the complex WKB method, whenever the corresponding quadrilateral is admissible. 
We are left therefore with the task of studying which quadrilaterals are admissible, as the parameters $E,\ell$ vary  in $\C \times \left\{ \Re \ell >\frac12\right\}$. 

The information on the admissible quadrilateral is encoded in the following admissibility graph.
\begin{definition}\label{def:inftygon}
Fixed the parameters $E,\ell$, the admissibility graph is the following graph embedded in $\{\WCs \setminus \{x \in \WCs, V(x;E,\ell)=0 \}\} \cup \{0\} \cup \{\infty_k, k\in \Z\}$: 
\begin{itemize}
\item The vertices of the graph are the boundary points $0$ and $\infty_k, k \in \Z$.
\item An edge connects two vertices if, chosen one of its two possible orientations, it is homotopic to a strictly admissible curve connecting the two vertices.
\end{itemize}
\end{definition}
\begin{remark}
Notice that if $\alpha \in \Q$, the admissibility graph is periodic, thus it can be reduced to a finite graph in the union of $\C$ and a finite set of boundary points. In particular, if $2 \alpha \in \N$, the graph has $2\alpha+2$ external vertices, $\infty_0, \dots, \infty_{2\alpha+1}$ and one internal vertex $0$.
\end{remark}

Computing the admissibility graph is quite a formidable task, even for the simple potential we are considering. A priori, the only information that we have is that the vertices $k$ and $k+1$ are connected by an edge, see Lemma \ref{lem:zkk1}.
 However, for the sake of studying the spectrum of the anharmonic oscillator when $\ell >-\frac12$, our task is greatly simplified.
 In fact, since the spectrum is real and positive, and it is encoded by the Fock-Goncharov coordinate $R_{\uo}$, we need only consider the admissibility graph for $E,\ell+\frac12$ real and positive, and restricted to the four vertices $0,\infty_{-1},\infty_0,\infty_1$.

It turns out that, with these restrictions, the admissibility graph is completely characterised by the nature of the positive roots of the potential $V(x;E,\ell)$ for $E>0, \ell>-\frac12$, whether there are no positive roots, or one positive double root, or two positive simple roots.
Letting
\begin{equation}
\label{eqn:e-min-x-min}
E_*(\ell)= \alpha^{-\frac{\alpha}{1+\alpha}}(1+\alpha) \left(\ell+\frac12\right)^{\frac{2\alpha}{1+\alpha}}, \quad x_*(\ell)= \alpha^{-\frac{1}{2+2\alpha}}  \left(\ell+\frac12\right)^{\frac{1}{1+\alpha}},
\end{equation}
it follows that
\begin{itemize}
\item[i)] If $E<E_*(\ell)$, then $V(x;E,\ell) >0$ for all $x>0$;
\item[ii)] If $E=E_*(\ell)$, then $x_*(\ell)$ is a double zero and $V(x;E,\ell) > 0$ for all $x >0,\, x \neq x_*(\ell)$;
\item[iii)] If $E>E_*(\ell)$ then $V(x;E,\ell)$ has two positive and simple zeroes $x_-, x_+$ (depending on $E,\ell$) with $x_- < x_+$; it follows that $V(x;E,\ell)>0$ for $x \in (0,x_-) \cup (x_+,\infty)$, and $V(x;E,\ell)<0$ for $x \in \left(x_-,x_+\right)$.
\end{itemize}
We notice that, by Theorem \ref{theorem:spec-properties}, if $E$ is a spectral point than $E>E_*$; hence, we actually need only study the third case.
Before addressing the admissibility graph for these three cases, we make a pause to introduce another embedded graph which is studied in relation in the realm of the complex WKB method, called the Stokes complex \cite{masoero2010poles} or Stokes graph \cite{iwaki2014}

\begin{figure}[H]
\includegraphics[width=3.5cm]{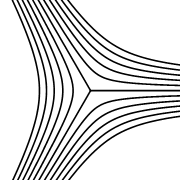}
\caption{Horizontal trajectories in a neighbourhood of a simple turning point.}
\label{fig:regularsimpleairy-tp}
\end{figure}

We first define the notion of turning point.
\begin{definition}
A (simple, double, etc...) zero of the reduced potential $V$ is called a (simple, double, etc...) turning point of the quadratic differential.
\end{definition}
As shown in \cite{strebel}[Theorem 7.2], if $x_0 \neq 0$ be a zero of order $\beta$, so that
$V(x)= a_0 (x-x_0)^{\beta} \left(1 + O(x) \right)$, there exists a holomorphic change of coordinate $x=\varphi(z)=x_0+ z + O(z^2)$ such that
\begin{equation*}
    \widetilde{S}(z)=S(\varphi(z))= \frac{2}{\beta+2}a_0^{\frac12} z^{\frac{\beta+2}{2}}.
\end{equation*}
Therefore, from a turning point $x_0$ of order $\beta$, there emanates exactly $\beta+2$ $\theta$-trajectories of the quadratic differential $V(x) dx^2$. If $V(x)=a_0 (x-x_0)^{\beta}+ O(|x-x_0|^{\beta+1})$, the horizontal trajectories emanates from $x_0$ in the directions $  \frac{2 \pi}{\beta+2} k - 2\theta+\arg a_0$, with $k=0,\dots,\beta+1$. See Figure \ref{fig:regularsimpleairy-tp}.

\begin{definition}[Stokes Complex] \label{def:stokescomplex}
We add to $\WCs$ the points $\left\{\infty_{k+\frac12}, k \in \Z\right\}$ by the following rule: a sequence $\{x_m\}_{m\in\mathbb{N}}$ is said to converge to  $\infty_{k+\frac12}$ if $|x_m| \to \infty$ and $\arg x_m \to \frac{(k+\frac12) \pi}{\alpha+1}$. Fixed $E,\ell$, the Stokes complex is the following graph embedded in  $\{\WCs\setminus \{V(x)=0\}\} \cup \{0\} \cup \left\{\infty_{k+\frac12}, k \in \Z\right\}$ :
\begin{itemize}
    \item The set of vertices of the graph is the union of the turning points $\{ x \in \WCs, V(x)=0\}$ and boundary points  $\{0\} \cup \left\{\infty_{k+\frac12}, k \in \Z\right\} $.
    \item The edges are vertical trajectories starting at a turning point.
\end{itemize}
\end{definition}
Notice that a turning point of order $\beta$ has valency $\beta+2$.

There is a vast literature on Stokes complexes, see e.g. \cite{iwaki2014} and references therein. 
In this lecture, we merely touch upon the notion of Stokes complex, but we make use of it to render graphically our computations. To give the reader a flavour of this subject, we present separately
the case $\alpha=1$, the isotropic harmonic oscillator, for which the Stokes complex is exactly known, see e.g. \cite{maro21}.

\subsubsection*{The isotropic harmonic oscillator}. We show the Stokes complex for the isotropic harmonic oscillator, namely the anharmonic oscillator with $\alpha=1$.
$$
V(x;E,\ell)=x^{2}+\frac{\left(\ell+\frac{1}{2}\right)^2}{x^2}-E, \; x_*=(\ell+\frac12)^{\frac12}, \; E_*=2\ell+1.
$$
Since the potential $V$ is meromorphic in $\C$, the Stokes complex is periodic, and it has four external vertices $\infty_{\frac12},\infty_{\frac32},\infty_{-\frac32},\infty_{-\frac12}$.
According to the trichotomy described above, there are three cases.\\
\paragraph{\textbf{Case I}: $\mathbf{0\leq E < E_*=2 \ell+1}$}
In this case, the potential has two complex conjugate pairs of zeroes. The Stokes complex and the admissibility graph are given in Figure \ref{figure:Stokes-complex-harmonic-3}. The quadrilateral $0\infty_{1-}\infty_0\infty_{1}$ is admissible and there is a horizontal trajectory connecting $0$ to $\infty_0$.

\begin{figure}[H]
\centering
\begin{tikzpicture}[decoration={markings, mark= at position 0.5 with {\arrow{stealth}}},scale=0.35, every node/.style={scale=0.65}]

\draw (0,0)  node[left] {$0$};

\draw[color=black!20,dashed] (0,0) -- (-45:11.5);
\draw[color=black!20,dashed] (0,0) -- (45:11.5);
\draw[color=black!20,dashed] (0,0) -- (135:11.5);
\draw[color=black!20,dashed] (0,0) -- (-135:11.5);

\fill (2.5,1.7) circle[radius=4pt];
\fill (2.5,-1.7) circle[radius=4pt];
\fill (-2.5,1.7) circle[radius=4pt];
\fill (-2.5,-1.7) circle[radius=4pt];

\draw (45:11.5)  node[right] {$\infty_{\frac12}$};
\draw (-45:11.5)  node[right] {$\infty_{-\frac12}$};
\draw (135:11.5)  node[left] {$\infty_{\frac32}$};
\draw (-135:11.5)  node[left] {$\infty_{-\frac32}$};

\draw (2.5,1.7) to [curve through={ (44.7:11.1) (44.7:11.3)}] (44.7:11.5);
\draw (2.5,-1.7) to [curve through={ (-44.7:11.1) (-44.7:11.3)}] (-44.7:11.5);
\draw (-2.5,1.7) to [curve through={ (135.3:11.1) (135.3:11.3)}] (135.3:11.5);
\draw (-2.5,-1.7) to [curve through={(-135.3:11.1) (-135.3:11.3)}] (-135.3:11.5);
\draw (2.5,1.7) -- (2.5,-1.7) -- (-2.5,-1.7) -- (-2.5,1.7) -- (2.5,1.7);

\fill (16.5,0) circle[radius=4pt];
\fill (11.5,0) circle[radius=4pt];
\fill (21.5,0) circle[radius=4pt];
\fill (16.5,5) circle[radius=4pt];
\fill (16.5,-5) circle[radius=4pt];
\fill (14.5,1.5) circle[radius=2pt];
\fill (14.5,-1.5) circle[radius=2pt];
\fill (18.5,1.5) circle[radius=2pt];
\fill (18.5,-1.5) circle[radius=2pt];
\draw (16.5,0) node[below left] {$0$};
\draw (21.5,0) node[right] {$\infty_0$};
\draw (16.5,5) node[above] {$\infty_1$};
\draw (16.5,-5) node[below] {$\infty_{-1}$};
\draw (11.5,0) node[left] {$\infty_2$};

\draw (11.5,0) -- (16.5,5);
\draw (11.5,0) -- (16.5,-5);
\draw (16.5,5) -- (21.5,0);
\draw (16.5,-5) -- (21.5,0);
\draw (11.5,0) -- (21.5,0);
\draw (16.5,5) -- (16.5,-5);
\end{tikzpicture}
\caption{\small  Isotropic harmonic oscillator potential with $\ell>-\frac{1}{2}$ and $E<2\ell+1$. Left: Topological representation of the Stokes complex. Right: Admissibility graph.}
\label{figure:Stokes-complex-harmonic-3}
\end{figure}
\paragraph{$\mathbf{E = E_*=2 \ell+1}$}
In this case, the  potential has two zeroes $ x=\pm x_* $, real and of multiplicity two. The Stokes complex and the admissibility graph are given in Figure \ref{figure:Stokes-complex-harmonic-2}. The quadrilateral $0\infty_{1-}\infty_0\infty_{1}$ is admissible and there are no admissible lines connecting $0$ to $\infty_0$ or $\infty_{-1}$ to $\infty_{1}$.

\begin{figure}[H]
\centering
\begin{tikzpicture}[decoration={markings, mark= at position 0.5 with {\arrow{stealth}}},scale=0.35, every node/.style={scale=0.65}]

\draw (0,0)  node[left] {$0$};

\draw[color=black!20,dashed] (0,0) -- (-45:11.5);
\draw[color=black!20,dashed] (0,0) -- (45:11.5);
\draw[color=black!20,dashed] (0,0) -- (135:11.5);
\draw[color=black!20,dashed] (0,0) -- (-135:11.5);

\fill (4.5,0) circle[radius=4pt];
\fill (-4.5,0) circle[radius=4pt];

\draw (45:11.5)  node[right] {$\infty_{\frac12}$};
\draw (-45:11.5)  node[right] {$\infty_{-\frac12}$};
\draw (135:11.5)  node[left] {$\infty_{\frac32}$};
\draw (-135:11.5)  node[left] {$\infty_{-\frac32}$};

\draw (-4.5,0) to [curve through={ (0,2.2)}] (4.5,0);
\draw (-4.5,0) to [curve through={ (0,-2.2)}] (4.5,0);
\draw (4.5,0) to [curve through={(42:9.4) (44.7:11.1) (44.7:11.3)}] (44.7:11.5);
\draw (4.5,0) to [curve through={(-42:9.4) (-44.7:11.1) (-44.7:11.3)}] (-44.7:11.5);
\draw (-4.5,0) to [curve through={(138:9.4) (135.3:11.1) (135.3:11.3)}] (135.3:11.5);
\draw (-4.5,0) to [curve through={(-138:9.4) (-135.3:11.1) (-135.3:11.3)}] (-135.3:11.5);

\fill (11.5,0) circle[radius=4pt];
\fill (16.5,0) circle[radius=4pt];
\fill (21.5,0) circle[radius=4pt];
\fill (16.5,5) circle[radius=4pt];
\fill (16.5,-5) circle[radius=4pt];
\fill (14.5,0) circle[radius=2pt];
\fill (18.5,0) circle[radius=2pt];
\draw (16.5,0) node[below left] {$0$};
\draw (11.5,0) node[left] {$\infty_2$};
\draw (21.5,0) node[right] {$\infty_0$};
\draw (16.5,5) node[above] {$\infty_1$};
\draw (16.5,-5) node[below] {$\infty_{-1}$};

\draw (11.5,0) -- (16.5,5);
\draw (11.5,0) -- (16.5,-5);
\draw (16.5,5) -- (21.5,0);
\draw (16.5,-5) -- (21.5,0);
\draw (16.5,5) --(16.5,-5);

\end{tikzpicture}
\caption{\small Isotropic harmonic oscillator potential with $\ell>-\frac{1}{2}$ and $E=2\ell+1$. Left: Topological representation of the Stokes complex. Right: Admissibility graph.}
\label{figure:Stokes-complex-harmonic-2}
\end{figure}

\paragraph{$\mathbf{E > E_*=2 \ell+1}$}
In this case, the potential has four simple zeroes, all of them being real, two positive $x_{-},x_+$ and two negative $-x_-,-x_+$.
The Stokes complex and the admissibility graph are given in Figure \ref{figure:Stokes-complex-harmonic}. The quadrilateral $0\infty_{1-}\infty_0\infty_{1}$ is admissible and there is a horizontal trajectory connecting $\infty_{-1}$ to $\infty_{1}$.

\begin{figure}[H]
\centering
\begin{tikzpicture}[decoration={markings, mark= at position 0.5 with {\arrow{stealth}}},scale=0.35, every node/.style={scale=0.65}]

\draw (0,0)  node[left] {$0$};

\draw[color=black!20,dashed] (0,0) -- (-45:11.5);
\draw[color=black!20,dashed] (0,0) -- (45:11.5);
\draw[color=black!20,dashed] (0,0) -- (135:11.5);
\draw[color=black!20,dashed] (0,0) -- (-135:11.5);

\fill (1.5,0) circle[radius=4pt];
\fill (4.5,0) circle[radius=4pt];
\fill (-1.5,0) circle[radius=4pt];
\fill (-4.5,0) circle[radius=4pt];

\draw (45:11.5)  node[right] {$\infty_{\frac12}$};
\draw (-45:11.5)  node[right] {$\infty_{-\frac12}$};
\draw (135:11.5)  node[left] {$\infty_{\frac32}$};
\draw (-135:11.5)  node[left] {$\infty_{-\frac32}$};

\draw (1.5,0) to [curve through={ (0,1.2) (-1.5,0) (0,-1.2)}](1.5,0);
\draw (1.5,0) -- (4.5,0);
\draw (-1.5,0) -- (-4.5,0);
\draw (4.5,0) to [curve through={(42:9.4) (44.7:11.1) (44.7:11.3)}] (44.7:11.5);
\draw (4.5,0) to [curve through={(-42:9.4) (-44.7:11.1) (-44.7:11.3)}] (-44.7:11.5);
\draw (-4.5,0) to [curve through={(138:9.4) (135.3:11.1) (135.3:11.3)}] (135.3:11.5);
\draw (-4.5,0) to [curve through={(-138:9.4) (-135.3:11.1) (-135.3:11.3)}] (-135.3:11.5);

\fill (11.5,0) circle[radius=4pt];
\fill (16.5,0) circle[radius=4pt];
\fill (21.5,0) circle[radius=4pt];
\fill (16.5,5) circle[radius=4pt];
\fill (16.5,-5) circle[radius=4pt];
\fill (15.5,0) circle[radius=2pt];
\fill (17.5,0) circle[radius=2pt];
\fill (13.5,0) circle[radius=2pt];
\fill (19.5,0) circle[radius=2pt];
\draw (11.5,0) node[left] {$\infty_2$};
\draw (21.5,0) node[right] {$\infty_0$};
\draw (16.5,5) node[above] {$\infty_1$};
\draw (16.5,-5) node[below] {$\infty_{-1}$};
\draw (16.5,0) node[below left] {$0$};

\draw (16.5,5) to [curve through={(14.5,0)}] (16.5,-5);
\draw (16.5,5) to [curve through={(18.5,0)}] (16.5,-5);

\draw (11.5,0) -- (16.5,5);
\draw (11.5,0) -- (16.5,-5);
\draw (16.5,5) -- (21.5,0);
\draw (16.5,-5) -- (21.5,0);
\draw (16.5,5) -- (16.5,-5);

\end{tikzpicture}
\caption{\small Isotropic harmonic oscillator potential with $\ell>-\frac{1}{2}$ and $E>2\ell+1$. Left: Topological representation of the Stokes complex. Right: Admissibility graph.}
\label{figure:Stokes-complex-harmonic}
\end{figure}
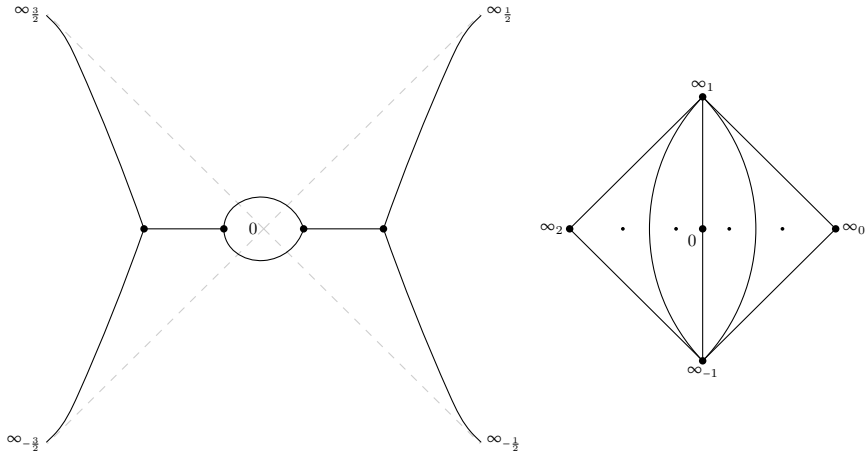
\begin{remark}
In all three cases above, if we erase the vertex $0$, the Stokes complex is connected. 
For this reason, the Riemann Surface $\left\{(x,y) \in \C^2, y^2=x^2-E+\frac{(\ell+\frac12)^2}{x^2}\right\}$ with $E>0, \ell >-\frac12$ is called a Boutroux curve (Boutroux curves play a prominent role in the WKB analysis of Painlevé transcendents, see \cite{masoero2010poles,maro18}).
This property does not extend to general $\alpha$. In fact, one can show that, if $\alpha \neq 1$, for every fixed $\ell>-\frac12$, the Stokes complex is connected when $E=0$, but it is not connected in a punctured neighbourhood of $E=0$.
\end{remark}
\subsubsection*{General potential: Turning points and admissible lines}

For general $\alpha$, we are not able to characterise fully the Stokes complex, but we are able to show that the quadrilateral $0\infty_{-1}\infty_0\infty_1$  is admissible and that 
$\infty_1$ and $\infty_{-1}$ are connected by a strictly admissible line if $E > E_*$.

The homotopy class of the admissible curves realizing the quadrilateral, the associated loop and the Stokes complex (restricted to the sector $|\arg x|\leq \frac{\pi}{2\alpha+2} $) are depicted in Figure \ref{fig:admissible-lines} and Figure
\ref{figure:topology-stokes-complex} below.  The proof of the content of Figure \ref{fig:admissible-lines} will appear in the more general Proposition \ref{prop:wkb-estimate-z-case-2}, that we will prove in the next and final lecture.

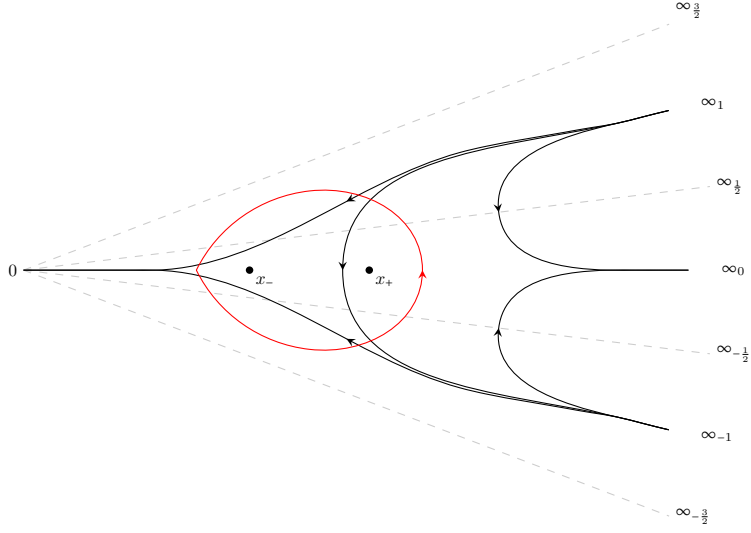
\begin{figure}[H]
\centering
\begin{tikzpicture}[decoration={markings, mark= at position 0.5 with {\arrow{stealth}}},scale=0.35, every node/.style={scale=0.65}]

\draw (0,0)  node[left] {$0$};

\fill (0:8.5) circle[radius=4pt];

\fill (0:13) circle[radius=4pt];

\draw (0:8.5) node[below right] {$x_-$};
\draw (0:13) node[below right] {$x_+$};

\draw (14:26) node[right] {$\infty_1$};

\draw (-14:26) node[right] {$\infty_{-1}$};

\draw (26,0) node[right] {$\infty_0$};

\draw[color=black!20,dashed] (0,0) -- (21:26);
\draw[color=black!20,dashed] (0,0) -- (-21:26);
\draw[color=black!20,dashed] (0,0) -- (-7:26);
\draw[color=black!20,dashed] (0,0) -- (7:26);

\draw[postaction=decorate] (14:25) to [curve through={(14:24) (14:23) (7:18) (23,0) (24,0)}] (25,0);

\draw[postaction=decorate] (-14:25) to [curve through={(-14:24) (-14:23) (-7:18) (23,0) (24,0)}] (25,0);

\draw[postaction=decorate] (14:25) to [curve through={(14:24) (14:23) (0:12) (-14:23) (-14:24)}] (-14:25);

\draw[postaction=decorate] (14:25) to [curve through={(14:24) (14:23) (14.2:22) (14.4:21) (15.1:18) (15.2:17) (5,0) (4,0) (3,0) (2,0) (1,0) }] (0,0);

\draw[postaction=decorate] (-14:25) to [curve through={(-14:24) (-14:23) (-14.2:22) (-14.4:21) (-15.1:18) (-15.2:17) (5,0) (4,0) (3,0) (2,0) (1,0) }] (0,0);

\draw[postaction=decorate, color=red] (6.5,0) to [curve through={(10.75,-3) (15,0) (10.75,3)}] (6.5,0);

\draw (-7:26)  node[right] {$\infty_{-\frac12}$};
\draw (7:26)  node[right] {$\infty_{\frac12}$};
\draw (-21:26)  node[right] {$\infty_{-\frac32}$};
\draw (21:26)  node[above right] {$\infty_{\frac32}$};

\end{tikzpicture}
\caption{\small With $E>E_*$. Black lines: Topological representation of admissible lines joining $\infty_{\pm1}$ to $0$ and $\infty_1$ to $\infty_{-1}$. Red line: Loop associated to the admissible quadrilateral $0\infty_{-1}\infty_0\infty_1$.}
\label{fig:admissible-lines}
\end{figure}

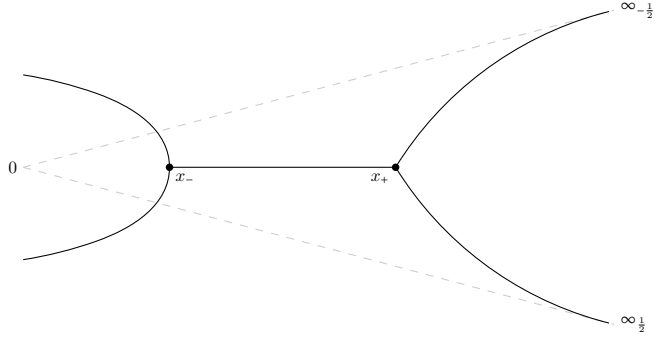
\begin{figure}[H]
\centering
\begin{tikzpicture}[decoration={markings, mark= at position 0.5 with {\arrow{stealth}}},scale=0.35, every node/.style={scale=0.65}]

\draw (0,0)  node[left] {$0$};

\fill (0:5.5) circle[radius=4pt] ;

\fill (0:14) circle[radius=4pt];

\draw (0:5.5) node[below right] {$x_-$};
\draw (0:14) node[below left] {$x_+$};

\draw[color=black!20,dashed] (0,0) -- (-15:23);
\draw[color=black!20,dashed] (0,0) -- (15:23);

\draw (5.5,0) -- (14,0);
\draw (0,-3.5) to [curve through={(2.75,-2.8) (5.5,0) (2.75,2.8)}] (0,3.5);
\draw (14,0) to [curve through={  (15:22.4) }] (15:22.8);
\draw (14,0) to [curve through={  (-15:22.4) }] (-15:22.8);

\draw (-15:23)  node[right] {$\infty_{\frac12}$};
\draw (15:23)  node[right] {$\infty_{-\frac12}$};

\end{tikzpicture}
\caption{\small With $E>E_*$. Topology of the Stokes complex, restricted to the sector $|\arg x|\leq \frac{\pi}{2\alpha+2} $.}
\label{figure:topology-stokes-complex}
\end{figure}

\subsection{Bohr-Sommerfeld Quantisation conditions}
By definition, see \eqref{eq:RW}, the WKB Fock-Goncharov coordinate $R^W_{\uo,-1,0,1}$ is given by
$$
R^W_{\uo}=-e^{\frac{\pi}{2}ind_{\gamma}(V)}e^{-\hbar^{-1}\oint_{\gamma}\sqrt{V(x)}dx},
$$
where $\gamma$ is the loop associated to the quadrilateral $\underline{0}\infty_{-1}\infty_0\infty_1$ and it is depicted in Figure \ref{fig:admissible-lines} above.
The loop $\gamma$ encircles the two real and positive roots $x_-,x_+$ and has index $2$. Moreover,
$-\oint\hbar^{-1}\oint_{\gamma}\sqrt{V(x)}dx=  2 i \int_{x_-}^{x_+}\sqrt{-V(x;E,\ell)} dx $, where the 
$\sqrt{-V(x;E,\ell)}$ is chosen to be positive on $x\in (x_-,x_+)$. Therefore,
\begin{equation}
\label{eqn:semicl-fock-goncharov}
R^W_{\uo}= \exp \left\{ 2 i \int_{x_-}^{x_+}\sqrt{-V(x;E,\ell)} dx \right\}.
\end{equation}
Recall that the spectral points are described by the implicit equation $R_{\uo}=-1$. By equating $R^W_{\underline{0}}=-1$, we obtain the following
Bohr-Sommerfeld quantisation conditions,
\begin{equation}
\label{eqn:bohr-sommerfeld}
\frac{1}{\pi}\int_{x_-}^{x_+} \sqrt{E-x^{2 \alpha}-\frac{\left(\ell+\frac{1}{2}\right)^2}{x^2}}dx=n+\frac{1}{2},\quad n\in\N,
\end{equation}
which yields the WKB approximation of the spectrum.
In the next and final lecture, we will show that the spectrum of the anharmonic oscillators is well-approximated by the solution to the Bohr-Sommerfeld conditions, in both asymptotic regimes that we are considering.

\paragraph{\textbf{Exercises. III}}

1. Prove that it is possible to deform strictly admissible trajectories as described in the surgery procedures, see Definition \ref{def:surgery}.

2. Show that the Stokes complex for the harmonic oscillator is as in \ref{figure:Stokes-complex-harmonic-2}, \ref{figure:Stokes-complex-harmonic-3}, and  Figure \ref{figure:Stokes-complex-harmonic}.

3. Show that the Stokes complex for general $\alpha$ is connected when $E=0$ and $\ell >-\frac12$.
 
\newpage
\section{Lecture IV. Asymptotic analysis of spectral determinants}

Throughout this lecture, we will use the following

\begin{notation} 
We use the notation $A \lesssim B$ (resp. $A \gtrsim B $ ) to indicate that $A \leq C\, B$ (resp. $A \geq C \, B $ ),
where $C >0$ is an absolute constant that only depends on fixed parameters.
We also write $A \lesssim_{k} B$ to indicate that the implicit constant depends on a parameter $k$.
\end{notation}

In this lecture, we prove that the spectrum of the anharmonic oscillator is well-approximated by the Bohr-Sommerfeld quantisation conditions \eqref{eqn:bohr-sommerfeld}. In view of the Bohr-Sommerfeld conditions, it is natural to introduce the following function,
\begin{equation}
\label{eq:WKBintegral}
\begin{aligned} 
& I\colon \R_{>0} \times  \R_{>-\frac{1}{2}} \ni (E,\ell)  \mapsto I(E,\ell)\in\R, \\
& I(E,\ell) := \begin{cases}
 \frac{1}{\pi}\int_{x_-}^{x_+}\sqrt{E -x^{2\alpha}-\frac{\left(\ell+\frac{1}{2}\right)^2}{x^2}} \, dx , & E>E_*(\ell) \\
 0, & E\leq E_*(\ell)
 \end{cases}
\end{aligned}
\end{equation}
where $E_*$ is as per \eqref{eqn:e-min-x-min}.
The Bohr-Sommerfeld quantization condition reads
\begin{equation}
\label{eqn:quantization-I-E-l}
I(E,\ell)=n+\frac{1}{2},\quad n\in\mathbb{Z}_{n\ge 0}.
\end{equation}
The following Proposition holds:
\begin{proposition}
\label{prop:solutions-bs-condition}
For any $n\in\mathbb{N}$, there exists one and only one solution to~\eqref{eqn:quantization-I-E-l}. This solution is denoted $\widehat{E}_n(\ell)$ and is a smooth function of $\ell\in\left(-\frac{1}{2},+\infty\right)$.
\end{proposition}
\begin{proof}
If $E>E_*$, then $\partial_E I(E,\ell)= \int_{x_-}^{x_+}\frac{1}{2\sqrt{E-x^{2\alpha}-\frac{(\ell+\frac12)^2}{x^2}}} dx >0$. The thesis follows from the implicit function theorem.
\end{proof}

We give now the precise statements of the main results of this lecture. Recall that, for any fixed $\ell>-\frac12$, the zeroes of $Q_+(E,\ell)$, namely the spectrum of the anharmonic oscillator,
are real positive and simple. Denoting them as $E_n(\ell)$, $n\in\mathbb{N}$, in such a way that $E_n(\ell)<E_{n+1}(\ell)$, the eigenfunction corresponding to $E_n$ has exactly $n$ zeros on $(0,+\infty)$.

The main results of this lecture are the following two theorems, that provide an asymptotic characterization of the spectral points $E_n(\ell)$ in the two asymptotic regimes that  we are considering.

In the first asymptotic regime, we let $E \to +\infty$ with $\ell$ fixed. This results in the study of the spectrum
$E_n(\ell)$ for $n$ large and $\ell$ fixed.
\begin{theorem}
\label{theorem:large-e-fixed-l}
Let us fix $\ell>-\frac{1}{2}$. There exists a positive integer $N\in\mathbb{N}$, depending on $\ell$, such that
\begin{equation}
\label{eqn:evalue-large-e-fixed-l}
\left|\left[\frac{\pi}{2} \frac{\Gamma\left(\frac{1+3\alpha}{2\alpha}\right)}{\Gamma\left(\frac{1+2\alpha}{2\alpha}\right)} \right]^{-\frac{2 \alpha}{\alpha+1}} (4n+2\ell+1)^{-\frac{2 \alpha}{\alpha+1}} E_{n}(\ell) -1 \right|\lesssim_{\ell}
\begin{cases}
\frac{1}{n} & \mbox{if }\alpha>\frac{1}{2}, \\
\frac{\log n}{n} & \mbox{if } \alpha=\frac{1}{2}, \\
\frac{1}{n^{\alpha+\frac{1}{2}}} & \mbox{if } 0<\alpha<\frac{1}{2}
\end{cases}
\end{equation}
holds for all $n\geq N$.
\end{theorem}

In the second asymptotic regime, we let $E ,\ell \to +\infty$ with $\nu=E\, \ell^{\frac{\alpha+1}{2\alpha}}$ fixed. This results in the study of the spectrum
$E_n(\ell)$ for $\ell \to +\infty$ and $n$ unrestricted.
\begin{theorem}
\label{theorem-large-e-l-bis}
For all $\ell$ sufficiently large and every $n \in \N$, the following estimate holds 
\begin{eqnarray}\label{eq:EnEhatn}
    \left| \frac{E_n(\ell)}{\widehat{E}_n(\ell)} -1 \right| \lesssim \ell^{-1} \frac{(2n+1)}{\ell}\left(G\left( \frac{2n+1}{\ell}\right)\right)^{-2}  .
\end{eqnarray}
In the above formula $G\colon [0,+\infty)\to (0,+\infty)$ is a smooth, strictly positive and strictly monotone function such that $G(0) = \frac{\alpha+1}{\alpha^{\frac{\alpha}{\alpha+1}}}$, $ G'(0) = \frac{\alpha^{-\frac{1}{\alpha+1}}}{2\sqrt{2\alpha+2}}, $ and $G(x)\sim A \, x^{\frac{\alpha+1}{2 \alpha}}$ as $x \to +\infty$, with $A=\frac{1}{2\sqrt{\pi}}\frac{\Gamma\left(\frac{1+2\alpha}{2\alpha}\right)}{\Gamma\left(\frac{1+3\alpha}{2\alpha}\right)}$.

In particular, there are two sub-regimes:
\begin{itemize}
\item[I)] For any fixed $n\in\mathbb{N}$, inequality
\begin{eqnarray}\label{eq:harmonicapproximation}
    \left|\frac{\alpha^{\frac{\alpha}{\alpha+1}}}{\alpha+1} \ell^{-\frac{2\alpha}{\alpha+1}} E_n(\ell) - 1 - \frac{2 \alpha \sqrt{2}}{\sqrt{\alpha+1}}\left(n+\frac{1}{2}\right) \ell^{-1}  \right| \lesssim_n  \ell^{-\frac{3}{2}} 
\end{eqnarray}
holds for all sufficiently big values of $\ell$ (depending on $n$).

\item[II)] If $\frac{2n+1}{\ell}\geq \mu $, for some $\mu>0$, then 
\begin{eqnarray}\label{eq:nlargerthanell}
    \left| \frac{E_n(\ell)}{\widehat{E}_n(\ell)} -1 \right| \lesssim_\mu \ell^{-1} \left(\frac{(2n+1)}{\ell}\right)^{-\frac1\alpha}.
\end{eqnarray}
\end{itemize}
\end{theorem}

\begin{remark}\label{rem:literaturethm}
The asymptotics of the spectrum for fixed $\ell$ is known in the physics literature, see e.g. \cite{dorey98}. Theorem \ref{theorem:large-e-fixed-l} is the first place, to our knowledge, where the error term is precisely estimated, and the result is actually proven in this generality.

The asymptotics of the spectrum for large $\ell$ is not so well studied. The asymptotics for the bottom of the spectrum, i.e. when $n$ is fixed, can be deduced by a standard method in quantum mechanics, the approximation of the potential at a non-degenerate minimum by a harmonic oscillator, see \cite{messiah14}; formula \eqref{eq:harmonicapproximation} indeed already appears, without proof, in \cite{coma20}.
The unrestricted formulae \eqref{eq:EnEhatn} and \eqref{eq:nlargerthanell} require the more specialised analysis that we have developed here. We notice that the same formulae were proven on the IM side of the ODE/IM correspondence in \cite{coma21}, as describing the large momentum asymptotics of Bethe roots for the Quantum KdV model. The coincidence of these asymptotic formulae constitutes the only known proof of the ODE/IM correspondence, see \cite{coma21,masoero2024qfunctions}.
\end{remark}

\subsection{WKB integrals}
\label{subsection:wkb-integrals}

We study here the asymptotics of the WKB integrals defined in~\eqref{eqn:reduced-wkbintegrals}. First of all, we notice that $I(E,\ell)$ enjoy the following covariance symmetry:
\begin{equation}
\label{eqn:covariance-wkbintegral}
 I\left(E \,\la^{\frac{2\alpha}{\alpha+1}},\ell \, \la -\frac{1}{2}\right)=\la \, I\left(E,\ell-\frac{1}{2} \right),
\end{equation}
for all $\lambda\in\mathbb{R}_{>0}$. Due to~\eqref{eqn:covariance-wkbintegral} we can fix the value of $E$ or $\ell$, therefore it is convenient to introduce the following quantities: 
\begin{equation}
\label{eqn:reduced-wkbintegrals}
 J_1(u):=I\left(1, u-\frac{1}{2} \right), \qquad J_2(u):=I\left(u,\frac{1}{2}\right)=u^{\frac{\alpha+1}{2\alpha}} J_1(u^{-\frac{\alpha+1}{2\alpha}}).
\end{equation}

The functions $J_1(u)$ and $J_2(u)$ are convenient to the aim of our asymptotic studies. Indeed, for the sake of our analysis, we need to study the small $u$ limit of the function $J_1(u)$ for the regime $E\to \infty$ with fixed $\ell$, while the function $J_2(u)$, for all values of $u$, characterizes the regime $E,\ell\to\infty$. More precisely, we have
\begin{equation}
\label{eqn:7-novembre-2024-1}
\left.I(E,\ell)\right|_{E=\hbar^{-\frac{2\alpha}{\alpha+1}}}=\hbar^{-1} J_1\left( \left(\ell+\frac{1}{2}\right)\hbar \right)
\end{equation}
and
\begin{equation}
\label{eqn:7-novembre-2024-2}
\left. I(E,\ell) \right|_{E\left(\ell+\frac{1}{2}\right)^{-\frac{2\alpha}{\alpha+1}}\equiv \nu}= \hbar^{-1} J_2(\nu).
\end{equation}

\begin{proposition}
\label{lemma:asymptotics-wkb-integrals-bis}
Let us fix a constant $C>0$. There exists a constant $\delta>0$ depending only on $C$ such that
\begin{equation}
\label{eq:J10bis}
\left|J_1\left( u \right)-\frac{1}{2 \sqrt{\pi}} \frac{\Gamma\left(\frac{1+2\alpha}{2\alpha}\right)}{\Gamma\left(\frac{1+3\alpha}{2\alpha}\right)}+\frac{1}{2}|u|\right|\lesssim_{C} 
\begin{cases}
|u|^2, & \mbox{for }\alpha>\frac{1}{2}, \\
|u|^2 \left|\log |u|\right| & \mbox{for }\alpha=\frac{1}{2}, \\
|u|^{2\alpha+1} & \mbox{for }0<\alpha<\frac{1}{2}
\end{cases}
\end{equation}
holds for all $|u|<\delta$, with $|\arg(u)|\le C |u|$. 
\end{proposition}

\begin{proof}
In the variable $u=\left(\ell+\frac{1}{2}\right)\hbar$, with small $|u|$, the end points $u_-,u_+$ of the WKB integral $J_1(u)$ have asymptotics
\[
u_-=u\left(1+\frac{u^{2\alpha}}{2}+O\left( u^{4 \alpha} \right)\right)
\]
and
\[
u_+=1-\frac{u^2}{2\alpha}+O(u^4),
\]
as $|u|\to 0$ (the proof of this asymptotics is given in Lemma \ref{turning-points-case1} of the subsequent section). Letting $r_-,r_+>0$ be fixed constants and defining $t_1:= |u|\left(1+r_- |u|^{2 \alpha}\right)$, $t_2:=1-r_+ |u|^2$, the path of integration is taken to be the composition of segments
\[
\left[u_-,t_1\right]*[t_1,t_2]*[t_2,u_+].
\]
Along the segments $[u_-,t_1]$ and $[t_2,u_+]$ the integral of $\sqrt{1-y^{2 \alpha}-u^2y^{-2}}dy$ is $O(|u|^{2 \alpha+1})$ and $O(|u|^3)$, respectively. On the segment $[t_1,t_2]$ we can write
\[
\int_{t_1}^{t_2}\sqrt{1-y^{2\alpha}-\frac{u^2}{y^2}} dy=\frac{\sqrt{\pi}}{4\alpha} \sum_{m,j\ge 0} \frac{(-1)^{m+j} u^{2m}}{m! j! \Gamma\left(\frac{3}{2}-(m+j)\right)} \int_{t_1^{2\alpha}}^{t_2^{2\alpha}} w^{\frac{1-2m}{2\alpha}+j-1} dw 
\]
(this series comes from two binomial expansions, which are uniformly and absolutely convergent, provided we take $r_-,r_+>0$ sufficiently large). From 
\[
\int_{{t_1}^{2 \alpha}}^{{t_2}^{2 \alpha}}  w^{\frac{1-2m}{2\alpha}+j-1} dw =
\begin{cases}
\frac{t_2^{1-2m+2\alpha j}-t_1^{1-2m+2\alpha j}}{\frac{1-2m}{2\alpha}+j} & \mbox{if } \frac{1-2m}{2\alpha}+j\neq 0 \\
2\alpha \log\frac{t_2}{t_1} & \mbox{if }\alpha=\frac{2s-1}{2 q},\,m=s,\,j=q\\
 & \mbox{for some }s,q\in\mathbb{Z}_{>0},
\end{cases}
\]
and using formulas
\[
\sum_{k\ge 0} \binom{\frac{1}{2}}{k} \frac{(-1)^k}{1+z k} =\frac{\sqrt{\pi}}{2+z} \frac{\Gamma\left(\frac{1}{z}\right)}{\Gamma\left(\frac{2+z}{2z}\right)},
\]
\[
\sum_{k\ge 0}  \binom{\frac{1}{2}}{k} \frac{(-1)^k}{1-2 k} z^k=\sqrt{1-z} +\sqrt{z} \arcsin{\sqrt{z}},\quad |z|<1,
\]
the result follows by extracting the first two leading terms in $|u|$, namely the terms involving $t_2$ with $m=0$ and summed over $j$ and the terms involving $t_1$ with $j=0$ and summed over $m$ (recall also that we require $|arg(u)|\le C |u|$).

\end{proof}

A direct corollary of \ref{lemma:asymptotics-wkb-integrals-bis} is the large $n$ asymptotics (with fixed $\ell>-\frac{1}{2}$) of the solution $\widehat{E}_n(\ell)$ to the Bohr-Sommerfeld quantisation conditions~\eqref{eqn:quantization-I-E-l}:

\begin{proposition}
\label{prop:large-n-semiclassical-E}
Let us fix $\ell>-\frac{1}{2}$. There exists a positive integer $N\in\mathbb{N}$, depending on $\ell$, such that
\begin{equation}
\label{eqn:large-n-semiclassical-E}
\left|\left[\frac{\pi}{2}\frac{\Gamma\left(\frac{1+3\alpha}{2\alpha}\right)}{\Gamma\left(\frac{1+2\alpha}{2\alpha}\right)}\right]^{-\frac{2\alpha}{\alpha+1}}(4n+2\ell+1)^{-\frac{2\alpha}{\alpha+1}} \widehat{E}_n(\ell)-1\right|\lesssim_{\ell}
\begin{cases}
\frac{1}{n} & \mbox{if }\alpha>\frac{1}{2}, \\
\frac{\log n}{n} & \mbox{if }\alpha=\frac{1}{2} \\
\frac{1}{n^{\alpha+\frac{1}{2}}} & \mbox{if }0<\alpha<\frac{1}{2}
\end{cases}
\end{equation}
holds for all $n\ge N$.
\end{proposition}

Before studying the function $J_2(\nu)$, it is convenient to introduce the following notation:
\begin{equation}
\label{eqn:critical-values}
\nu_*:=\frac{\alpha+1}{\alpha^{\frac{\alpha}{\alpha+1}}},\quad\mbox{and}\quad y_*:=\alpha^{-\frac{1}{2\alpha+2}},
\end{equation}
being $\nu=\nu_*$ the critical value at which the rescaled potential $y^{2 \alpha}+\frac{1}{y^2}-\nu$ has a double turning point at $y=y_*$.

\begin{proposition}
\label{lemma:asymptotics-wkb-integrals-ter}
\begin{itemize}
\item[I)] There exists a constant $\delta>0$ such that
\begin{equation}
\label{J2u*bis}
\left|J_2(\nu)-\frac{\alpha^{-\frac{1}{\alpha+1}}}{2\sqrt{2 \alpha+2}}(\nu-\nu_*)\right| \lesssim |\nu-\nu_*|^{\frac{3}{2}}
\end{equation}
holds for all  $0<\nu-\nu_*<\delta$;
\item[II)] There exists a constant $M>0$ such that
\begin{equation}
\label{J2u*ter}
\left|J_2\left( \nu \right)-\frac{\nu^{\frac{\alpha+1}{2\alpha}}}{2 \sqrt{\pi}} \frac{\Gamma\left(\frac{1+2\alpha}{2\alpha}\right)}{\Gamma\left(\frac{1+3\alpha}{2\alpha}\right)}+\frac{1}{2}\right|\lesssim 
\begin{cases}
\nu^{-\frac{\alpha+1}{2\alpha}}, & \mbox{for }\alpha>\frac{1}{2}, \\
\nu^{-\frac{3}{2}} \left|\log \nu \right| & \mbox{for }\alpha=\frac{1}{2}, \\
\nu^{-(\alpha+1)} & \mbox{for }0<\alpha<\frac{1}{2}
\end{cases}
\end{equation}
holds for all $\nu \ge M$;
\item[III)] There exist $C_2>C_1 >0$ such that
\begin{equation}
\label{eqn:bounds-der-J2}
C_1 \,\nu^{\frac{1-\alpha}{2\alpha}} \leq J_2'(\nu) \leq C_2 \, \nu^{\frac{1-\alpha}{2\alpha}}
\end{equation}
for all $\nu\ge \nu_*$.
\end{itemize}
\end{proposition}

To prove point I) of Proposition \ref{lemma:asymptotics-wkb-integrals-ter} we need to know the asymptotics of the real turning points $\hat{y}_-(\nu)$, $\hat{y}_+(\nu)$ of the rescaled potential $y^{2 \alpha}+\frac{1}{y^2}-\nu$ as $\nu-\nu_*\to 0^+$:

\begin{lemma}
\label{lemma:7-novembre-2024-1}
There exists a constant $\delta>0$ such that the positive real turning points $\hat{y}_-(\nu)$, $\hat{y}_+(\nu)$ of the rescaled potential $y^{2 \alpha}+\frac{1}{y^2}-\nu$ satisfy
\begin{equation}
\label{eqn:coalescing-tps-nu}
\begin{aligned}
&  \left|\hat{y}_-(\nu)-y_*+\alpha^{-\frac{1}{\alpha+1}}(2\alpha+2)^{-\frac{1}{2}}(\nu-\nu_*)^{\frac{1}{2}}+\alpha^{-\frac{1}{2\alpha+2}}\frac{2\alpha-5}{12} (\nu-\nu_*)\right|\lesssim \left|\nu-\nu_*\right|^{\frac{3}{2}} \\
& \left| \hat{y}_+(\nu)-y_*-\alpha^{-\frac{1}{\alpha+1}}(2\alpha+2)^{-\frac{1}{2}}(\nu-\nu_*)^{\frac{1}{2}} +\alpha^{-\frac{1}{2\alpha+2}}\frac{2\alpha-5}{12}(\nu-\nu_*) \right|\lesssim \left|\nu-\nu_*\right|^{\frac{3}{2}}
\end{aligned}
\end{equation}
for all $0<\nu-\nu_*<\delta$ ($y_*$ and $\nu_*$ are defined in~\eqref{eqn:critical-values}).
\end{lemma}
\begin{proof}
First of all, for all $\nu$ sufficiently close to $\nu_*$ we can write
\[
\nu-y^{2 \alpha}-\frac{1}{y^2}=\nu-\nu_*+f(y),
\]
where $f(y)$ is a holomorphic function of $y$ in a neighbourhood of $y=y_*$ (not depending on $\nu$) such that
\[
\left|f(y)+2(\alpha+1)^{\frac{2}{\alpha+1}}(y-y_*)^2+\frac{2}{3} \alpha^{\frac{5}{2\alpha+2}}(\alpha+1)(2\alpha-5)(y-y_*)^3\right|\lesssim |y-y_*|^4.
\]
The equation for the turning points can be written as
\[
y=y_*\pm \sqrt{\frac{\alpha^{-\frac{2}{\alpha+1}}}{2\alpha+2}(\nu-\nu_*)+ \left(\frac{\alpha^{-\frac{2}{\alpha+1}}}{2\alpha+2} f(y)+(y-y_*)^2\right)};
\]
for each choice of the sign in the previous equation, existence and uniqueness of a solution $\hat{y}_\pm(\nu)$ for all $\nu$ sufficiently close to $\nu_*$ (from the right), as well as a (convergent) Puiseux expansion in powers of $(\nu-\nu_*)^{\frac{1}{2}}$, is established by the general theory (see \cite{kirwan}, Section 7.2). The computation of the coefficients of the Puiseux expansion is done by iteration method (see also \cite{hinch}, Section 1.3, for simple computational recipes).
\end{proof}

\begin{proof}[Proof of Proposition \ref{lemma:asymptotics-wkb-integrals-ter}]
Inequality~\eqref{J2u*bis} of point I) is obtained by Taylor expanding the integrand around $(y,\nu)=(y_*,\nu_*)$ and taking into account the asymptotics of the real, coalescing turning points as $\nu-\nu_*\to0^+$ given in Lemma \ref{lemma:7-novembre-2024-1} (see also \cite{coma21}, equation 2.22, and \cite{masoero2024qfunctions}, Lemma 6.7). Point II) is just a consequence of the relation between $J_1$ ad $J_2$ (see~\eqref{eqn:reduced-wkbintegrals}) and an application of inequalities~\eqref{eq:J10bis} of Proposition \ref{lemma:asymptotics-wkb-integrals-bis}. To prove~\eqref{eqn:bounds-der-J2} of point III), we have
\[
J_2'(\nu)=\frac{1}{\pi}\int_{\hat{y}_-(\nu)}^{\hat{y}_+(\nu)} \left(\nu-y^{2 \alpha}-\frac{1}{y^2}\right)^{-\frac{1}{2}} dy,
\]
we can (uniformly and absolutely) Taylor expand the integrand around the point of minimum $y=y_*$ and, taking into account the estimates $r_- \nu^{-\frac{1}{2}}<\hat{y}_-(\nu)<R_- \nu^{-\frac{1}{2}}$, $r_+ \nu^{\frac{1}{2\alpha}}<\hat{y}_+(\nu)<R_+ \nu^{\frac{1}{2\alpha}} $, for some $0<r_{\pm}<1$ and $R_\pm>1$ not depending on $\nu$ (these estimates are proved in Lemma \ref{turning-points-nu} of Section \ref{section:theorem2}), the result follows.
\end{proof}

As a direct corollary of  Proposition \ref{lemma:asymptotics-wkb-integrals-ter}, point I), we obtain also the asymptotics of the solution $\widehat{E}_n(\ell)$ to the Bohr-Sommerfeld quantisation conditions~\eqref{eqn:quantization-I-E-l} for fixed $n\in\mathbb{N}$ and large $\ell$:
\begin{proposition}
\label{prop:fixed-n-large-l-hatE}
Let us fix $n\in\mathbb{N}$. There exists a constant $M>0$ depending on $n$ such that
\begin{equation}
\label{eqn:fixed-n-large-l-hatE}
\left|\frac{\alpha^{\frac{\alpha}{\alpha+1}}}{\alpha+1}\ell^{-\frac{2\alpha}{\alpha+1}} \widehat{E}_n(\ell)-1-\frac{2 \alpha \sqrt{2}}{\sqrt{\alpha+1}}\left(n+\frac{1}{2}\right)\frac{1}{\ell}\right|\lesssim_n \frac{1}{\ell^{\frac{3}{2}}}
\end{equation}
holds for all $\ell>M$.
\end{proposition}

\subsection{Proof of Theorem \ref{theorem:large-e-fixed-l}}
\label{subsec:WKB-approximations}

The proof of Theorem \ref{theorem:large-e-fixed-l} is rather long, hence we will break it into several steps. First, we need to show existence of admissible lines joining the vertices of the quadrilateral $\underline{0}\infty_{-1}\infty_0\infty_1$ belonging to the homotopy classes of the curves represented in Figure \ref{fig:admissible-lines}. We have to show then that the quantities $\left|z_{\pm 1,\underline{0}}-1\right|$ and $\left|z_{1,-1}-1\right|$, where the quantity $z_{a,b}$ is as in~\eqref{eq:zab} and is evaluated on the homotopy classes of the curves of Figure \ref{fig:admissible-lines}, go to zero as $\hbar\to 0$, obtaining in this way an approximation of the Fock-Goncharov coordinate $R_{\underline{0}}\left(\hbar^{-\frac{2\alpha}{\alpha+1}};\ell\right)$ in terms of the semiclassical Fock-Goncharov coordinate $R_{\underline{0}}^W\left(\hbar^{-\frac{2 \alpha}{\alpha+1}};\ell\right)$ given in~\eqref{eqn:semicl-fock-goncharov}. This is the content of Proposition \ref{prop:wkb-estimate-z-case-1}. This result requires the study of the turning points in the sector $-\frac{3\pi}{2\alpha+2}<\arg(y)<\frac{3\pi}{2\alpha+2}$, which is given in Lemma \ref{turning-points-case1}. Eventually, the proof of Theorem \ref{theorem:large-e-fixed-l} will be an application of these results and Rouché's Theorem \ref{theorem:rouche}.   

\begin{proposition}
\label{prop:wkb-estimate-z-case-1}
Let us fix $\ell>-\frac{1}{2}$ and a constant $C>0$. There exists a constant $\delta>0$ depending only on $\ell,C$ such that
\begin{align}
\label{eqn:estimate-z-case-1}
& \left|z_{\pm 1,\underline{0}} - 1\right|\lesssim_{\ell,C} |\hbar|  \\
& \left|z_{1,-1}-1 \right| \lesssim_{\ell,C} |\hbar| \label{eqn:estimate-z-case-1b}
\end{align}
hold for all $|\hbar|<\delta$, with $|\arg(\hbar)|\leq C |\hbar|$. Here $z_{a,b}$ is the symbol defined in~\eqref{eq:zab} and it is evaluated on the homotopy classes of curves joining the vertices of the quadrilateral $\underline{0}\infty_{-1}\infty_0\infty_1$ represented in Figure \ref{fig:admissible-lines}.

In particular, it follows from \eqref{eqn:estimate-z-case-1b} that the Sibuya solutions $\Psi_1$ and $\Psi_{-1}$ are linearly independent if $|\hbar|$ is small enough.
\end{proposition}

In order to prove Proposition \ref{prop:wkb-estimate-z-case-1}, we need to locate the turning points of the rescaled potential $y^{2\alpha}+\frac{\left(\ell+\frac{1}{2}\right)^2 \hbar^2}{y^2}-1$:

\begin{lemma}
\label{turning-points-case1}
Let us fix $\ell>-\frac{1}{2}$. The following holds:
\begin{itemize}
\item[I)] There are precisely two simple turning points in the sector $-\frac{\pi}{2\alpha+2} \le \arg(y) \le \frac{\pi}{2\alpha+2}$, call them $y_-(\hbar;\ell)$ and $y_+(\hbar;\ell)$; furthermore, there exists a constant $\delta>0$, depending on $\ell$, such that
\begin{equation}
\label{asymptotics-xplus}
\left| \frac{1}{\hbar}\left(\ell+\frac{1}{2}\right)^{-1} y_-(\hbar;\ell)-1-\frac{1}{2} \left(\ell+\frac{1}{2}\right)^{2\alpha} \hbar^{2 \alpha} \right| \lesssim_{\ell} |\hbar|^{4 \alpha}
\end{equation}
and
\begin{equation}
\label{asymptotics-xzero}
\left| y_+(\hbar;\ell)-1+\frac{1}{2\alpha}\left(\ell+\frac{1}{2}\right)^2 \hbar^2 \right| \lesssim_{\ell} |\hbar|^4
\end{equation}
holds for all $|\hbar|<\delta$ and $|\arg(\hbar)|<\frac{\pi}{2\alpha+2}$.
\item[II)] In each sector $-\frac{3 \pi}{2\alpha+2}<\arg(y)<-\frac{\pi}{2\alpha+2}$ and $\frac{\pi}{2\alpha+2}<\arg(y)<\frac{3\pi}{2\alpha+2}$ there exists at most one turning point, in which case they are denoted $y_{\pm 1}(\hbar;\ell)$; furthermore, there exists a constant $\delta>0$, depending on $\ell$, such that
\begin{equation}
\label{asymptotics-xplusminus1}
\left| y_{\pm 1}(\hbar;\ell) - e^{\pm\frac{i\pi}{\alpha}} +\frac{e^{\mp \frac{i\pi}{\alpha}}}{2\alpha}\left( \ell+\frac{1}{2} \right)^2 \hbar^2 \right| \lesssim_{\ell} |\hbar|^4
\end{equation}
holds for all $|\hbar|<\delta$, $|\arg(\hbar)|<\frac{\pi}{2 \alpha+2}$.
\item[III)] There are no other turning points in the sector $-\frac{3 \pi}{2\alpha+2}<\arg(y)<\frac{3 \pi}{2\alpha+2}$.
\end{itemize}
\end{lemma}

\begin{proof}

The equation for the turning points of the rescaled potential is
\[
y^{2\alpha}-y^2+\left(\ell+\frac{1}{2}\right)^2 \hbar^2=0.
\]
Let us introduce for convenience the parameter $\eta:=\left(\ell+\frac{1}{2}\right)^2 \hbar^2$ and the function $G(y,\eta):=y^{2\alpha}-y^2+\eta$. For any $k\in\mathbb{Z}$ we have
\[
G\left(e^{i\frac{k\pi}{\alpha}},0\right)=0,\quad \frac{\partial }{\partial y} G\left(e^{i\frac{k\pi}{\alpha}},0\right)=2 \alpha e^{i\frac{k\pi}{\alpha}}.
\]
By the implicit function theorem, for any $k\in\mathbb{Z}$ we receive a holomorphic function $x_k(\eta)$ such that $G(y,\eta)=0$ in a neighbourhood of $(y,\eta)=\left(e^{i\frac{k\pi}{\alpha}},0\right)$ if and only if $y=x_k(\eta)$. Plugging the Taylor series of $x_k(\eta)$ into the equation $G\left(x_k^{(\alpha)}(\eta),\eta\right)=0$ we can compute its coefficients up to any fixed order and by holomorphicity of $x_k(\eta)$ we know the order of the remainder as $\eta\to 0$. In particular, we find
\[
\left|x_k(\eta)-e^{i\frac{k\pi}{\alpha}}+\frac{e^{-i\frac{k\pi}{\alpha}}}{2 \alpha} \eta\right|\lesssim_{k} |\eta|^2.
\]
Specializing to $k=0,\pm 1$ and coming back to the original parameters, we obtain inequalities~\eqref{asymptotics-xzero} and~\eqref{asymptotics-xplusminus1}. 

The proof of inequality~\eqref{asymptotics-xplus} is left as an exercise for the reader. The proof of point III) is just a direct checking. 

\end{proof}

\begin{figure}[H]
\begin{tikzpicture}[decoration={markings, mark= at position 0.5 with {\arrow{stealth}}},scale=0.35, every node/.style={scale=0.65}]

\draw[dashed] (-60:3) arc[start angle=-60, end angle=60, radius=3];
\draw[dashed] (3,0) -- (10,0);
\draw[dashed] (10,0) -- (30:15); 
\fill (0:2.2) circle[radius=4pt];
\fill (9.83,0.87) circle[radius=4pt];
\fill (0,0) circle[radius=4pt];
\draw (0,0) node[left] {$0$};
\draw (0:2.2) node[below] {$\lambda_1$};
\draw (9.83,0.87) node[above left] {$\lambda_2$};
\draw (9.0,7.5) node[above] {$\infty_1$};
\draw[thick] (2.2,0) arc[start angle=180, end angle=0, radius=0.8];
\draw[thick] (0,0) -- (2.2,0);
\draw[thick] (3.8,0) -- (9.1,0);
\draw[thick] (9.1,0) arc[start angle=180, end angle=100, radius=0.9];
\draw[thick] (9.83,0.87) to [curve through={(9.8,0.97)}] (9.0,7.5);
\end{tikzpicture}
\caption{\small The admissible path $\gamma_{1,\underline{0}}$.}
\label{fig:4-novembre-2024-1}
\end{figure}
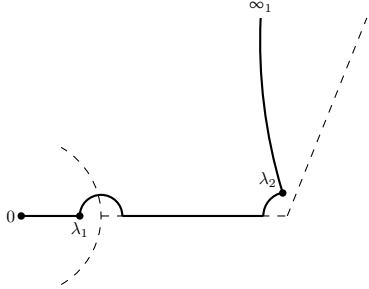

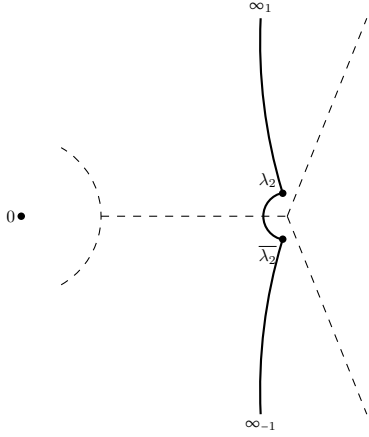
\begin{figure}[H]
\begin{tikzpicture}[decoration={markings, mark= at position 0.5 with {\arrow{stealth}}},scale=0.35, every node/.style={scale=0.65}]

\draw[dashed] (-60:3) arc[start angle=-60, end angle=60, radius=3];
\draw[dashed] (3,0) -- (10,0);
\draw[dashed] (10,0) -- (-30:15); 
\draw[dashed] (10,0) -- (30:15); 
\fill (9.83,-0.87) circle[radius=4pt];
\fill (9.83,0.87) circle[radius=4pt];
\fill (0,0) circle[radius=4pt];
\draw (0,0) node[left] {$0$};
\draw (9.83,-0.87) node[below left] {$\overline{\lambda_2}$};
\draw (9.83,0.87) node[above left] {$\lambda_2$};
\draw (9.0,-7.5) node[below] {$\infty_{-1}$};
\draw (9.0,7.5) node[above] {$\infty_{1}$};
\draw[thick] (9.1,0) arc[start angle=180, end angle=260, radius=0.9];
\draw[thick] (9.1,0) arc[start angle=180, end angle=100, radius=0.9];
\draw[thick] (9.83,-0.87) to [curve through={(9.8,-0.97)}] (9.0,-7.5);
\draw[thick] (9.83,0.87) to [curve through={(9.8,0.97)}] (9.0,7.5);
\end{tikzpicture}
\caption{\small The admissible path $\gamma_{1,-1}$.}
\label{fig:4-novembre-2024-1-bis}
\end{figure}

\begin{proof}[Proof of Proposition \ref{prop:wkb-estimate-z-case-1}]
Let us start with inequality~\eqref{eqn:estimate-z-case-1}. From Theorem \ref{thm:fundamental-theorem} (see also Remark \ref{remark:refined-estimate}) it follows that it is sufficient to find an admissible curve $\gamma_{\pm 1,\underline{0}}$ belonging to the homotopy class of the curves joining $\infty_{\pm 1}$ to $\underline{0}$ represented in Figure \ref{fig:admissible-lines}  such that 
\begin{equation}
\label{eqn:13-dicembre-2024-1}
\sup_{y\in\gamma_{\pm 1,\underline{0}}}\left|\int_{\infty_1,\gamma_{1,\underline{0}}}^y \left(\frac{e^{-2\left( \Re S(y;\hbar,\ell)-\Re S(t;\hbar,\ell) \right)}-1}{2}\right) F\left(t;\hbar,\ell\right)dt\right|\lesssim_{\ell,C} |\hbar|,
\end{equation}
for all sufficiently small values of $|\hbar|$ and $|\arg(\hbar)|\le C |\hbar|$, where
\[
S(y;\hbar,\ell):=\int_{y_-(\hbar;\ell)}^y\hbar^{-1}\sqrt{t^{2\alpha}+\frac{\left(\ell+\frac{1}{2}\right)^2\hbar^2}{t^2}-1}dt
\]
($y_-(\hbar;\ell)$ is the turning point of point I) of Lemma \ref{turning-points-case1}) and
\[
F(y;\hbar,\ell):=\frac{\hbar}{\sqrt{y^{2 \alpha}+\frac{\left(\ell+\frac{1}{2}\right)^2 \hbar^2}{y^2}-1}}\left[\frac{1}{4 y^2}+\frac{1}{2}\left\{S(y;\hbar,\ell),y\right\}\right],
\]
being $\left\{S(y;\hbar,\ell),y\right\}$ the Schwarzian derivative
\footnote{See \eqref{eq:schwarziander} for the definition of the Schwarzian derivative.}
of $S(y;\hbar,\ell)$ 

Letting $r_-,r_+>0$ be two fixed, sufficiently big numbers, depending on $\ell$, we construct a path $\gamma_{1,\underline{0}}$ as the following composition of paths:
\[
\gamma_{1,\underline{0}}:=\Lambda_1*\mathcal{C}_-*\mathcal{I}*\mathcal{C}_+*\Lambda_2,
\]
where $\Lambda_1$ is the horizontal trajectory starting at $0$ and ending at $\lambda_1:= |\hbar|\left(\ell+\frac{1}{2}\right) \left(1- r_- |\hbar|^{2 \alpha}\right)$ (i.e. $\left\{\Im S(y;\hbar,\ell)=\Im S(\lambda_1;\hbar,\ell)\right\}$), $\mathcal{C}_-$ is the arc starting at $\lambda_1$, encircling the turning point $y_-(\hbar;\ell)$ and ending at $|\hbar|\left(\ell+\frac{1}{2}\right) \left(1+ r_- |\hbar|^{2 \alpha}\right)$, $\mathcal{I}$ is the segment on the real line starting at $|\hbar|\left(\ell+\frac{1}{2}\right) \left(1+ r_- |\hbar|^{2 \alpha}\right)$ and ending at $1-r_+ |\hbar|^{2}$, $\mathcal{C}_+$ is the arc starting at $1-r_+ |\hbar|^{2}$, encircling the turning point $y_+(\hbar;\ell)$ and ending at $\lambda_2:=1-r_+ |\hbar|^{2} e^{i \phi}$ (for some small $\phi>0$), and $\Lambda_2$ is the horizontal trajectory starting at $\lambda_2$ and ending at $\infty_1$ (i.e. $\left\{\Im S(y;\hbar,\ell)=\Im S(\lambda_2;\hbar,\ell)\right\}$). This path is depicted in Figure \ref{fig:4-novembre-2024-1}. 

Along the path $\Lambda_2$ the function $\Re S(y;\hbar,\ell)$ is monotonic and $\left|F(y;\hbar,\ell)\right|\lesssim_{\ell}\frac{|\hbar|}{|y|^{\alpha+2}}$; along the arc $\mathcal{C}_+$ the upper bounds $\left|\Re S(y;\hbar,\ell)\right| \lesssim_{\ell,C} \left| \hbar \right|^3$ and $\left| F(y;\hbar,\ell) \right|\lesssim_{\ell} |\hbar|^{-\frac{3}{2}}$ hold; along the segment $\mathcal{I}$ we have $\left|\Re S(y;\hbar,\ell)\right|\lesssim_{\ell,C} |\hbar|$ and $\left| F(y;\hbar_1,\ell) \right|\lesssim_{\ell} |\hbar|$; along the arc $\mathcal{C}_-$ we have $\left|\Re S(y;\hbar,\ell) \right| \lesssim_{\ell,C} |\hbar|^{2\alpha+1}$ and $\left| F(y;\hbar,\ell) \right|\lesssim_{\ell} 1$; finally, along $\Lambda_1$ the function $\Re S(y;\hbar,\ell)$ is monotonic and $\left|F(y;\hbar,\ell)\right|\lesssim_{\ell} 1$. Putting together these estimates (and taking into account also the length of each finite piece of $\gamma_{1,\underline{0}}$) we obtain inequality~\eqref{eqn:13-dicembre-2024-1}. For the path $\gamma_{-1,\underline{0}}$ we just take the complex conjugate of $\gamma_{1,\underline{0}}$ and inequality~\eqref{eqn:13-dicembre-2024-1} still holds.

To prove~\eqref{eqn:estimate-z-case-1b} we follow the same technique by constructing an admissible line $\gamma_{1,-1}$ starting at $\infty_1$ and ending at $\infty_{-1}$ by joining the curves $\gamma_{1,\underline{0}}$ and $\gamma_{-1,\underline{0}}$ on the real line, where they meet, as depicted in Figure \ref{fig:4-novembre-2024-1-bis}.
\end{proof}

At this point, we have all the elements to prove Theorem \ref{theorem:large-e-fixed-l}:

\begin{proof}[Proof of Theorem \ref{theorem:large-e-fixed-l}]
We consider here the case $\alpha>\frac{1}{2}$, being the other cases analogous.

From Proposition \ref{prop:wkb-estimate-z-case-1} and \ref{lemma:asymptotics-wkb-integrals-bis}, it follows that
\begin{equation}
\label{eqn:6-novembre-2024-2}
\left|R_{\underline{0}}\left(\hbar^{-\frac{2\alpha}{\alpha+1}};\ell\right)\exp\left(i \hbar^{-1}\frac{\sqrt{\pi}}{2} \frac{\Gamma\left(\frac{1+2\alpha}{2\alpha}\right)}{\Gamma\left(\frac{1+3\alpha}{2\alpha}\right)}- i\frac{\pi}{2}\left(\ell+\frac{1}{2}\right)\right)+1\right|\lesssim_{\ell,C} |\hbar|^{-1}
\end{equation}
for all sufficiently small $|\hbar|$, with $|\arg(\hbar)|<C|\hbar|$. 

Let $\mathcal{D}_n$ be the disc in the complex $\hbar$-plane whose boundary is
\begin{equation}
\label{eqn:22-novambre-2024-1}
\partial\mathcal{D}_n:=\left\{ \hbar-\frac{2}{\sqrt{\pi}}\frac{\Gamma\left(\frac{1+2\alpha}{2\alpha}\right)}{\Gamma\left(\frac{1+3\alpha}{2\alpha}\right)}(4n+2\ell+1)^{-1}= \frac{R}{n^2} e^{i \theta},\, 0\le \theta< 2\pi \right\},
\end{equation}
where $R>0$ is a positive number (depending on $\ell$) to be fixed. For any sufficiently big $n\in\mathbb{N}$, from inequality~\eqref{eqn:6-novembre-2024-2} we can find a constant $K_1>0$ (depending on $\ell$) such that
\[
\left|R_{\underline{0}}\left(\hbar^{-\frac{2\alpha}{\alpha+1}};\ell\right)+ \exp\left(-i \hbar^{-1} \frac{\sqrt{\pi}}{2} \frac{\Gamma\left(\frac{1+2\alpha}{2\alpha}\right)}{\Gamma\left(\frac{1+3\alpha}{2\alpha}\right)}+i\frac{\pi}{2}\left(\ell+\frac{1}{2}\right)\right)\right|_{\hbar\in\partial \mathcal{D}_n}\le \frac{K_1}{n}.
\]
Furthermore, for any sufficiently big $n\in\mathbb{N}$, we can find a constant $K_2>0$ (depending on $\ell$) such that
\[
\left|1-\exp\left(-i \hbar^{-1} \frac{\sqrt{\pi}}{2} \frac{\Gamma\left(\frac{1+2\alpha}{2\alpha}\right)}{\Gamma\left(\frac{1+3\alpha}{2\alpha}\right)}+i\frac{\pi}{2}\left(\ell+\frac{1}{2}\right)\right)\right|_{\hbar\in\partial \mathcal{D}_n}\ge \frac{R K_2}{n}.
\]
Choosing $R>\frac{K_1}{K_2}$, it follows that
\begin{equation}
\label{eqn:22-novembre-2024-2}
\begin{aligned}
& \left|R_{\underline{0}}\left(\hbar^{-\frac{2\alpha}{\alpha+1}};\ell\right)+\exp\left(-i \hbar^{-1} \frac{\sqrt{\pi}}{2} \frac{\Gamma\left(\frac{1+2\alpha}{2\alpha}\right)}{\Gamma\left(\frac{1+3\alpha}{2\alpha}\right)}+i\frac{\pi}{2}\left(\ell+\frac{1}{2}\right)\right)\right|_{\hbar\in\partial \mathcal{D}_n}\\
& < \left|1-\exp\left(-i \hbar^{-1} \frac{\sqrt{\pi}}{2} \frac{\Gamma\left(\frac{1+2\alpha}{2\alpha}\right)}{\Gamma\left(\frac{1+3\alpha}{2\alpha}\right)}+i\frac{\pi}{2}\left(\ell+\frac{1}{2}\right)\right)\right|_{\hbar\in\partial \mathcal{D}_n}.
\end{aligned}
\end{equation}
By Rouché's Theorem \ref{theorem:rouche}, we conclude that, for all sufficiently big $n$, in each disc $\mathcal{D}_n$ we have precisely one zero of $R_{\underline{0}}\left(\hbar^{-\frac{2\alpha}{\alpha+1}};\ell\right)+1$. Let us take now a sufficiently big positive integer $n$, let us consider the discs $\mathcal{D}_n$ and $\mathcal{D}_{n+1}$, and let us consider the region depicted in Figure \ref{figure:two-discs-ngbh} below.

\begin{figure}[H]
\centering
\begin{tikzpicture}[decoration={markings, mark= at position 0.5 with {\arrow{stealth}}},scale=0.35, every node/.style={scale=0.65}] 

\draw[thick] (0,1.6) arc[start angle=90, end angle=270, radius=1.6];
\draw[color=black!20,dashed] (0,-1.6) arc[start angle=-90, end angle=90, radius=1.6];

\draw[color=black!20,dashed] (5.6,2.1) arc[start angle=90, end angle=270, radius=2.1];
\draw[thick] (5.6,-2.1) arc[start angle=-90, end angle=90, radius=2.1];

\draw[thick] (0,1.6) -- (5.6,2.1);
\draw[thick] (0,-1.6) -- (5.6,-2.1);

\end{tikzpicture}
\caption{\small Neighbourhood constructed from two consecutive discs $\mathcal{D}_n$ and $\mathcal{D}_{n+1}$.}
\label{figure:two-discs-ngbh}
\end{figure}
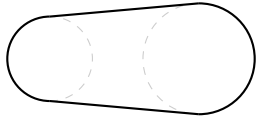
By eventually redefining the constant $R>0$ appearing in~\eqref{eqn:22-novambre-2024-1} (depending on $\ell$), we can show that inequality~\eqref{eqn:22-novembre-2024-2} still holds on the boundary of the region of Figure \ref{figure:two-discs-ngbh}. Rouché's Theorem \ref{theorem:rouche} implies that there are precisely two zeros of $R_{\underline{0}}\left(\hbar^{-\frac{2\alpha}{\alpha+1}};\ell\right)+1$ in that region, which in turn are those contained in $\mathcal{D}_{n}$ and $\mathcal{D}_{n+1}$. Proceeding by induction, we can conclude that there exists some positive integer number $n_0$ such that all the real zeros of $R_{\underline{0}}\left(\hbar^{-\frac{2\alpha}{\alpha+1}};\ell\right)+1$ in a neighbourhood of $\hbar=0$ are contained in
\[
\bigcup_{n\ge n_0} \mathcal{D}_n,
\]
where each $\mathcal{D}_n$ contains precisely one zero. As a consequence, and by construction of the discs $\mathcal{D}_n$, there exists a (a priori unknown) positive integer $k\in\mathbb{N}$ such that
\[
\left|\frac{\sqrt{\pi}}{2}\frac{\Gamma\left(\frac{1+3\alpha}{2\alpha}\right)}{\Gamma\left(\frac{1+2\alpha}{2\alpha}\right)}(4 n+2 \ell+1)\hbar_{n+k}(\ell)-1\right|\lesssim_\ell \frac{1}{n},
\]
where $\hbar_{n+k}(\ell)$ is the unique zero of $R_{\underline{0}}\left(\hbar^{-\frac{2\alpha}{\alpha+1}};\ell\right)+1$ contained in $\mathcal{D}_n$, for all sufficiently big $n\in\mathbb{N}$.

We make now the following 
\begin{claim}
\label{claim-zeros-efunction}
Sibuya's solution $\left.\Psi_0\left(\hbar^{-\frac{1}{\alpha+1}}y; \hbar^{-\frac{2\alpha}{\alpha+1}},\ell\right)\right|_{\hbar=\hbar_{n+k}}$ has precisely $n$ real zeros, for all sufficiently big values of $n$.
\end{claim}
From the general Sturm-Liouville theory, in particular Theorem \ref{theorem:spec-properties}, we know that the eigenfunction corresponding to the $n$-th eigenvalue has precisely $n$ positive real zeros. From the previous claim, it follows then that $k=0$. Finally, coming back to the original parameter $E=\hbar^{-\frac{2\alpha}{\alpha+1}}$ we obtain the statement of the Theorem.

\end{proof}

To complete the proof of Theorem \ref{theorem:large-e-fixed-l} we need to prove Claim \ref{claim-zeros-efunction}. To requires a WKB analysis of Sibuya's or Frobenius solutions at the turning points, which goes beyond the purposes of these introductory lectures. Here we merely give some ideas and refer the reader to \cite{degano2024} for an example of a detailed WKB analysis at the turning points.

\begin{proof}[Sketch of the proof of Claim \ref{claim-zeros-efunction}]
From the standard WKB asymptotics we notice that the eigenfunction $\left.\Psi_0\left(\hbar^{-\frac{1}{\alpha+1}}y; \hbar^{-\frac{2\alpha}{\alpha+1}},\ell\right)\right|_{\hbar=\hbar_{n+k}}$ does not have zeros on $(0,y_-(\hbar;\ell)]$ and $[y_+(\hbar;\ell),+\infty)$ since it has asymptotically an exponential, monotonic behaviour. Defining
\[
\frac{2}{3}\left[-S_-(y;\hbar,\ell)\right]^{\frac{3}{2}}:=\int_{y_-(\hbar;\ell)}^y \sqrt{t^{2 \alpha}+\frac{\left(\ell+\frac{1}{2}\right)^2\hbar^2}{t^2} -1} dt
\]
and 
\[
\frac{2}{3}\left[-S_+(y;\hbar,\ell)\right]^{\frac{3}{2}}:=\int_{y_+(\hbar;\ell)}^y \sqrt{t^{2 \alpha}+\frac{\left(\ell+\frac{1}{2}\right)^2\hbar^2}{t^2} -1} dt,
\]
it can be shown that, on some interval $[y_-(\hbar_{n+k};\ell),\tilde{y}_-]$ and $[\tilde{y}_+,y_+(\hbar_{n+k};\ell)]$, for some $\tilde{y}_-$ and $\tilde{y}_+$ such that $y_-(\hbar_{n+k};\ell)<\tilde{y}_+<\tilde{y}_-<y_+(\hbar_{n+k};\ell)$, we have the following asymptotics
\[
\left.\Psi_0\left(\hbar^{-\frac{1}{\alpha+1}}y; \hbar^{-\frac{2\alpha}{\alpha+1}},\ell\right)\right|_{\hbar=\hbar_{n+k}}\sim \mathcal{C}_n^- \left[\frac{d}{dy} S_-(y;\hbar_{n+k},\ell)\right]^{-\frac{1}{2}}\operatorname{Ai} \left(-\hbar_{n+k}^{-\frac{2}{3}}S_-(y;\hbar_{n+k},\ell)\right),
\]
as $n\to+\infty$, uniformly with respect to $y\in[y_-(\hbar_{n+k},\ell),\tilde{y}_-]$, and for some normalization constant $\mathcal{C}_n^-$, and
\[
\left.\Psi_0\left(\hbar^{-\frac{1}{\alpha+1}}y; \hbar^{-\frac{2\alpha}{\alpha+1}},\ell\right)\right|_{\hbar=\hbar_{n+k}}\sim \mathcal{C}_n^+ \left[\frac{d}{dy} S_+(y;\hbar_{n+k},\ell)\right]^{-\frac{1}{2}}\operatorname{Ai} \left(-\hbar_{n+k}^{-\frac{2}{3}}S_+(y;\hbar_{n+k},\ell)\right),
\]
as $n\to+\infty$, uniformly with respect to $y\in[\tilde{y}_+,y_+(\hbar_{n+k},\ell)]$, and for some normalization constant $\mathcal{C}_n^+$. Using the asymptotics~\eqref{eq:J10bis} for the WKB integral $J_1$, the asymptotics for the eigenvalue $\hbar_{n+k}$ as $n\to+\infty$ and from the fact that the Airy function $\operatorname{Ai}$ has only (negative) real zeros, we can argue that the eigenfunction corresponding to the eigenvalue $\hbar_{n+k}$ has precisely $n$ real zeros.
\end{proof}

\begin{remark}
From the standard WKB asymptotics of Sibuya's/Frobenius solutions,  we can actually prove that the eigenfunction $\left.\Psi_0\left(\hbar^{-\frac{1}{\alpha+1}}y; \hbar^{-\frac{2\alpha}{\alpha+1}},\ell\right)\right|_{\hbar=\hbar_{n+k}}$, for all sufficiently big values of $n$, has $n$ zeros in a small rectangle containing the segment $\left[y_-(\hbar;\ell),y_+(\hbar;\ell)\right]$ by computing the integral along the boundary of the rectangle of the logarithmic derivative, namely
\[
\frac{1}{2\pi i}\oint \left.\frac{\Psi_0'\left(\hbar^{-\frac{1}{\alpha+1}}y; \hbar^{-\frac{2\alpha}{\alpha+1}},\ell\right)}{\Psi_0\left(\hbar^{-\frac{1}{\alpha+1}}y; \hbar^{-\frac{2\alpha}{\alpha+1}},\ell\right)} \right|_{\hbar=\hbar_{n+k}} dy = n,
\]
if $n$ is large enough. However, the complex WKB method we have developed in these lectures does not allow concluding that these zeros are real, and a more refined analysis at the turning points cannot be avoided.
\end{remark}

\subsection{Proof of Theorem \ref{theorem-large-e-l-bis}}
\label{section:theorem2}
We follow here the same scheme of the previous section. We first prove that $\left|z_{\pm 1,\underline{0}}-1\right|$ goes to $0$ as $\hbar\to 0$, for all $\nu\ge\nu_*$, and $\left|z_{-1,1}-1\right|$ goes to $0$ as $\hbar\to 0$ for all $\nu\ge\tilde{\nu}$, for any fixed $\tilde{\nu}>\nu_*$. This is the content of Proposition \ref{prop:wkb-estimate-z-case-2}. To this aim we need to locate the turning points in the sector $-\frac{3\pi}{2\alpha+2}<\arg(y)<\frac{3\pi}{2\alpha+2}$ (see Lemma \ref{turning-points-nu}) and we need to prove existence of admissible lines joining the vertices of the quadrilateral $\underline{0}\infty_{-1}\infty_0\infty_1$ belonging to the homotopy class of the curves represented in Figure \ref{fig:admissible-lines}. Eventually, we have to give an estimate of the quantity appearing on the left-hand side of~\eqref{eq:ztuurho} along such curves. This is performed in the proof of Proposition \ref{prop:wkb-estimate-z-case-2}. Using these results together with Rouché's Theorem we will solve asymptotically equation $R_{\underline{0}}\left(\nu\hbar^{-\frac{2\alpha}{\alpha+1}};\hbar^{-1}-\frac{1}{2}\right)+1=0$ with respect to $\hbar$ and eventually the proof of Theorem \ref{theorem-large-e-l-bis} will consist in the inversion of the formula we obtain with respect to the parameter $\nu$.

\begin{proposition}
\label{prop:wkb-estimate-z-case-2} 
\begin{itemize}
\item[I)] Let us fix a constant $C>0$. There exists a constant $\delta>0$ depending only on $C$ such that
\begin{equation}
\label{eqn:estimate-z-case-2}
\left| z_{\pm 1,\underline{0}} - 1 \right|\lesssim |\hbar|
\end{equation}
holds for all $|\hbar|<\delta$ with $|\arg(\hbar)|\le C |\hbar|$ and all $\nu\in[\nu_*,+\infty)$;
\item[II)] Let us fix a constant $C>0$ and $\tilde{\nu}>\nu_*$. There exists a constant $\delta>0$ depending on $C,\tilde{\nu}$ such that
\begin{equation}
\label{eqn:estimate-z-case-2b}
\left|z_{1,-1}-1\right|\lesssim_{\tilde{\nu}} |\hbar|
\end{equation}
holds for all $|\hbar|<\delta$, with $|\arg(\hbar)|\le C|\hbar|$, and all $\nu\ge\tilde{\nu}$.
\end{itemize}
In~\eqref{eqn:estimate-z-case-2} and~\eqref{eqn:estimate-z-case-2b} the quantities $z_{a,b}$ as those defined in~\eqref{eq:zab} and are evaluated on the homotopy classes of admissible lines joining the vertices of the quadrilateral $\underline{0}\infty_{-1}\infty_0\infty_1$ represented in Figure \ref{fig:admissible-lines}.
\end{proposition}

In order to prove Proposition \ref{prop:wkb-estimate-z-case-2} we need also some estimates for the position of the positive real, simple turning points $\hat{y}_-(\nu)<\hat{y}_+(\nu)$ of the rescaled potential $y^{2\alpha}+\frac{1}{y^2}-\nu$, which is the following

\begin{lemma}
\label{turning-points-nu}
The following holds:
\begin{itemize}
\item[I)] The real, simple turning points $\hat{y}_-(\nu)<\hat{y}_+(\nu)$ are the only turning points in the closed sector $-\frac{\pi}{2\alpha+2}\le \arg(y)\le \frac{\pi}{2\alpha+2}$, furthermore, inequalities
\begin{equation}
\label{eqn:bounds-real-tp-nu}
r_- \nu^{-\frac{1}{2}}<\hat{y}_-(\nu)<R_- \nu^{-\frac{1}{2}},\quad r_+ \nu^{\frac{1}{2\alpha}}<\hat{y}_+(\nu)<R_+ \nu^{\frac{1}{2\alpha}},
\end{equation}
hold for some constants $0<r_-,r_+<1$, $R_-,R_+>1$ independent of $\nu$ and all $\nu>\nu_*$;
\item[II)] There is at most one turning point in each sector $-\frac{3\pi}{2\alpha+2}<\arg(y)<-\frac{\pi}{2\alpha+2}$ and $\frac{\pi}{2\alpha+2}<\arg(y)<\frac{3\pi}{2\alpha+2}$ for all $\nu\ge\nu_*$, in which case they are denoted $\hat{y}_{\pm 1}(\nu)$;
\item[III)] There are no other turning points in the sector $-\frac{3\pi}{2\alpha+2}<\arg(y)<\frac{3\pi}{2\alpha+2}$.
\end{itemize}
\end{lemma}

\begin{proof}
Let us consider the regions
\[
\mathcal{D}_-:=\left\{r e^{i\theta}|\, r_- \nu^{-\frac{1}{2}}< r< R_- \nu^{-\frac{1}{2}},\, \phi_1\le\theta\le\phi_2\right\},
\]
where $0<r_-<1$, $R_->1$ and $-\frac{\pi}{2\alpha+2}<\phi_1<\phi_2<\frac{\pi}{2\alpha+2}$ are fixed constants not depending on $\nu$, and
\[
\mathcal{D}_+:=\left\{r e^{i \theta}|,\, r_+ \nu^{\frac{1}{2\alpha}} <r<R_+ \nu^{\frac{1}{2\alpha}},\,\phi_1\le\theta\le \phi_2  \right\},
\]
where $0<r_+<1$, $R_+>1$ are fixed constants not depending on $\nu$ and $\phi_1,\phi_2$ are as before. We can check that
\[
\left|1-\nu y^2\right|_{\partial \mathcal{D}_-}>\left| y^{2 \alpha+2} \right|_{\partial \mathcal{D}_-}
\]
and
\[
\left|y^{2 \alpha}-\nu\right|_{\partial \mathcal{D}_+} > \left|\frac{1}{y^2}\right|_{\partial \mathcal{D}_+},
\]
provided $r_-,r_+$ are chosen sufficiently close to $0$ and $R_-,R_+$ are chosen sufficiently big (not depending on $\nu$). Due to Rouché's Theorem \ref{theorem:rouche}, we can conclude that the rescaled potential $y^{2 \alpha}+\frac{1}{y^2}-\nu$ has precisely one simple zero in both $\mathcal{D}_-$ and $\mathcal{D}_+$, which must be $\hat{y}_-(\nu)$ and $\hat{y}_+(\nu)$, respectively. Similarly, we can consider the regions
\[
\mathcal{D}_{ -1}:=\left\{r e^{i \theta}|\, r_+ \nu^{\frac{1}{2\alpha}}<r <R_+ \nu^{\frac{1}{2\alpha}},\,\phi_1^-\le\theta\le\phi_2^- \right\}
\]
and
\[
\mathcal{D}_{+ 1}:=\left\{r e^{i \theta}|\, r_+ \nu^{\frac{1}{2\alpha}}<r <R_+ \nu^{\frac{1}{2\alpha}},\,\phi_1^+\le\theta\le\phi_2^+ \right\},
\]
where $0<r_+<1$, $R_+>1$ are chosen as before, while $-\frac{3\pi}{2\alpha+2}<\phi_1^-<\phi_2^-<-\frac{\pi}{2\alpha+2}$ and $\frac{\pi}{2\alpha+2}<\phi_1^+<\phi_2^+<\frac{3\pi}{2\alpha+2}$ are again fixed constants not depending on $\nu$. We find again 
\[
\left|y^{2 \alpha}-\nu\right|_{\partial \mathcal{D}_{\pm 1}} > \left|\frac{1}{y^2}\right|_{\partial \mathcal{D}_{\pm 1}},
\]
from which we can infer that the rescaled potential has at most one simple zero in $\mathcal{D}_{\pm 1}$, which is the case when $\alpha>2$. To prove that there are no other turning points in the sector $-\frac{3\pi}{2\alpha+2}<\arg(y)<\frac{3\pi}{2\alpha+2}$ we repeatedly apply Rouché's Theorem in the remaining regions.
\end{proof}

We can now prove Proposition \ref{prop:wkb-estimate-z-case-2}:

\begin{proof}[Proof of Proposition \ref{prop:wkb-estimate-z-case-2}]
By Theorem \ref{thm:fundamental-theorem} it is sufficient to prove that there exist lines $\gamma_{\pm 1,\underline{0}}$, $\gamma_{1,-1}$ belonging to the homotopy classes of the lines represented in Figure \ref{fig:admissible-lines} such that
\begin{itemize}
\item[i)] $\Re S(y;\hbar,\nu)$ is monotonic along $\gamma_{\pm 1,\underline{0}}$, for all $\nu\in[\nu_*,+\infty)$, and along $\gamma_{1,-1}$ for all $\nu\ge\tilde{\nu}$,
\item[ii)] $\rho_{\gamma_{\pm 1,\underline{0}}}(\nu)\lesssim |\hbar|$, for all $\nu\in[\nu_*,+\infty)$, and $\rho_{\gamma_{1,-1}}(\nu)\lesssim |\hbar|$ for all $\nu\ge\tilde{\nu}$,
\end{itemize}
where 
\[
S(y;\hbar,\nu):=\int_{\hat{y}_-(\nu)}^y \hbar^{-1} \sqrt{\nu-t^{2\alpha}-\frac{1}{t^2}} dt.
\]
Notice also that in order to prove boundedness of $|\hbar|^{-1} \rho_{\gamma_{\pm 1,\underline{0}}}(\nu) $ for all $\nu\in[\nu_*,+\infty)$ and of $|\hbar|^{-1}\rho_{\gamma_{1,-1}}(\nu)$ for all $\nu\ge\tilde{\nu}$ it is sufficient to prove it for all sufficiently big values of $\nu$ (independent of $\hbar$), then, by the continuous dependence of $\rho_{\gamma_{\pm 1,\underline{0}}}(\nu)$ and $\rho_{\gamma_{1,-1}}(\nu)$ with respect to $\nu$, the conclusion follows.

We construct a path $\gamma_{1,\underline{0}}$ as a composition of horizontal and vertical trajectories. Let us choose points $\lambda_1:= \nu^{-\frac{1}{2}}(s_1+i s_2)$ and $\lambda_2:= \nu^{\frac{1}{2\alpha}} (t_1+i t_2)$, for some fixed numbers $s_1, s_2, t_1, t_2>0$. We define
\[
\begin{aligned}
& \Lambda_1:=\left\{\Im S(y;\hbar,\nu)= \Im S(\lambda_1;\hbar,\nu)\right\},\quad \Lambda_2:=\left\{ \Re S(y;\hbar,\nu)= \Re S(\lambda_1;\hbar,\nu) \right\}, \\
& \Lambda_3:=\left\{ \Im S(y;\hbar,\nu) = \Im S(\lambda_2;\hbar,\nu) \right\}.
\end{aligned}
\]
Choosing $s_2$ and $t_2$ sufficiently small (not depending on $\nu$ and $\hbar$) we can show that the horizontal trajectory $\Lambda_1$ has $y=0$ as end point and that the vertical trajectory $\Lambda_2$ meets the horizontal trajectory $\Lambda_3$ at $\lambda_2$ for all $\nu\in [\nu_*,+\infty)$ and all sufficiently small (not depending on $\nu$) $|\hbar|$ with $|\arg(\hbar)|\le C |\hbar|$. The computations needed to see these facts consist in studying the sign of the components of $\operatorname{grad}_{y_1,y_2} Re S(y;\hbar,\nu)$ and $\operatorname{grad}_{y_1,y_2} \Im S(y;\hbar,\nu)$, where we set $y=y_1+i y_2$, so to have a description of the tangent field of the trajectories, namely $\left(-\frac{\partial}{\partial y_2}  \Re S(y;\hbar,\nu),\frac{\partial}{\partial y_1}  \Re S(y;\hbar,\nu) \right)$ for the tangent field of the vertical trajectories and $\left(-\frac{\partial}{\partial y_2}  \Im S(y;\hbar,\nu),\frac{\partial}{\partial y_1}  \Im S(y;\hbar,\nu) \right)$ for the tangent field of the horizontal trajectories. This description allows us to follow a trajectory starting and ending at the chosen fixed points. Furthermore, with these choices of $s_2, t_2$ the parameters $s_1,t_1$ are determined (and strictly greater than $1$), for all $\nu\in[\nu_*,+\infty)$. Finally, we denote $\widetilde{\Lambda}_1$ the piece of $\Lambda_1$ starting at $0$ and ending at $\lambda_1$, $\widetilde{\Lambda}_2$ the piece of $\Lambda_2$ starting at $\lambda_1$ and ending at $\lambda_2$, $\widetilde{\Lambda}_3$ the piece of $\Lambda_3$ starting at $\lambda_2$ and ending at $\infty_{1}$. We construct $\gamma_{1,\underline{0}}$ as the composition
\[
\gamma_{1,\underline{0}}=\widetilde{\Lambda}_1*\widetilde{\Lambda}_2*\widetilde{\Lambda}_3
\]
(see Figure \ref{figure:gamma10-case2}). By construction, along $\gamma_{1,\underline{0}}$ the function $\Re S(y;\hbar,\nu)$ is monotonic for all $\nu\in[\nu_*,+\infty)$ and all sufficiently small (not depending on $\nu$) $|\hbar|$ with $|\arg(\hbar)|\le C |\hbar|$. Finally, letting
\[
F(y;\nu)=\frac{\hbar}{\sqrt{y^{2\alpha}+\frac{1}{y^2}-\nu}}\left[\frac{1}{4 y^2}+\frac{1}{2} \left\{S(y;\hbar,\nu),y\right\}\right]
\]
($\left\{S(y;\hbar,\nu),y\right\}$ denotes the Schwarzian derivative of $S(y;\hbar,\nu)$ with respect to $y$), we can check that
\[
|\hbar|^{-1}\int_{\widetilde{\Lambda}_1,0}^{\lambda_1}\left|F(y;\nu)\right| |dy| \lesssim 1,
\]
for all $\nu\in[\nu_*,+\infty)$,
\[
|\hbar|^{-1}\int_{\widetilde{\Lambda}_2,\lambda_1}^{\lambda_2}\left|F(y;\nu)\right| |dy| \lesssim \nu^{-\frac{\alpha+1}{2\alpha}}
\]
and
\[
|\hbar|^{-1}\int_{\widetilde{\Lambda}_3,\lambda_2}^{\infty}\left|F(y;\nu)\right| |dy| \lesssim \nu^{-\frac{\alpha+1}{2\alpha}}
\]
for all sufficiently big (not depending on $\hbar$) $\nu$. The conclusion follows. For the path $\gamma_{-1,\underline{0}}$ we just take the complex conjugate of $\gamma_{1,\underline{0}}$, while the path $\gamma_{1,-1}$ is constructed by joining the paths $\gamma_{1,\underline{0}}$ and $\gamma_{-1,\underline{0}}$ through a horizontal line emanating from a point $\hat{y}_-(\nu)<\lambda_3<\hat{y}_+(\nu)$ as depicted in Figure \ref{figure:gamma1-1-case2} (in order to choose such a point $\lambda_3$ we need the estimates for the turning points given in Lemma \ref{turning-points-nu}.)

\end{proof}

\begin{figure}[H]
\centering
\begin{tikzpicture}[decoration={markings, mark= at position 0.5 with {\arrow{stealth}}},scale=0.35, every node/.style={scale=0.65}]

\draw[dashed] (-60:3) arc[start angle=-60, end angle=60, radius=3];
\draw[dashed] (3,0) -- (10,0);
\draw[dashed] (10,0) -- (30:15); 
\fill (30:4) circle[radius=4pt];
\fill (14:10.6) circle[radius=4pt];
\fill (0,0) circle[radius=4pt];
\draw[thick] (0,0) -- (30:4);
\draw[thick] (30:4) to [curve through={(27.14:4.0) (9:7.3)}] (14:10.6); 
\draw[thick] (14:10.6) to [curve through={(15.08:10.6) (9.2,6)}] (9.0,7.5);
\draw (0,0) node[left] {$0$};
\draw (30:4) node[above] {$\lambda_1$};
\draw (14:10.6) node[above left] {$\lambda_2$};
\draw (9.0,7.5) node[above] {$\infty_1$};

\end{tikzpicture}
\caption{\small The admissible path $\gamma_{1,\underline{0}}$.}
\label{figure:gamma10-case2}
\end{figure}
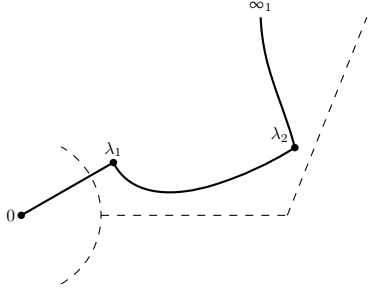
\begin{figure}[H]
\centering
\begin{tikzpicture}[decoration={markings, mark= at position 0.5 with {\arrow{stealth}}},scale=0.35, every node/.style={scale=0.65}]

\draw[dashed] (-60:3) arc[start angle=-60, end angle=60, radius=3];
\draw[dashed] (3,0) -- (10,0);
\draw[dashed] (10,0) -- (30:15); 
\draw[dashed] (10,0) -- (-30:15);
\fill (14:10.6) circle[radius=4pt];
\fill (-14:10.6) circle[radius=4pt];
\fill (0,0) circle[radius=4pt];
\fill (0:7.2) circle[radius=4pt];
\draw[thick] (30:4) to [curve through={(27.14:4.0) (9:7.3)}] (14:10.6); 
\draw[line width=5pt, color=white] (30:4) to [curve through={(27.14:4.0) (9:7.2)}] (9:7.25); 
\draw[line width=5pt, color=white] (-30:4) to [curve through={(-27.14:4.0) (-9:7.2)}] (-9:7.25);
\draw[thick] (14:10.6) to [curve through={(15.08:10.6) (9.2,6)}] (9.0,7.5);
\draw[thick] (9:7.3) -- (-9:7.3);
\draw[thick] (-30:4) to [curve through={(-27.14:4.0) (-9:7.3)}] (-14:10.6); 
\draw[line width=5pt, color=white] (-30:4) to [curve through={(-27.14:4.0) (-9:7.2)}] (-9:7.23);
\draw[thick] (-14:10.6) to [curve through={(-15.08:10.6) (9.2,-6)}] (9.0,-7.5);
\draw (0,0) node[left] {$0$};
\draw (14:10.6) node[above left] {$\lambda_2$};
\draw (-14:10.6) node[below left] {$\overline{\lambda_2}$};
\draw (9.0,7.5) node[above] {$\infty_1$};
\draw (9.0,-7.5) node[below] {$\infty_{-1}$};
\draw (0:7.3) node[below left] {$\lambda_3$};

\end{tikzpicture}
\caption{\small The admissible path $\gamma_{1,-1}$.}
\label{figure:gamma1-1-case2}
\end{figure}
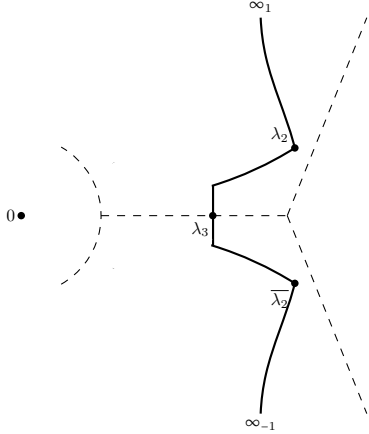

\begin{remark}
If $\hbar$ is taken to be real, $\Re S(y;\hbar,\nu)$ is monotonic along the lines $\gamma_{\pm 1,\underline{0}}$ and $\gamma_{-1,1}$ constructed in the proof of Proposition \ref{prop:wkb-estimate-z-case-2} for all $\hbar>0$.
\end{remark}

Before going to the proof of Theorem \ref{theorem-large-e-l-bis}, we need one more preliminary result. Let us consider the equation
\[
R_{\underline{0}}\left(\nu\hbar^{-\frac{2\alpha}{\alpha+1}};\hbar^{-1}-\frac{1}{2}\right)+1=0.
\]
We denote by $\hbar_{n}(\nu)$ the solutions obtained by solving with respect to $\hbar$, and we denote by $\nu_n(\hbar)$ the solutions obtained by solving with respect to $\nu$. Notice that we have $\hbar_n\left(\nu_n(\hbar)\right)=\hbar$ and $\nu_n\left(\hbar_n(\nu)\right)=\nu$. Analogously, considering the Bohr-Sommerfeld quantisation conditions
\[
\hbar^{-1} J_2(\nu)=n+\frac{1}{2},
\]
we denote by $\hat{\hbar}_n(\nu)=\frac{2 J_2(\nu)}{2n+1}$ its solutions obtained by solving with respect to $\hbar$, and we denote by $\hat{\nu}_n(\hbar)$ its solutions obtained by solving with respect to $\nu$. Again, we have $\hat{\hbar}_n\left(\hat{\nu}_n(\hbar)\right)=\hbar$ and $\hat{\nu}_n\left(\hbar_n(\nu)\right)=\nu$.

\begin{lemma}
\label{lemma:hbarn-approximation}
There exists a constant $\delta>0$ such that
\begin{equation}
\label{eqn:estimate-hn-hathn}
\left|\frac{\hbar_n(\nu)}{\hat{\hbar}_n(\nu)}-1\right|\lesssim \hat{\hbar}_n(\nu) \nu^{-\frac{\alpha+1}{2\alpha}}
\end{equation}
holds for all $(n,\nu)\in \mathbb{N}\times [\nu_*,+\infty)$ such that $\hat{\hbar}_n(\nu) \nu^{-\frac{\alpha+1}{2\alpha}}<\delta$.
\end{lemma}

\begin{proof}
The proof is similar to the one given for Theorem \ref{theorem:large-e-fixed-l}. Let us fix a constant $C>0$. From Theorem \ref{theorem:wkb-cross-ratio}, we have
\[
\left|R_{\underline{0}}\left(\nu \hbar^{-\frac{2 \alpha}{\alpha+1}};\hbar^{-1}-\frac{1}{2}\right) e^{2\pi i \hbar^{-1} J_2(\nu)}-1\right|\lesssim_{C}|\hbar|,
\]
for all $|\hbar|$ sufficiently small (independent of $\nu\ge \nu_*$), with $|\arg(\hbar)|\le C |\hbar|$. Let us consider the discs $\mathcal{D}_n$ whose boundary is
\begin{equation}
\label{eqn:discs-Dn-case2}
\partial \mathcal{D}_n:=\left\{ \hat{\hbar}_n(\nu)+\left(\hat{\hbar}_n(\nu)\right)^2 \nu^{-\frac{\alpha+1}{2\alpha}} R e^{i \theta},\,0\le \theta<2\pi \right\},
\end{equation}
where $R>0$ is a constant independent of $\nu$ to be chosen. For all $(n,\nu)\in\mathbb{N}\times [\nu_*,+\infty)$ so that $\hat{\hbar}_n(\nu)\nu^{-\frac{\alpha+1}{2 \alpha}}$ is sufficiently small, the following estimates hold:
\[
\left|R_{\underline{0}}\left(\nu \hbar^{-\frac{2 \alpha}{\alpha+1}};\hbar^{-1}-\frac{1}{2}\right)-  e^{-2\pi i \hbar^{-1} J_2(\nu)}\right|_{\hbar\in\partial \mathcal{D}_n}\le K_1 \hat{\hbar}_n(\nu)\nu^{-\frac{\alpha+1}{2\alpha}}
\]
and 
\[
\left|1- e^{-2\pi i \hbar^{-1} J_2(\nu)} \right|_{\hbar\in\partial \mathcal{D}_n}\ge R K_2 \hat{\hbar}_n(\nu)\nu^{-\frac{\alpha+1}{2\alpha}},
\]
for some constants $K_1,K_2>0$ independent of $\nu$. Choosing $R>\frac{K_1}{K_2}$, we obtain 
\[
\begin{aligned}
& \left|R_{\underline{0}}\left(\nu \hbar^{-\frac{2 \alpha}{\alpha+1}};\hbar^{-1}-\frac{1}{2}\right)-  e^{-2\pi i \hbar^{-1} J_2(\nu)}\right|_{\hbar\in\partial \mathcal{D}_n} \\
& < \left|1- e^{-2\pi i \hbar^{-1} J_2(\nu)} \right|_{\hbar\in\partial \mathcal{D}_n}.
\end{aligned}
\]
Hence, by Rouché's Theorem \ref{theorem:rouche} there exists a zero $\hbar_n(\nu)$ of $R_{\underline{0}}\left(\nu \hbar^{-\frac{2 \alpha}{\alpha+1}};\hbar^{-1}-\frac{1}{2}\right)$ inside $\mathcal{D}_n$ and, by construction, inequality
\[
\left|\frac{\hbar_n(\nu)}{\hat{\hbar}_n(\nu)}-1\right|\lesssim \hat{\hbar}_n(\nu) \nu^{-\frac{\alpha+1}{2\alpha}}
\]
holds. At this point, we follow the same reasoning of the proof of Theorem \ref{theorem:large-e-fixed-l} considering a region containing two consecutive discs $\mathcal{D}_n$ and $\mathcal{D}_{n+1}$, as depicted in Figure \ref{figure:two-discs-ngbh}, and by using again Rouché's Theorem we conclude.
\end{proof}

We have now all the tools to prove Theorem \ref{theorem-large-e-l-bis}:

\begin{proof}[Proof of Theorem \ref{theorem-large-e-l-bis}]

First of all, notice that
\begin{equation}
\label{preliminary}
\begin{aligned}
\left| \hbar_n\left(\hat{\nu}_n(\hbar)\right) - \hat{\hbar}_n\left(\hat{\nu}_n(\hbar)\right) \right| & \ge \left| \nu_n(\hbar)-\hat{\nu}_n(\hbar)\right| \inf_{\nu\ge \hat{\nu}_n(\hbar)} \hat{\hbar}_n'(\nu) \\
&\gtrsim  \left| \nu_n(\hbar)-\hat{\nu}_n(\hbar)\right| \frac{\left(\hat{\nu}_n(\hbar)\right)^{\frac{1-\alpha}{2\alpha}}}{2n+1},  
\end{aligned}
\end{equation}
where in the last inequality we have used estimate~\eqref{eqn:bounds-der-J2} of Lemma \ref{lemma:asymptotics-wkb-integrals-ter} for $J_2'(\nu)$. Furthermore, notice that for all $\delta>0$, condition
\[
\left.\hat{\hbar}_n(\nu) \nu^{-\frac{\alpha+1}{2\alpha}} \right|_{\nu=\hat{\nu}_n(\hbar)}<\delta
\]
is satisfied for all $n\in\mathbb{N}$ and all sufficiently small values of $|\hbar|$ (independent of $n$). As a consequence, we can evaluate inequality~\eqref{eqn:estimate-hn-hathn} of Lemma \ref{lemma:hbarn-approximation} at $\nu=\hat{\nu}_n(\hbar)$ and, using estimate~\eqref{preliminary}, we obtain
\[
\left|\frac{\nu_n(\hbar)}{\hat{\nu}_n(\hbar)}-1\right|\lesssim |\hbar|^2 (2n+1) \left(\hat{\nu}_n(\hbar)\right)^{-\frac{\alpha+1}{\alpha}},
\]
for all $n\in\mathbb{N}$ and all sufficiently small $|\hbar|$ (not depending on $n$). Coming back to the original parameters, we obtain the statement.

\end{proof}

\section*{Appendix I. Complex Analysis}
Here we collect a few results in analysis and complex analysis that we used in the main text.

\begin{definition}
Given a closed subset $G$ of $\R^n$ or $\C^n$ and a measurable complex-valued function $f$ over it,  we denote by $\mathcal{C}(G)$ the space of continuous bounded functions on $G$. Moreover, we denote by
\begin{equation}\label{eq:supnorm}
    \| f\|_{\infty}= \sup_{x \in G } |f(x)|,
\end{equation}
the sup or $L^{\infty}$ norm.
\end{definition}

\begin{definition}
   Let $\mathcal{U} \subset \C^M$ be the product of $M$ domains $D_1 \times D_2 \times \dots D_M$ with $D_i \subset \C$ and by $\overline{U}$ its closure. We denote by $\mathcal{H}(\mathcal{U})$ the space of continuous bounded function on $\overline{\mathcal{U}}$ which, restricted to $\mathcal{U}$, are holomorphic.
    
    We denote by $\mathcal{H}([0,1]\times \mathcal{U})$ the space of continuous bounded functions $f$ on
    $[0,1] \times \overline{\mathcal{U}}$ such that, for every $t \in[0,1]$, the restriction $f(t;\cdot): \mathcal{U} \to \C$ is holomorphic.
\end{definition}
We know from standard analysis courses that, equipped with the sup norm, $\mathcal{C}(G)$ is a Banach space, namely a complete normed linear space.
It slightly less well-known that $\mathcal{H}(\mathcal{U})$ (and $\mathcal{H}([0,1]\times \mathcal{U})$), equipped with the same norm, are Banach spaces too. To prove this fact, we need two important results in complex analysis, namely  Morera's theorem, and Hartog's Theorem on Separate Holomorphicity
\begin{theorem}[Morera's Theorem]
Let $D \subset \C$ be a domain and $f(z): D \to \C$ be continuous . If $\int_{\gamma} f(z) \, dz=0$ for every closed contractible curve $\gamma$ in $D$ then $f$ is holomorphic.
\end{theorem}

\begin{theorem}[Hartog's Theorem on Separate Holomorphicity] \label{thm:hartog}
Let $D \subset \C^N$ be a domain. If $f:D \to \C$ is separately holomorphic at every point $z \in D$, then $f$  is holomorphic.
\end{theorem}

 \begin{proposition}\label{prop:closure}
$\mathcal{H}(\mathcal{U})$ and $\mathcal{H}([0,1]\times \mathcal{U})$, equipped with the sup norm, are Banach spaces.
  \begin{proof}
  We show that  $\mathcal{H}(\mathcal{U})$ is a Banach space and let the case of the space $\mathcal{H}([0,1]\times \mathcal{U})$ to the reader.

We need to show that  $\mathcal{H}(\mathcal{U})$ is a closed subset of $\mathcal{C}(\mathcal{U})$. We consider then a sequence 
$f_n \in \mathcal{H}(\mathcal{U})$ such that $f_n \to f \in  \mathcal{C}(\mathcal{U})$.

  With $\uu =(u_1,\dots, u_M) \in \mathcal{U}$, we let $\gamma_i$ be a contractible loop in the domain
  $D_i$. Since  $f_n \to f$  in  $\mathcal{C}(\mathcal{U})$ then
\begin{align*}
\oint_{\gamma_i} f(c_1,\dots, u_i, \dots, c_N) du_i= \lim_{n \to \infty} \oint_{\gamma_i} f_n(c_1,\dots, u_i, \dots, c_N) du_i= \lim_{n \to \infty} 0=0,
 \end{align*}
 for any choice of $c_j \in D_j$, $j \neq i$.
Therefore, by Morera's Theorem, $f$ is separately holomorphic with respect to each variable $u_i$ and by Hartog's Theorem $f$ is holomorphic when restricted to the interior of $\mathcal{U}$. Thus, $\mathcal{H}(\mathcal{U})$ is a closed subset of $\mathcal{C}(\mathcal{U})$.
  \end{proof}
 \end{proposition}
 \begin{proposition}\label{prop:dominatedintegral}
     Let $f\colon[0,1] \times [0,1] \times \overline{\mathcal{U}} \to \mathbb{C}$ be a measurable function such that
     \begin{itemize}
 \item[I)] For almost every $s$: $f(s,\cdot;\cdot)\colon [0,1] \times \overline{\mathcal{U}} \to \C$ is continuous and, fixed $t \in [0,1]$, $f(s,t;\cdot)\colon \mathcal{U} \to \C$ is holomorphic. 
 \item[II)] $f(\cdot,t;\uu)$ is uniformly integrable in the following sense: there exists a measurable positive function $h \in L^1([0,1])$ such that
 \begin{equation}
 \sup_{(t,\uu) \in [0,1] \times \overline{\mathcal{U}}}|f(s,t;\uu)|\leq h(s).
 \end{equation}
 \end{itemize}
 The function $F\colon [0,1] \times \overline{\mathcal{U}} \to \C$ defined by $
 F(t;\uu):=\int_0^1 f(s,t;\uu) dt $
 is continuous and, fixed $t \in [0,1]$, $F(t;\cdot): \mathcal{U} \to \C$ is holomorphic.
 \begin{proof}
 Continuity of the function $F\colon [0,1] \times \overline{\mathcal{U}} \to \C$ is a standard result in measure theory, see \cite{malliavin} 7.8.3. Holomorphicity of $F$ with respect to $\uu$ follows from Morera's and Hartog's Theorems, by means of the same considerations as in the proof of the proposition above.
 \end{proof}
 \end{proposition}
 
Rouché's Theorem is also used in Lecture 4.
\begin{theorem}[Rouché's Theorem]
\label{theorem:rouche}
Let $f$ and $g$ be holomorphic functions inside and on a simple closed contour $\Gamma$. If the strict inequality
\[
|g(z)|<|f(z)|
\]
holds for each $z\in \Gamma$, then $f$ and $f+g$ have the same number of zeros (counting multiplicities) inside $\Gamma$.
\end{theorem}

\subsubsection*{Linear ODES with analytic coefficients}

Let $A\colon D \subset \C \to Mat(N)$ be a holomorphic function with values in the space of $N \times N$ matrices. We call
\begin{equation}\label{eq:linearODE}
\frac{d Y(x)}{d x}=A(x) Y(x)
\end{equation}
a first order linear ODE of rank $N$, with analytic coefficients, or simply a linear analytic ODE.
As it is well-known, the second order scalar linear ODE, i.e. Schr\"odinger equation,
\begin{equation*}
	\psi''(x)=U(x)\psi(x)
\end{equation*}
is equivalent to
the rank $2$ linear ODE, via the following transformation
\begin{equation}
	Y'(x)=\left(\begin{matrix} 0 & 1 \\
		U(x) & 0
	\end{matrix}\right) Y(x), \qquad Y(x)=\left(\begin{matrix} \psi(x) \\
		\psi'(x)
	\end{matrix}\right).
\end{equation}
It is well-known that linear ODEs in the complex plane are well-posed on simply-connected domains.
\begin{theorem}[Global well-posedness and analytic dependence on the parameters]\label{thm:genwelposedness}
 Let $A\colon D \to \C$ be a holomorphic function.
 If the domain $D$ is simply connected, for every $x_0 \in D$ and $Y_0 \in \C^N$ the Cauchy problem
	\begin{equation}\label{eq:CauchylinearODE}
\frac{d Y(x)}{d x}=A(x) Y(x), \; Y(x_0)=Y_0
	\end{equation}
admits a unique solution $Y(x;Y_0)\colon D \to \C^N$.

The solution $Y$ depends holomorphically on $Y_0$, and on any additional parameters, provided the matrix $A$ depends holomorphically on these parameters.

\begin{proof}
See \cite{Ilya08}.
\end{proof}

\end{theorem}
We have the following corollaries of well-posedness.
\begin{corollary}\label{cor:extension}
Let $D,D' \subset \C$ be domains such that $D \subset D'$, let $\mathcal{U} \subset \C^M$ be a poly-disc, let $A\colon D'\times U \to Mat(N)$ be analytic and 
 $Y\colon D\times U \to \C^N$ be an analytic function that solves $Y'(x;\uu)=A(x;\uu)Y(x;\uu)$.
 
For every $\uu \in \mathcal{U}$, the function $Y(\cdot;\uu)$ has a unique extension $\widetilde{Y}(\cdot;\uu)\colon D' \to \C^N$, and the function
$\widetilde{Y}:D' \times \mathcal{U} \to \mathbb{C}^N$ is analytic and solves the equation $Y'(x;\uu)=A(x;\uu)Y(x;\uu)$.
\end{corollary}

\begin{corollary}[Solution on a curve extends to a simply connected domain]
Let $\gamma\colon (0,1) \to D$ be a smooth curve and
$y\colon (0,1) \to C^N$ be a solution of $Y'(x)=A(x)Y(x)$ along $\gamma$, namely a differentiable function such that $\dot{y}(t)= A(\gamma(t)) \dot{\gamma}(t) y(t) $ for all $t \in (0,1)$.

There exists a unique $Y\colon D \to \C^N$ solution of $Y'(x)=A(x)Y(x)$ such that $y(t)=Y(\gamma(t))$.

If moreover $y\colon (0,1) \times \mathcal{U} \to C^N$ and $\gamma\colon (0,1) \times \mathcal{U} \to D$ depend analytically on the parameters $\uu$ then $Y\colon D\times \mathcal{U} \to \C^N$
is analytic.
\end{corollary}

\subsubsection*{Transformation laws}
In the theory developed above, a global coordinate $x$ on $\C$ was fixed, hence the coefficients as well as the solutions of a linear ODE are functions.
It is however often necessary to understand how the coefficients and the solutions transform under change of coordinates.
The adequate complex geometric framework for the coefficients of a differential equation and for its solutions are well-known, see e.g. \cite{frenkel2007langlands}; we review it briefly here.

Let $x=\varphi(z)$, with $\varphi\colon  D' \to D$, be a bi-holomorphic map.
The transformation law for the first order ODE is rather neat:
If $Y$ satisfies the linear ODE \eqref{eq:linearODE}, $Y'(x)=A(x) Y(x)$, then
\begin{align}
\widetilde{Y}: D' \to \C, \widetilde{Y}(z)= Y(\varphi(z))
\end{align}
satisfies the linear ODE $\widetilde{Y}'(z)=\widetilde{A}(z) \widetilde{Y}(z)$, with
\begin{align}
	 \widetilde{A}\colon  D' \to \C^N,\quad  \widetilde{A}(z)=\left(\varphi'(z)\right) \widetilde{A}(\varphi(z)).
\end{align}
Hence, we say that $Y$ is a holomorphic function, i.e. a section of the trivial line-bundle, and $A$ is a holomorphic one-form, i.e. a section of the canonical line bundle  $K_X$.

The situation is different in the case of a Schr\"odinger equation:
If $\psi$ satisfies the Schr\"odinger equation \eqref{eq:schr}, $\psi''(x)=U(x) \psi(x)$, then
\begin{equation}\label{eq:psichange}
\widetilde{\psi}\colon  D' \to D , \quad	\widetilde{\psi}(z)=  \left(\varphi'(z)\right)^{-\frac12} \psi(\varphi(z))
\end{equation}
satisfies the Schr\"odinger equation $\widetilde{\psi}''(z)= \widetilde{U}(z) \widetilde{\psi}(z)$, with
\begin{align}
& \label{eq:potchange}
	\widetilde{U}(z)= \left(\varphi'(z)\right)^{2} U(\varphi(z)) - \frac12 \left\lbrace \varphi(z), z \right\rbrace.
\end{align}
where $\left\lbrace \varphi(z), z \right\rbrace$ is the Schwarzian derivative of $\varphi$ \footnote{The Schwarzian derivative is a magic object, see \cite{ovsienko2009}},
\begin{equation}\label{eq:schwarziander}
	\left\lbrace \varphi(z), z \right\rbrace = \frac{\varphi'''(z)}{\varphi'(z)} -\frac32  \left(\frac{\varphi''(z)}{\varphi'(z)} \right)^2.
\end{equation}
Notice that the transformation law for $\psi$ and $U$ are not tensorial:
 $\psi$ is a density of weight $-\frac12$
 and $U$ is a \textit{projective connection}.

Let us use the above law to write the anharmonic oscillator equation \eqref{eq:anharmonic} with respect to a global coordinate $z$ on $\WCs \cong \C$.
To this aim, we use the projection as the change of variable  $\varphi\colon \C \to \widetilde{\C^*},\, \varphi(z)=e^z $.
Under the map $\varphi$, using the transformation law \eqref{eq:potchange}, equation \eqref{eq:anharmonic} reads
\begin{equation}\label{eq:anharmonicexp}
-\psi''(z)= \widetilde{U}(z) \psi(z), \quad  \widetilde{U}(z)=  e^{2(\alpha+1)z} - E e^{2z} + \left(\ell+\frac12\right)^2 .
\end{equation}
\begin{remark}
Notice that, due to the transformation law \eqref{eq:potchange}, positivity is not a property of potentials invariant under change of coordinates. In particular, the anharmonic potential
$\widetilde{U}(z)$ is positive on the real $z$ axis for every $\ell$ real, but
the anharmonic potential $U(x)$ is  positive on the positive $x$ semi-axis 
only when $\ell\geq0$ or $\ell\leq-1$.
The former property implies that the spectrum of the anharmonic oscillator is (strictly) positive for every real $\ell$ even though for some real $\ell$ the potential $U(x)$ is not positive on $\Re x>0$.    
\end{remark}

\section*{Appendix II. Anharmonic oscillators and the ODE/IM correspondence}
In this Appendix, we collect some general results on the anharmonic oscillator
\begin{equation}
	\frac{d^2\psi(x)}{dx^2}= \left(x^{2 \alpha}+\frac{\ell(\ell+1)}{ x^2}-E\right) \psi, \quad x\in\WCs,
\end{equation}
which are not proven in the main text. We follow \cite{masoero2024qfunctions}. We warn the readers that in \textit{loc. cit.}, the anharmonic oscillators appear in the form
\begin{equation}
\tilde{\psi}''(y)=  \left( \la^2 y^k(y-1)+ \frac{\tilde{l}(\tilde{l}+1)}{y^2} \right) \tilde{\psi}(y),
\end{equation}
which is equivalent to the form we use under the following change of coordinates and parameters
\begin{equation}\label{eq:xtoyparameters}
 x= \left(\frac{2 \, \la}{k+2}\right)^{\frac{k+2}{k+3}}y^{\frac{k+2}{2}},\quad \alpha=\frac{1}{k+2}, \quad
 \ell+\frac{1}{2}=\frac{2\left(\tilde{l}+\frac{1}{2}\right)}{k+2}, \quad E=\left(\frac{2\la}{k+2}\right)^{\frac{2}{k+3}}.
\end{equation}

Fixed $\alpha>0,\, \ell$, we are interested in solutions of \eqref{eq:anharmonic} which are analytic function of the energy parameter $E$.
For this reason, we define
\begin{equation}
\mathcal{A}_{\ell}= \left\{ \psi: \WCs \times \C \to \C, \mbox{holomorphic}, \psi \mbox{ satisfies } \eqref{eq:anharmonic} \right\}.
\end{equation}

The following is a corollary of Theorem \ref{thm:genwelposedness}.
\begin{corollary}
$\mathcal{A}_{\ell}$ is a free module of rank $2$ over the ring $\mathcal{O}_E$ of entire functions in the variable $E$.
\end{corollary}

\subsubsection*{Monodromy and Generalised Frobenius solutions}
It is easily seen that the following operator, called Quantum Monodromy, is well-defined on $\mathcal{A}_{\ell}$:
\begin{equation}
	\big(\mathcal{M}\psi\big)(x,E)= \psi(q x, q^{-2} E), \quad q=e^{\frac{\pi i}{\alpha+1}}.
\end{equation}

An eigenbasis of $\mathcal{M}$ is found via a generalised Frobenius expansion, which produces two solutions, the subdominant generalised Frobenius solution $\chi_+$ and the dominant generalised Frobenius solution $\chi_+$.
The solution $\chi_+ \in \mathcal{A}_{\ell}$ is defined via the convergent expansion
\begin{equation}\label{eq:chi+ser}
	\chi_+(x,E):= x^{\ell+1} \left( 1 + \sum_{m,n  \in \N^2 \setminus  (0,0)} c_{m,n} \big(x^2 E\big)^m \big(x^{2\alpha+2}\big)^n \right).
\end{equation}
The solution $\chi_- \in \mathcal{A}_{\ell}$ is defined via the convergent expansion
\begin{equation}\label{eq:chi-ser}
	\chi_-(x,E):= x^{-\ell} \left( 1 + \sum_{m,n  \in \N^2 \setminus  (0,0)} c_{m,n} \big(x^2 E\big)^m \big(x^{2\alpha+2}\big)^n \right),
\end{equation} 
 whenever $\ell+\frac12 \notin \{ i+ \alpha j, (i,j) \in \N \times \N \} $
\footnote{If on the contrary $\ell+\frac12 \in \{ i+ \alpha j, (i,j) \in \N \times \N \}$, a logarithmic term must be added to $\chi_-$. This corresponds to the fact that the quantum monodromy is not diagonalisable in these cases}.
Given the two expansions, one checks directly that
\begin{equation}\label{eq:moneigen}
	\mathcal{M}\chi_{+}= q^{2 \ell+1}\chi_{+}, \quad \mathcal{M}\chi_{-}= q^{-\ell}\chi_{-},
\end{equation}
and that $\{ \chi_+,\chi_-\}$ is a basis of 
$\mathcal{A}_{\ell}$, whenever  $\ell+\frac12 \notin \{ i+ \alpha j, (i,j) \in \N \times \N \} $.

Notice finally that $\chi_+$ can also be characterized as the 
unique solution that satisfies the asymptotics
\begin{equation}\label{eq:chi+as}
	\forall E \in \C, \; \lim_{x \to 0^+} x^{-\ell-1}\chi_+(x,E)=1.
\end{equation}

\subsubsection*{Spectral determinants}
Denote by $\Psi_{k}(x;E,\ell), \, k \in \Z$, the $k-th$ Sibuya solution, defined in the main text.
As we proved in Theorem \ref{thm:sibuya}, $\Psi_k$ is an entire function of $\ell$ and $E$, hence,  fixed $\ell$, $\Psi_k \in \mathcal{A}_{\ell}$.

It is customary to define three spectral determinants $Q_{\pm}(E;\ell)$ and $T(E;\ell)$, via the following expression
\begin{align}\label{eq:QTdef}
&    Q_{\pm}(E;\ell)= Wr_x\left[\chi_{\pm}(x;E,\ell),\Psi_0(x;E,\ell)\right] \\ \nonumber
&  T(E;\ell)=Wr_x\left[\Psi_{-1}(x;E,\ell), \Psi_{1}(x;E,\ell)\right].
\end{align}
In the case of $Q_-$, the formula above is valid whenever $\ell+\frac12 \notin \{ i+ \alpha j, (i,j) \in \N \times \N \}  $. Notice that $T$ coincides with the Stokes multiplier $\sigma_0$ up to constant factor, see \cite{masoero2024qfunctions}.

The following relations were discovered by Dorey and Tateo, in the seminal papers \cite{dorey98,bazhanov01},
that marked the beginning of the ODE/IM correspondence:
\begin{theorem}
  Let $\ell$ be fixed. $Q_{\pm}(E;\ell)$ and $T(E;\ell)$ are entire functions of $E$.

Moreover, the following identities hold
    \begin{align}\label{eq:QQTQ}
   & \gamma \, Q_+(q E) Q_-(q^{-1} E) - \gamma^{-1} \,  Q_+(q^{-1} E) Q_-( E)=1 \\ \nonumber
    & T(E) Q_{\pm}(E)= \gamma^{\pm1} Q_{\pm} ( q^2 E) + \gamma^{\mp1} Q_{\pm} ( q^{-2} E) \\ \nonumber
    & q= e^{i\frac{\pi }{\alpha+1}} , \quad \gamma=e^{ i \frac{ \ell+\frac12}{\alpha+1}\pi} .
    \end{align}
    Whenever $Q_-$ appears, the statements and equations above are valid if $\ell +\frac{1}{2} \notin  \{ i+ \alpha j, (i,j) \in \N \times \N \} $.
\end{theorem}
Restricting our attention to the spectral determinant $Q_+$ and to the case $\ell \in \R$, we have the following characterization of the spectrum.
\begin{theorem}
\label{theorem:spec-properties}
Let $\ell>-\frac12$ be fixed.

The zeros of $Q_+(E)$ are simple, discrete, real and strictly greater than $E_*=\alpha^{-\frac{\alpha}{1+\alpha}}(1+\alpha) \left(\ell+\frac12\right)^{\frac{2\alpha}{1+\alpha}}$.
They form an increasing infinite sequence $\{E_n\}_{n \in \N}$ with the property that the eigenvector associated to the eigenvalue $E_n$ has $n$ simple zeroes on $(0,+\infty)$.
\begin{proof}
We consider the operator $\mathcal{L}=-\partial_x+x^{2\alpha}+ \frac{\ell(\ell+1)}{x^2}$ on $L^2(\R_{\geq0})$. This operator belongs to a class of operators studied in \cite{resi75}, Appendix to X.1, by means of Sturm-Liouville methods.
It is there shown that if $\ell\geq \frac12$, the operator $\mathcal{L}$ is itself self-adjoint. If on the contrary,
$-\frac12 < \ell < \frac12$, the operator $\mathcal{L}$ is symmetric but not self-adjoint; however, the boundary conditions that we have chosen, namely  $\psi \propto \chi_+$ as $x \to 0^+$, select a self-adjoint extension of $\mathcal{L}$. In both cases, 
eigenvalues are simple, since the kernel of $\mc{L} -E$ has at most dimension $1$. 

The fact that the spectrum is greater than $E_*$ follows from studying the anharmonic oscillator equation in the exponential coordinate \eqref{eq:anharmonicexp},
$$\psi''(z)=\left(e^{2(\alpha+1)z}-E e^{2z}+ \left(\ell+\frac12 \right)^2 \right) \psi(z), \quad E\in \R, z \in \R .$$
A necessary conditions for a solution that vanishes both at $-\infty$ and at $+\infty$ to exists  is that the potential $e^{2(\alpha+1)z}-E e^{2z}+ \left(\ell+\frac12\right)^2$ is negative on a non-empty segment of the real axis. In fact, if that does not happen, the solution vanishing at $-\infty$ and positive in a neighbourhood of $-\infty$ would be increasing on $\R$.
A simple computation shows the potential is non-negative when $E\leq E_*$.
 
\end{proof}
\end{theorem}


\def\cprime{$'$} \def\cydot{\leavevmode\raise.4ex\hbox{.}} \def\cprime{$'$}


\begin{thebibliography}{10}

\bibitem{bazhanov01}
V.~Bazhanov, S.~Lukyanov, and A.~Zamolodchikov.
\newblock {Spectral determinants for Schrodinger equation and Q operators of
  conformal field theory}.
\newblock {\em J.Statist.Phys.}, 102:567--576, 2001.

\bibitem{bender69}
C.~Bender and T.~Wu.
\newblock Anharmonic oscillator.
\newblock {\em Physical Review}, 184(5):1231, 1969.

\bibitem{bender13}
Carl~M Bender and Steven~A Orszag.
\newblock {\em Advanced mathematical methods for scientists and engineers I:
  Asymptotic methods and perturbation theory}.
\newblock Springer Science \& Business Media, 2013.

\bibitem{bergweiler95}
W.~Bergweiler and A.~Eremenko.
\newblock On the singularities of the inverse to a meromorphic function of
  finite order.
\newblock {\em Revista Matem{\'a}tica Iberoamericana}, 11(2):355--373, 1995.

\bibitem{bertola24}
M.~Bertola, E.~Chavez-Heredia, and T.~Grava.
\newblock Exactly solvable anharmonic oscillator, degenerate orthogonal
  polynomials and painlev{\'e} {II}.
\newblock {\em Communications in Mathematical Physics}, 405(2):52, 2024.

\bibitem{borrego24}
J.~Borrego-Morell and B.~Shapiro.
\newblock Semiclassical expansion for exactly solvable differential operators.
\newblock {\em arXiv preprint arXiv:2402.19087}, 2024.

\bibitem{bridgeland-masoero-2023}
T.~Bridgeland and D.~Masoero.
\newblock On the monodromy of the deformed cubic oscillator.
\newblock {\em Math. Ann.}, 385:193–258, 2023.

\bibitem{coma20}
R.~Conti and D.~Masoero.
\newblock Counting monster potentials.
\newblock {\em JHEP}, 02:059, 2021.

\bibitem{coma21}
R.~Conti and D.~Masoero.
\newblock On solutions of the bethe ansatz for the quantum kdv model.
\newblock {\em Communications in Mathematical Physics}, pages 1--56, 2023.

\bibitem{costin20}
O.~Costin and G.~Dunne.
\newblock Physical resurgent extrapolation.
\newblock {\em Physics Letters B}, 808:135627, 2020.

\bibitem{cotti23}
G.~Cotti, D.~Guzzetti, and D.~Masoero.
\newblock Asymptotic solutions for linear odes with not-necessarily meromorphic
  coefficients: a levinson type theorem on complex domains, and applications.
\newblock {\em arXiv preprint arXiv:2310.19739}, 2023.

\bibitem{cinese}
D.~De~Martino and D.~Masoero.
\newblock Asymptotic analysis of noisy fitness maximization, applied to
  metabolism \& growth.
\newblock {\em Journal of Statistical Mechanics: Theory and Experiment},
  2016(12):123502, 2016.

\bibitem{degano2024}
G.~Degano.
\newblock {ODE}/{IM} correspondence in the semiclassical limit: {L}arge degree
  asymptotics of the spectral determinants for the ground state potential.
\newblock arXiv:2409.07866, 2024.

\bibitem{delabaere97}
E.~Delabaere, H.~Dillinger, and F.~Pham.
\newblock Exact semiclassical expansions for one-dimensional quantum
  oscillators.
\newblock {\em Journal of Mathematical Physics}, 38(12):6126--6184, 1997.

\bibitem{dorey98}
P.~Dorey and R.~Tateo.
\newblock On the relation between {S}tokes multipliers and the {T}-{Q} systems
  of conformal field theory.
\newblock {\em Nuclear Phys. B}, 563(3):573--602, 1999.

\bibitem{ecalle81}
J.~{\'E}calle.
\newblock Les fonctions resurgentes, i, ii, iii.
\newblock {\em Publications Mathematiques d’Orsay. Paris}, 1985, 1981.

\bibitem{bateman2}
A.~Erd{\'e}lyi, W.~Magnus, F.~Oberhettinger, and F.~Tricomi.
\newblock {\em Higher transcendental functions. {V}ol. {II}}.
\newblock McGraw-Hill Book Company, Inc., New York-Toronto-London, 1953.

\bibitem{erdelyi-book-wkb}
A.~Erdélyi.
\newblock {\em Asymptotic expansions}.
\newblock Dover Publications, 1956.

\bibitem{eremenko18}
A.~Eremenko and A.~Gabrielov.
\newblock Pt-symmetric eigenvalues for homogeneous potentials.
\newblock {\em Journal of Mathematical Physics}, 59(5), 2018.

\bibitem{fedoryuk93}
M.~Fedoryuk.
\newblock {\em Asymptotic Analysis}.
\newblock Springer, 1993.

\bibitem{fock2006moduli}
V.~Fock and A.~Goncharov.
\newblock Moduli spaces of local systems and higher teichm{\"u}ller theory.
\newblock {\em Publications Math{\'e}matiques de l'IH{\'E}S}, 103:1--211, 2006.

\bibitem{frenkel2007langlands}
E.~Frenkel.
\newblock {\em Langlands correspondence for loop groups}, volume 103.
\newblock Cambridge University Press Cambridge, 2007.

\bibitem{hinch}
E.~J. Hinch.
\newblock {\em {P}erturbation {M}ethods}.
\newblock Cambridge University Press, 1991.

\bibitem{Ilya08}
Y.~Ilyashenko and S.~Yakovenko.
\newblock {\em Lectures on analytic differential equations}, volume~86 of {\em
  Graduate Studies in Mathematics}.
\newblock American Mathematical Society, Providence, RI, 2008.

\bibitem{iwaki2014}
K.~Iwaki and T.~Nakanishi.
\newblock Exact wkb analysis and cluster algebras.
\newblock {\em Journal of Physics A: Mathematical and Theoretical},
  47(47):474009, 2014.

\bibitem{kawai05}
T.~Kawai and Y.~Takei.
\newblock {\em Algebraic analysis of singular perturbation theory}, volume 227.
\newblock American Mathematical Soc., 2005.

\bibitem{kirwan}
F.~Kirwan.
\newblock {\em {C}omplex {A}lgebraic {C}urves}.
\newblock Cambridge University Press, 1992.

\bibitem{malliavin}
P.~Malliavin.
\newblock {\em {I}ntegration and {P}robability}.
\newblock Springer New York, NY, 1995.

\bibitem{masoero10th}
D.~Masoero.
\newblock {\em Essays on the Painlev{\'e} first equation and the cubic
  oscillator}.
\newblock PhD thesis, SISSA, 2010.

\bibitem{masoero2010poles}
D.~Masoero.
\newblock Poles of int{\'e}grale tritronqu{\'e}e and anharmonic oscillators. a
  wkb approach.
\newblock {\em Journal of Physics A: Mathematical and Theoretical},
  43(9):095201, 2010.

\bibitem{masoero10non}
D.~Masoero.
\newblock Poles of int{\'e}grale tritronqu{\'e}e and anharmonic oscillators.
  asymptotic localization from wkb analysis.
\newblock {\em Nonlinearity}, 23(10):2501, 2010.

\bibitem{dtba}
D.~Masoero.
\newblock Y-{S}ystem and {D}eformed {T}hermodynamic {B}ethe {A}nsatz.
\newblock {\em Lett. Math. Phys.}, 94(2):151--164, 2010.

\bibitem{masoero2024qfunctions}
D.~Masoero, E.~Mukhin, and A.~Raimondo.
\newblock {$Q$}-functions for lambda opers.
\newblock arXiv:2312.08842, 2024.

\bibitem{maro18}
D.~Masoero and P.~Roffelsen.
\newblock Poles of {P}ainlev{\'e} {IV} rationals and their distribution.
\newblock {\em SIGMA. Symmetry, Integrability and Geometry: Methods and
  Applications}, 14:002, 2018.

\bibitem{maro21}
D.~Masoero and P.~Roffelsen.
\newblock Roots of generalised hermite polynomials when both parameters are
  large.
\newblock {\em Nonlinearity}, 34(3):1663, 2021.

\bibitem{messiah14}
A.~Messiah.
\newblock {\em Quantum mechanics}.
\newblock Dover Publications, 2014.

\bibitem{olver97}
Frank Olver.
\newblock {\em Asymptotics and special functions}.
\newblock AK Peters/CRC Press, 1997.

\bibitem{ovsienko2009}
V.~Ovsienko and S.~Tabachnikov.
\newblock What is ... the schwarzian derivative?
\newblock {\em Notices of the American Mathematical Society}, 56(1):34--36,
  2009.

\bibitem{resi75}
M.~Reed and B.~Simon.
\newblock {\em Methods of modern mathematical physics. {II}. {F}ourier
  {A}nalysis, {S}elf-{A}djointness}.
\newblock Academic Press, New York, 1975.

\bibitem{sibuya75}
Y.~Sibuya.
\newblock {\em Global theory of a second order linear ordinary differential
  equation with a polynomial coefficient}.
\newblock Elsevier, 1975.

\bibitem{strebel}
K.~Strebel.
\newblock {\em {Q}uadratic {D}ifferentials}.
\newblock Springer Berlin, Heidelberg, 1984.

\bibitem{voros83}
A.~Voros.
\newblock The return of the quartic oscillator. the complex wkb method.
\newblock In {\em Annales de l'IHP Physique th{\'e}orique}, volume~39, pages
  211--338, 1983.

\end{thebibliography}
\end{document}